\documentclass[aps,groupedaddress, superscriptaddress,showpacs,amsmath,amssymb,twocolumn,pra]{revtex4-1}
\usepackage{bm}
\usepackage[dvips]{graphicx}
\usepackage{wrapfig,subfigure}
\usepackage{amssymb}
\usepackage{color}
\usepackage[normalem]{ulem} \usepackage{soul,xcolor} \setstcolor{black} \setul{}{2.4pt}
\usepackage{amsmath,bm}
\usepackage{epstopdf}
\usepackage{mathtools} 
\usepackage{extarrows} 
 \usepackage{bbm}
\usepackage{stmaryrd}
\usepackage{graphicx}
\usepackage{hhline}

\usepackage[pdfpagemode=UseNone,pdfstartview=FitH,colorlinks=true,linkcolor=blue,citecolor=blue,urlcolor=blue]{hyperref}
\usepackage[all]{hypcap}

\def\be{\begin{equation}}
\def\ee{\end{equation}}

\newcommand{\inum}{\iota}

\begin{document}
\title{Error correcting Bacon-Shor code with continuous measurement of noncommuting operators}
\author{Juan Atalaya}
\affiliation{Department of Chemistry, University of California, Berkeley, CA 94720, USA}
\author{Alexander N.\ Korotkov}
\affiliation{Google Inc., 340 Main Street, Venice, CA 90291, USA}
\affiliation{Department of Electrical and Computer Engineering, University of California, Riverside, CA 92521, USA}
\author{K. Birgitta Whaley}
\affiliation{Department of Chemistry, University of California, Berkeley, CA 94720, USA}
\date{\today}

\begin{abstract}
We analyze the continuous operation of the nine-qubit error correcting Bacon-Shor code with all noncommuting gauge operators measured at the same time. The error syndromes are continuously monitored using cross-correlations of sets of three measurement signals. We calculate the logical error rates due to $X$, $Y$ and $Z$ errors in the physical qubits and compare the continuous implementation with the discrete operation of the code. We find that both modes of operation exhibit similar performances when  the measurement strength from continuous measurements is sufficiently strong. We also estimate the value of the crossover error rate of the physical qubits, below which continuous error correction gives smaller logical error rates. Continuous operation has the advantage of passive monitoring of errors and avoids the need for additional circuits involving ancilla qubits. 
\end{abstract}
\maketitle

\section{Introduction}
Quantum error correction (QEC) is one of the most active research areas in the quantum computing field. Fault-tolerant quantum computing~\cite{Shor1995,Steane1996,Gottesman97,Gottesman2010,Nielsen-Chuang-Book,LidarBook} features QEC as an essential ingredient {to enable robust computation in noisy environments and to further} achieve the system size scalability that is necessary to show the quantum advantage over classical algorithms. Significant experimental efforts have been devoted to implement quantum error correcting codes in current quantum computer hardwares~\cite{Cory1998,Wineland2004,Schindler2011,Schoelkopf2012,Wrachtrup2014,Hanson2014,Nigg2014,Martinis2015,DiCarlo2015,Chow2014}. In particular, surface codes~\cite{Fowler2012,Raussendorf2007,Kitaev98,Terhal2015} have recently drawn considerable attention because of their comparatively high noise threshold,  {while Bacon-Shor codes~\cite{Bacon2006,Cross2007,Poulin2005} have attracted study both on account of a favorable noise threshold~\cite{Cross2007} and because,} regardless of the code distance, they only require  measurement of two-qubit operators on neighboring qubits~\cite{Monroe2017}. 
 
Continuous QEC has been theoretically investigated for a long time~\cite{PazZurek,Ahn2002,Ahn2003,Ahn2004,Sarovar2004,Sarovar2005,Brun2007,Landahl2008,Mabuchi2009,Brun2016,JustincQEC2019}.
{Recent work in this direction has focused on schemes in which} the error syndrome operators {of a QEC code that is defined for discrete error and recovery operations} are monitored in real time using continuous quantum measurements~\cite{KrausBook,Diosi1988,Molmer1992,Belavkin1992,CarmichaelBook,Milburn1993,Korotkov1999,Gambetta2008,Korotkov2011,Korotkov2016} instead of {the} projective measurements {that are} used in conventional QEC. Most previous works have focused on the  continuous operation of stabilizer quantum error correcting codes, where the measured operators commute with each other~\cite{Gottesman97}. {In contrast,} continuous operation of subsystem codes {such as} the Bacon-Shor codes is a relatively unexplored subject~\cite{Atalaya2017}. {Analysis of  subsystem codes} is complicated by the fact that the {measured operators} do not commute. Renewed interest in continuous QEC has been triggered by the rapid experimental progress 
{in continuous quantum measurement in the context of circuit QED} setups~\cite{Katz2006,Palacios-Laloy2010,Hatridge2013,Murch2013,Riste2013,Hacoen-Gourgy2016,Huard2018,Siddiqi2014} together with the realization of quantum feedback technologies~\cite{Siddiqi2012,DiCarlo2014} with superconducting qubits. {These therefore constitute  a promising testbed for implementation of continuous QEC.}

In this work we theoretically analyze the continuous operation of the nine-qubit Bacon-Shor code, which is the smallest quantum error correcting code from the family of Bacon-Shor codes~\cite{Bacon2006, Jiang2017}. We extend here  
{the} previous work  {of two of us} on the continuous operation of the four-qubit Bacon-Shor code~\cite{Atalaya2017}, which is the smallest error detecting code from such family of codes. 

\begin{figure}[t!]
\centering
\includegraphics[width=0.9\linewidth, trim =4.5cm 4.75cm 4cm 1.5cm,clip=true]{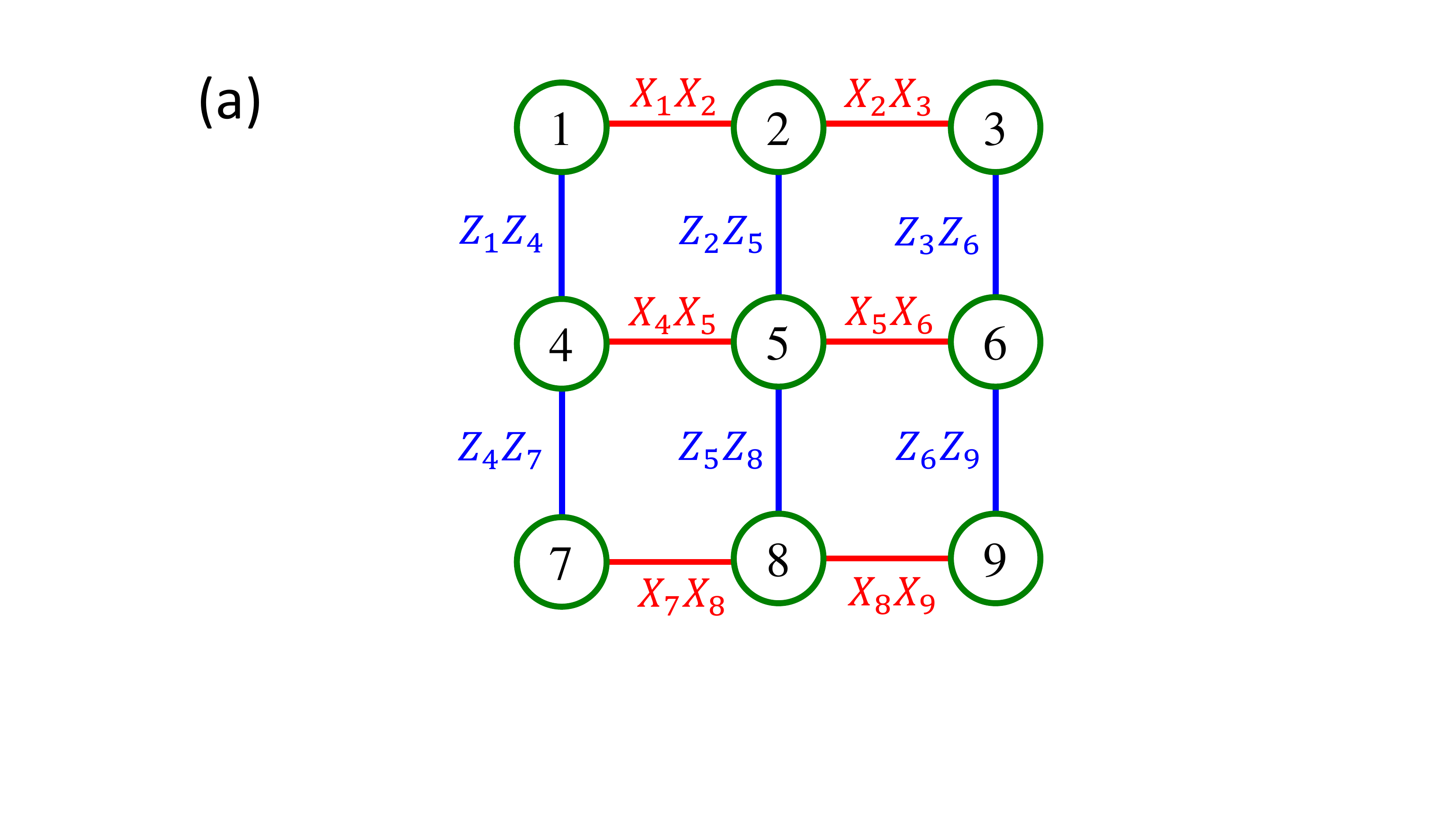}
\includegraphics[width=0.9\linewidth, trim =4.5cm 4.65cm 4cm 5cm,clip=true]{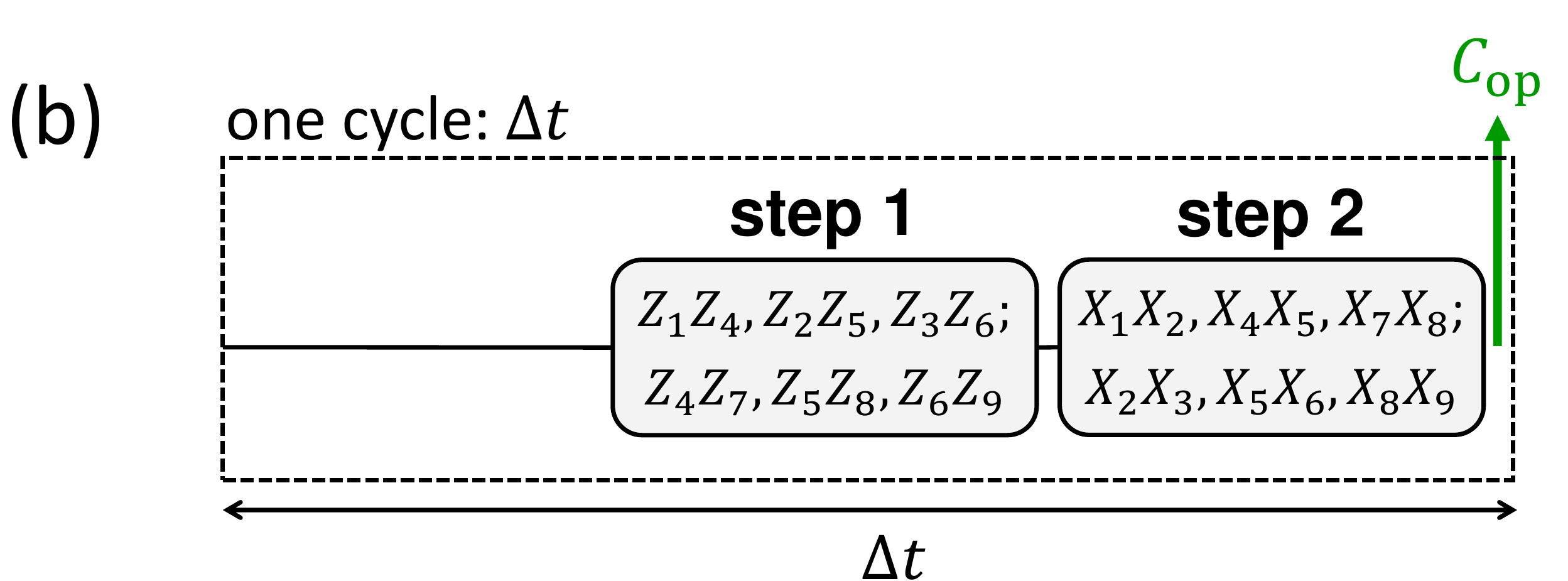}
\includegraphics[width=0.9\linewidth, trim =6.5cm 2cm 6.5cm 1.5cm,clip=true]{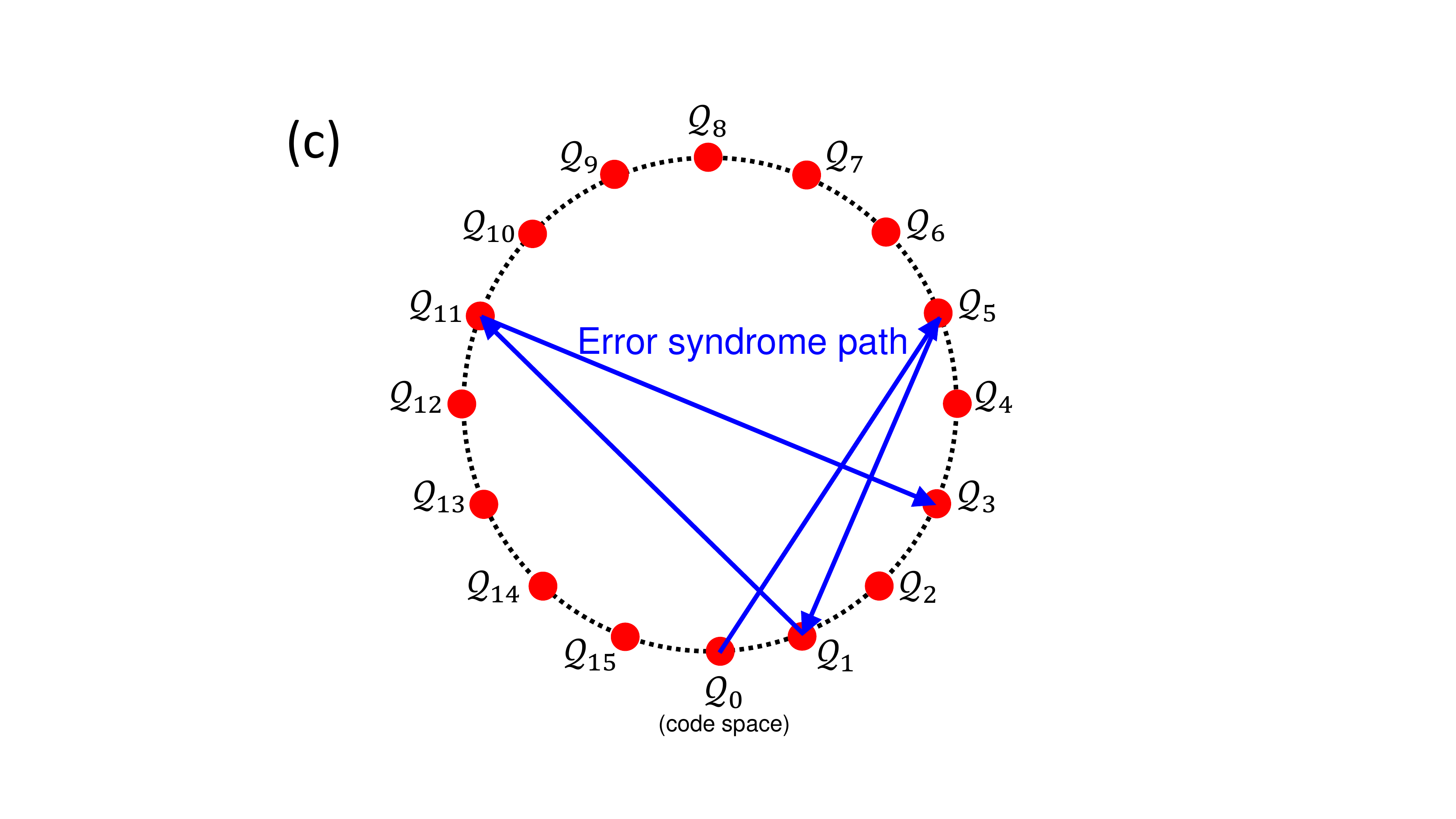}
\caption{(a) The nine-qubit Bacon-Shor code. The code operation is based on measurement of 12 gauge operators: $Z_1Z_4$, $Z_2Z_5$, $Z_3Z_6$, $Z_4Z_7$, $Z_5Z_8$, $Z_6Z_9$,  $X_1X_2$, $X_4X_5$, $X_7X_8$, $X_2X_3$, $X_5X_6$ and $X_8X_9$. Circles in this panel indicate the physical qubits. Panel (b) shows the conventional (discrete) operation of the code where the gauge operators are projectively measured in two steps. The cycle ends with the application of a discrete (instantaneous) error correcting operation $C_{\rm op}$ that depends on the error syndrome---see Table~\ref{table-I}. Cycle duration is $\Delta t$. Panel (c) shows the error syndrome path.  Arrows indicate state transitions between subspaces  ($\mathcal{Q}_\ell$) due to errors.} \vspace{-0.5cm}
\label{fig:fig1}
\end{figure}

The nine-qubit Bacon-Shor code encodes one logical qubit into nine physical qubits, which are conveniently arranged in a square lattice as shown in Fig.~\ref{fig:fig1}~(a). The error syndrome is defined in terms of the values of four stabilizer generators: $Z_1Z_4\,Z_2Z_5\,Z_3Z_6$, $Z_4Z_7\,Z_5Z_8\,Z_6Z_9$, $X_1X_2\,X_4X_5\,X_7X_8$ and $X_2X_3\,X_5X_6\,X_8X_9$. However, instead of directly measuring such multi-qubit Pauli operators, their values are obtained from the measurement of twelve noncommuting two-qubit operators (the so-called gauge operators): $Z_1Z_4$, $Z_2Z_5$, $Z_3Z_6$, $Z_4Z_7$, $Z_5Z_8$, $Z_6Z_9$, $X_1X_2$, $X_4X_5$, $X_7X_8$, $X_2X_3$, $X_5X_6$ and $X_8X_9$. In the conventional operation, the gauge operators are projectively measured in two sequential steps---see Fig.~\ref{fig:fig1}~(b), since  they do not commute.
The values of the stabilizer generators {and hence the error syndromes are then} obtained from the product of three discrete measurement outcomes (e.g., the value of $Z_1Z_4\,Z_2Z_5\,Z_3Z_6$ is obtained from the product of outcomes $\pm1$ of $Z_1Z_4$, $Z_2Z_5$ and $Z_3Z_6$). The {value of the} error syndrome determines the {specific} error correcting operation $C_{\rm op}$ that ought to be applied {to} a physical qubit at the end of each operation cycle. 

The main question we address in this paper is how to achieve continuous operation of the nine-qubit Bacon-Shor code, where all noncommuting gauge operators are continuously measured at the same time. The quantum backaction induced by such noncommuting measurements makes the nine-qubit state evolve diffusively {in the 512-dimensional Hilbert space}. A useful description is achieved by parameterizing the nine-qubit state in terms of probability amplitudes of one logical qubit and four effective qubits that we refer to as the gauge qubits~\cite{Terhal2015}. In this description, state diffusion of the full state can be seen as state diffusion of the gauge qubits due to simultaneous continuous measurement of twelve (effective) noncommuting operators. 
The gauge qubits dynamics plays an important role in the error analysis of the continuous operation of the nine-qubit Bacon-Shor code~\cite{Atalaya2017}. 
A {related} measurement-induced state evolution has been theoretically studied in Refs.~\cite{Korotkov2010,Atalaya2018b,Atalaya2018c} and recently observed in Ref.~\cite{Hacoen-Gourgy2016} for a single qubit subject to simultaneous continuous measurement of the noncommuting observables $\sigma_x$ and $\sigma_z$. 

In our continuous QEC protocol, stabilizer generators are monitored in real time using time-averaged cross-correlators of three measurement signals (e.g., $Z_1Z_4\,Z_2Z_5\,Z_3Z_6$ is continuously monitored via the triple correlator of the measurement signals from continuous measurement of $Z_1Z_4$, $Z_2Z_5$ and $Z_3Z_6$). Time averaging is necessary because the measurement signals are noisy and their product is even noisier~\cite{Atalaya2018a}. In our protocol, active correction of errors is only performed {at the end of}  the continuous operation and no  realizations are discarded. In the presence of errors, the system state jumps from the code space to {one of} the error subspaces, {or between error subspaces}. This evolution over {multiple} subspaces is characterized by the {\it error syndrome path}, which is shown to uniquely determine the errors, modulo {the action of} gauge operators. This error syndrome path is the central object in our continuous QEC protocol, see Fig.~\ref{fig:fig1}~(c).  We {track this path} using a simple two-error-threshold algorithm applied to the time-averaged cross-correlators. The {path} monitoring is, however, not perfect, since the cross-correlators  {are noisy and require time averaging}, {which slows down their response to errors}. The discrepancy between the actual and the monitored error syndrome paths leads to finite logical error rates, which {we} calculate both analytically and numerically. We also find the optimal {values of the four parameters} for {this} continuous QEC protocol, namely two integration time parameters and two error threshold parameters. 

{Our main conclusion is}
that continuous operation of the nine-qubit Bacon-Shor code is indeed possible and {that} its performance can be comparable to that of {the conventional QEC approach of discrete, projective measurements onto ancillas, followed by discrete state recovery operations {at each operation cycle}}. The main advantage of the continuous operation is the passive monitoring of errors, {with consequent avoidance of ancilla circuits}. We also determine the crossover value of the physical qubit error rate below which the error rate of the {corrected} logical qubit is smaller than {that} 
of the physical qubits. 

The {remainder of the paper} is organized as follows. In Section~\ref{sec:conventional-operation}, we briefly discuss the conventional operation of the nine-qubit Bacon-Shor code; we introduce the orthonormal bases for the code space and the error subspaces and derive formulas for the logical error rates that are later used to compare the conventional and continuous implementations. In Section~\ref{sec:continuous-operation}, we derive our main results for the continuous operation. We introduce the idea of the gauge and logical qubits in the code space and error subspaces. We present {a continuous quantum measurement}
model for the evolution of the gauge qubits and discuss how to account  for decoherence. We then present our continuous QEC protocol, which is based on continuous monitoring of the error syndrome path, and calculate the logical error rates for this protocol. In Section~\ref{cont-operation-optimization}, we find the optimal parameters of the continuous QEC protocol, {estimate the crossover error rate for the physical qubits, and} compare the performances of the continuous and conventional operations. {Section~\ref{conclusions} presents a discussion and conclusions.} 

\section{Nine-qubit Bacon-Shor code with projective measurements}
\label{sec:conventional-operation}
\subsection{System, code space, and discrete QEC protocol}
\label{subsec:discrete-qec-protocol}
The nine-qubit Bacon-Shor code encodes one logical qubit into nine physical qubits, labeled $1-9$ in Fig.~\ref{fig:fig1}~(a).
{The conventional discrete operation of the code} is based on projective measurement of two-qubit operators, {which are referred to as} gauge operators and {are} indicated in Fig.~\ref{fig:fig1}~(a) by the vertical and horizontal edges. {These gauge operators} are denoted  {by}
\begin{subequations}
\label{eq:Gauge-operators}
\begin{eqnarray}
Z_1Z_4&=&\, Z_{14}=\, G_1,\;\;\; Z_4Z_7=\, Z_{47}=\, G_{4},\nonumber\\
Z_2Z_5&=&\, Z_{25}=\,G_2 ,\;\;\;Z_5Z_8=\, Z_{58}=\, G_{5},\nonumber\\
Z_3Z_6&=&\, Z_{36}=\, G_3,\;\;\;Z_6Z_9=\, Z_{69}=\, G_{6},
\label{eq:Z-gauges}
\end{eqnarray}
{\rm and }
\begin{eqnarray}
X_1X_2&=&\, X_{12}=\, G_7,\;\;\;X_2X_3=\, X_{23}=\,G_{10} ,\nonumber \\
X_4X_5&=&\, X_{45}=\, G_8,\;\;\;X_5X_6=\, X_{56}=\, G_{11},\nonumber \\
X_7X_8&=&\, X_{78}=\, G_9,\;\;\;X_8X_9=\, X_{89}=\, G_{12},
\label{eq:X-gauges}
\end{eqnarray}
\end{subequations}
where $Z_j$ and $X_j$ are Pauli operators that act on the $j$th physical qubit. {For} example, $Z_1 = \sigma_z\varotimes \openone_{256}$, $X_1 = \sigma_x\varotimes \openone_{256}$, etc., $\sigma_z=|0\rangle\langle0| - |1\rangle\langle1|$ and {$\sigma_x = |0\rangle\langle1| + |1\rangle\langle0|$} are the conventional Pauli matrices, and $\openone_{256}$ is the $256\times256$ identity matrix. The two-qubit operators~\eqref{eq:Z-gauges} and~\eqref{eq:X-gauges} are referred to as the $Z$- and $X$-gauge operators, respectively. Since such groups of gauge operators do not commute {with each other}, they are sequentially measured in steps 1 and 2 {respectively, using} projective measurements---see Fig.~\ref{fig:fig1}~(b).  The projective measurements are assumed to be instantaneous. A discrete (instantaneous) error correcting operation $C_{\rm op}$ ({including the identity if no error is detected}) is then applied to a specific physical qubit {whose identity is determined by the joint values of all} {step-1 and step-2} measurement outcomes, which are $\pm1$ since the gauge operators are Pauli operators.  

The group generated by all gauge operators $G_k$ has an Abelian subgroup, {referred to as the 'stabilizer'}, with four generators 
{that commute with each other:}
\begin{eqnarray}
S_z^{(1)} &=&\, Z_{14}\,Z_{25}\, Z_{36}=\, G_1\,G_2\,G_3,\nonumber\\
S_z^{(2)} &=&\, Z_{47}\,Z_{58}\, Z_{69}=\, G_4\,G_5\,G_6, \nonumber\\
S_x^{(1)} &=&\, X_{12}\,X_{45}\,X_{78}=\, G_7\,G_8\,G_9,\nonumber \\
S_x^{(2)} &=&\, X_{23}\,X_{56}\,X_{89}=\, G_{10}\,G_{11}\,G_{12},
\label{eq:stabilizers}
\end{eqnarray}
This property allows us to divide the full Hilbert space into 16 {32-dimensional eigenspaces in which} the stabilizer generators have definite values. The values of $S_x^{(1)}$,  $S_z^{(1)}$, $S_x^{(2)}$ and $S_z^{(2)}$  determine the error syndrome pattern in each of these eigenspaces, 
{as summarized in Table~\ref{table-I}. We shall employ this ordering throughout the remainder of the paper.}  As usual, the eigenspace where all stabilizer generator eigenvalues are +1 is referred to as the code space, which is denoted by $\mathcal{Q}_0$. In the code space the product of the outcomes of $\{Z_{14}, Z_{25},Z_{36}\}$ is $+1$, and the same holds for the product of the outcomes of $\{Z_{47}, Z_{58}, Z_{69}\}$, $\{X_{12},X_{45}, X_{78}\}$, and $\{X_{23}, X_{56}, X_{89}\}$. If at least one of these products is $-1$, the system state is in one of the other 15 eigenspaces that are referred to as the error subspaces and {are} denoted by $\mathcal{Q}_{\ell}$ with $\ell=1$ to $15$. Note that all gauge operators commute with the stabilizer generators, so step-1 and step-2 measurements do not change the error syndrome pattern.
  
{We} now introduce the following orthonormal basis for the 32-dimensional code space $\mathcal{Q}_0$:
\begin{widetext}
\begin{eqnarray}
\label{eq:psi-basis}
|\phi_1\rangle&=&\left(|000\, 000\, 000\rangle     + |110\, 110\, 110\rangle + |101\, 101\, 101\rangle + |011\, 011\, 011\rangle\right)/2, \nonumber \\
|\phi_2\rangle&=&\left(|000\, 000\, 011\rangle     + |110\, 110\, 101\rangle + |101\, 101\, 110\rangle + |011\, 011\, 000\rangle\right)/2,\nonumber \\
|\phi_3\rangle&=&\left(|000\, 000\, 110\rangle     + |110\, 110\, 000\rangle + |101\, 101\, 011\rangle + |011\, 011\, 101\rangle\right)/2, \nonumber \\
|\phi_4\rangle&=&\left(|000\, 000\, 101\rangle     + |110\, 110\, 011\rangle + |101\, 101\, 000\rangle + |011\, 011\, 110\rangle\right)/2, \nonumber \\
|\phi_5\rangle&=&\left(|000\, 011\, 011\rangle     + |110\, 101\, 101\rangle + |101\, 110\, 110\rangle + |011\, 000\, 000\rangle\right)/2, \nonumber \\
|\phi_6\rangle&=&\left(|000\, 011\, 000\rangle     + |110\, 101\, 110\rangle + |101\, 110\, 101\rangle + |011\, 000\, 011\rangle\right)/2, \nonumber \\
|\phi_7\rangle&=&\left(|000\, 011\, 101\rangle     + |110\, 101\, 011\rangle + |101\, 110\, 000\rangle + |011\, 000\, 110\rangle\right)/2, \nonumber \\
|\phi_8\rangle&=&\left(|000\, 011\, 110\rangle     + |110\, 101\, 000\rangle + |101\, 110\, 011\rangle + |011\, 000\, 101\rangle\right)/2, \nonumber \\
|\phi_9\rangle&=&\left(|000\, 110\, 110\rangle     + |110\, 000\, 000\rangle + |101\, 011\, 011\rangle + |011\, 101\, 101\rangle\right)/2, \nonumber \\
|\phi_{10}\rangle&=&\left(|000\, 110\, 101\rangle + |110\, 000\, 011\rangle + |101\, 011\, 000\rangle + |011\, 101\, 110\rangle\right)/2, \nonumber \\
|\phi_{11}\rangle&=&\left(|000\, 110\, 000\rangle + |110\, 000\, 110\rangle + |101\, 011\, 101\rangle + |011\, 101\, 011\rangle\right)/2, \nonumber \\
|\phi_{12}\rangle&=&\left(|000\, 110\, 011\rangle + |110\, 000\, 101\rangle + |101\, 011\, 110\rangle + |011\, 101\, 000\rangle\right)/2, \nonumber \\
|\phi_{13}\rangle&=&\left(|000\, 101\, 101\rangle + |110\, 011\, 011\rangle + |101\, 000\, 000\rangle + |011\, 110\, 110\rangle\right)/2, \nonumber \\
|\phi_{14}\rangle&=&\left(|000\, 101\, 110\rangle + |110\, 011\, 000\rangle + |101\, 000\, 011\rangle + |011\, 110\, 101\rangle\right)/2, \nonumber \\
|\phi_{15}\rangle&=&\left(|000\, 101\, 011\rangle + |110\, 011\, 101\rangle + |101\, 000\, 110\rangle + |011\, 110\, 000\rangle\right)/2, \nonumber \\
|\phi_{16}\rangle&=&\left(|000\, 101\, 000\rangle + |110\, 011\, 110\rangle + |101\, 000\, 101\rangle + |011\, 110\, 011\rangle\right)/2, \nonumber \\ 
|\phi_{16+j}\rangle &=& X_1X_2X_3X_4X_5X_6X_7X_8X_9\,|\phi_j\rangle,\;\;\;\;j=1,2,...16. 
\end{eqnarray}
\end{widetext}
It is straightforward to check that each of the nine-qubit states~\eqref{eq:psi-basis} is an eigenstate of all four stabilizer generators with eigenvalue $+1$. The procedure to obtain the states in the computational basis {as written above} is described in Appendix~\ref{Appendix-A}. As we shall see below, analysis of the logical errors in the continuous operation is most conveniently performed in terms of the evolution of the system wavefunction due to errors and the continuous measurements. Therefore we need to specify a particular basis for the code and error subspaces. 

An orthonormal basis for each error subspace can be constructed from the  orthonormal basis vectors $|\phi_j\rangle$ of the code space. However, this construction is not unique. For instance, let us consider the orthonormal basis for $\mathcal{Q}_1$, where the error syndrome values are $S_x^{(1)}=S_z^{(1)}=S_x^{(2)}=1$ and $S_z^{(2)}=-1$ (see~Table~\ref{table-I}). Indeed, we can choose  orthonormal basis vectors for $\mathcal{Q}_1$ either as $X_7|\phi_j\rangle$ or $X_8|\phi_j\rangle$ or $X_9|\phi_j\rangle$  (with $j=1$ to $32$), since $X_7$, $X_8$ or $X_9$ anticommute with $S_z^{(2)}$ and commute with the other stabilizer generators. The reason for this freedom is that these orthonormal bases are equivalent  modulo a gauge operator or product of gauge operators. {For example},  the orthonormal basis vectors $X_7|\phi_j\rangle$ and $X_8|\phi_j\rangle$ are equivalent modulo $G_9=X_{7}X_{8}$. We will choose $Q_1|\phi_j\rangle$ with $Q_1=X_9$ as the orthonormal basis vectors for the error subspace $\mathcal{Q}_1$. There is also similar freedom in choosing the orthonormal basis for the other error subspaces. Our choice for the orthonormal bases used {in this work} for the error subspaces is specified in Table~\ref{table-I}. The ordering of the error syndrome in this table is not binary but set by the Pauli operators ($Q_{\ell}=X_9$, $Y_9$, $Z_9$, $X_1$, $X_9X_1$, etc.) that define the orthonormal basis vectors of subspaces $\mathcal{Q}_\ell$. 

\begin{table}[t]
\caption{\label{table-I} Error syndrome, orthonormal basis vectors and error correcting operations for the error subspaces $\mathcal{Q}_{\ell\neq0}$ (code space is denoted by $\mathcal{Q}_0$). The error syndrome is defined as the values of the stabilizer generators $S_x^{(1)}$, $S_z^{(1)}$, $S_x^{(2)}$  and $S_z^{(2)}$ (in this order) and the orthonormal basis vectors in the error subspaces are obtained by applying operators $Q_\ell$ to the basis vectors $|\phi_j\rangle$ of the code space, given in Eq.~\eqref{eq:psi-basis}. $Q_0=\openone$. } 
\begin{ruledtabular}
\begin{tabular}{c c c c c c c c}
&\hspace{-1.9cm}sub-& \hspace{-0.8cm}error & \hspace{-0.7cm}synd&\hspace{-0.95cm} rome & {basis vectors}& \hspace{0.25cm}error correcting&\\ 
space  & $\,S_x^{(1)}$ & $\,S_z^{(1)}$ & $\,S_x^{(2)}$ & $\,S_z^{(2)}$ & $\,{Q}_\ell|\phi_j\rangle\,$& \hspace{0.25cm} operation ($C_{\rm op}$)&\\ \hline 
$\mathcal{Q}_0$ & $+1$ & $+1$ & $+1$ & $+1$ & $|\phi_j\rangle$ & $\openone$ (identity) \\
$\mathcal{Q}_1$ & $+1$ & $+1$ & $+1$ & $-1$ & $X_9|\phi_j\rangle$ & $X_7,X_8$ or $X_9$ \\
$\mathcal{Q}_2$ & $+1$ & $+1$ & $-1$ & $-1$ & $Y_9|\phi_j\rangle$ & $Y_9$ \\
$\mathcal{Q}_3$ & $+1$ & $+1$ & $-1$ & $+1$ & $Z_9|\phi_j\rangle$ & $Z_3,Z_6$ or $Z_9$ \\

$\mathcal{Q}_4$ & $+1$ & $-1$ & $+1$ & $+1$ & $X_1|\phi_j\rangle$ & $X_1,X_2$ or $X_3$ \\
$\mathcal{Q}_5$ & $+1$ & $-1$ & $+1$ & $-1$ & $X_9X_1|\phi_j\rangle$ & $X_4,X_5$ or $X_6$ \\
$\mathcal{Q}_6$ & $+1$ & $-1$ & $-1$ & $-1$ & $Y_9X_1|\phi_j\rangle$ & $Y_6$ \\
$\mathcal{Q}_7$ & $+1$ & $-1$ & $-1$ & $+1$ & $Z_9X_1|\phi_j\rangle$ & $Y_3$ \\

$\mathcal{Q}_8$ & $-1$ & $-1$ & $+1$ & $+1$ & $Y_1|\phi_j\rangle$ & $Y_1$ \\
$\mathcal{Q}_9$ & $-1$ & $-1$ & $+1$ & $-1$ & $X_9Y_1|\phi_j\rangle$ & $Y_4$ \\
$\mathcal{Q}_{10}$ & $-1$ & $-1$ & $-1$ & $-1$ & $Y_9Y_1|\phi_j\rangle$ & $Y_5$ \\
$\mathcal{Q}_{11}$ & $-1$ & $-1$ & $-1$ & $+1$ & $Z_9Y_1|\phi_j\rangle$ & $Y_2$ \\

$\mathcal{Q}_{12}$ & $-1$ & $+1$ & $+1$ & $+1$ & $Z_1|\phi_j\rangle$ & $Z_1,Z_4$ or $Z_7$ \\
$\mathcal{Q}_{13}$ & $-1$ & $+1$ & $+1$ & $-1$ & $X_9Z_1|\phi_j\rangle$ & $Y_7$ \\ 
$\mathcal{Q}_{14}$ & $-1$ & $+1$ & $-1$ & $-1$ & $Y_9Z_1|\phi_j\rangle$ & $Y_8$ \\
$\mathcal{Q}_{15}$ & $-1$ & $+1$ & $-1$ & $+1$ & $Z_9Z_1|\phi_j\rangle$ & $Z_2,Z_5$ or $Z_8$ \\
\end{tabular}
\end{ruledtabular}
\end{table} 

The nine-qubit state is initially prepared in the code space at the beginning of the code operation. Then step-1 and step-2 measurements will not kick the state out of the code space and, in the absence of decoherence, the state will always remain in the code space. In this ideal situation, the measurement outcomes for $\{Z_{47},Z_{58},Z_{69}\}$ can be $\{+1,+1,+1\}$, $\{+1,-1,-1\}$, $\{-1,+1,-1\}$ or $\{-1,-1,+1\}$ (note that the product of the three numbers in each group is $+1$), and the same ``good'' outcomes can also be obtained for measurement of  $\{Z_{14}, Z_{25}, Z_{36}\}$. There are thus $4\times4=16$ ``good'' outcome configurations for step-1 measurements. The same outcomes ($\{+1,+1,+1\}$, $\{+1,-1,-1\}$, $\{-1,+1,-1\}$, $\{-1,-1,+1\}$) can also be obtained for measurements of $X_{12},X_{45}$ and $X_{78}$  as well as for $X_{23}, X_{56}$ and $X_{89}$, so there are also 16 ``good'' outcome configurations for step-2 measurements. 

If some of the values of the stabilizer generators are $-1$, the  conventional QEC protocol dictates that we apply {an} error correcting operation $C_{\rm op}$ at the end of the cycle [see Fig.~\ref{fig:fig1}~(b)]; the specific $C_{\rm op}$ depends on the error syndrome as indicated in Table~\ref{table-I}. After applying $C_{\rm op}$, the system state is returned to the code space; however, the logical state can be 
{degraded} if several errors happen within a cycle (as discussed in Section~\ref{Section:Operation-with-errors}).

\subsection{Operation without errors}
\label{sec:operation-w/o-errors}
In the absence of errors, step-1 measurements collapse the state to one of the following states (for simplicity of notation, we {write step-1} measurement results $\pm1$ as $\pm$)  
%
\begin{eqnarray}
\label{eq:Z-states}
|Z+++,+++\rangle &= {\alpha}|\phi_1\rangle + {\beta}|\phi_{17}\rangle, \nonumber\\
|Z+++,+--\rangle &= {\alpha}|\phi_2\rangle + {\beta}|\phi_{18}\rangle,\nonumber\\
|Z+++,--+\rangle &= {\alpha}|\phi_3\rangle + {\beta}|\phi_{19}\rangle,\nonumber\\
|Z+++,-+-\rangle &= {\alpha}|\phi_4\rangle + {\beta}|\phi_{20}\rangle,\nonumber\\
|Z+--,+++\rangle &= {\alpha}|\phi_5\rangle + {\beta}|\phi_{21}\rangle, \nonumber\\
|Z+--,+--\rangle &= {\alpha}|\phi_6\rangle + {\beta}|\phi_{22}\rangle,\nonumber\\
|Z+--,--+\rangle &= {\alpha}|\phi_7\rangle + {\beta}|\phi_{23}\rangle,\nonumber\\
|Z+--,-+-\rangle &= {\alpha}|\phi_8\rangle + {\beta}|\phi_{24}\rangle,\nonumber\\
|Z--+,+++\rangle &= {\alpha}|\phi_{9}\rangle + {\beta}|\phi_{25}\rangle, \nonumber\\
|Z--+,+--\rangle &= {\alpha}|\phi_{10}\rangle + {\beta}|\phi_{26}\rangle,\nonumber\\
|Z--+,--+\rangle &= {\alpha}|\phi_{11}\rangle + {\beta}|\phi_{27}\rangle,\nonumber\\
|Z--+,-+-\rangle &= {\alpha}|\phi_{12}\rangle + {\beta}|\phi_{28}\rangle,\nonumber\\
|Z-+-,+++\rangle &= {\alpha}|\phi_{13}\rangle + {\beta}|\phi_{29}\rangle, \nonumber\\
|Z-+-,+--\rangle &= {\alpha}|\phi_{14}\rangle + {\beta}|\phi_{30}\rangle,\nonumber\\
|Z-+-,--+\rangle &= {\alpha}|\phi_{15}\rangle + {\beta}|\phi_{31}\rangle,\nonumber\\
|Z-+-,-+-\rangle &= {\alpha}|\phi_{16}\rangle + {\beta}|\phi_{32}\rangle,
\end{eqnarray}
%
where $|Z\, g_1g_2g_3, g_4g_5g_6\rangle$ denotes the nominal step-1 collapse state that corresponds to the ``good'' outcome configuration $g_1,g_2,...g_6$ for the $Z$-gauge operators $G_1, ... G_6$, respectively. Each of these collapse states are parametrized by the complex-valued variables $\alpha$ and $\beta$ that represent the probability amplitudes to be in the zero ($|0_{\rm L}\rangle$) or one ($|1_{\rm L}\rangle$) logical states, respectively. The state of the {\it logical qubit} is defined as 
\begin{align}
\label{eq:logical-state}
|\Psi_{\rm L}\rangle = \alpha\, |0_{\rm L}\rangle + \beta\, |1_{\rm L}\rangle.
\end{align}
Similarly, step-2 measurements collapse the state to one of 16 possible states that are denoted by $|X\, g_7g_8g_9, g_{10}g_{11}g_{12}\rangle$ (with $g_7,g_8,... g_{12}$ being also a ``good'' outcome configuration). These nominal step-2 collapse states can be expressed as a linear combination of all 16 nominal step-1 collapse states of Eq.~\eqref{eq:Z-states} (and {\it vice versa}) with coefficients $\pm1/4$, so $|X\, g_7g_8g_9, g_{10}g_{11}g_{12}\rangle$ is parametrized by the {same} logical state $(\alpha,\beta)$. {Thus, the logical state is immune to measurement of the gauge operators}. The probability that any of the nominal step-2 collapse states occurs after step-1 measurements is $1/16$. In the absence of errors, no error correction is needed so $C_{\rm op}=\openone$. Then, step-1 measurements of the next cycle collapse the state $|X\, g_7g_8g_9, g_{10}g_{11}g_{12}\rangle$ to any of the states of Eq.~\eqref{eq:Z-states} with probability $1/16$, and so on. The real unitary matrix that relates the nominal step-1 and step-2 collapse states is given in Appendix~\ref{Appendix-A}.  

{Since we focus here} on the performance of the nine-qubit Bacon-Shor code against errors, we {may} assume that the encoding step (i.e., preparation of an initial wavefunction such as $|Z+++,+++\rangle$ with a given logical state $\alpha,\beta$) is perfect. At the end of the code operation there is also a decoding step to obtain the logical state from the code space; {this} decoding step is also assumed to be perfect. 

{\it The gauge qubits.} A general nine-qubit state $|\Psi_{\mathcal{Q}_0}\rangle$ in the code space can be written as 
\begin{align}
\hspace{-0.5cm}|\Psi_{\mathcal{Q}_0}\rangle =&\;c_{0000}\,|Z\!+\!++,\!+\!+\!+\rangle + c_{0001}\,|Z\!+\!++,\!+\!-\! -\rangle  \nonumber \\
&\hspace{-0.75cm}+ c_{0010}\,|Z\!+\!++,\!-\!-\!+\rangle  +  c_{0011}\,|Z\!+\!++,\!-\!+\!-\rangle \nonumber \\
&\hspace{-0.75cm}+c_{0100}\,|Z\!+\!--,\!+\!+\!+\rangle  +  c_{0101}\,|Z\!+\!--,\!+\!-\!-\rangle  \nonumber \\ &\hspace{-0.75cm}+c_{0110}\,|Z\!+\!--,\!-\!-\!+\rangle  +  c_{0111}\,|Z\!+\!--,\!-\!+\!-\rangle \nonumber \\
&\hspace{-0.75cm}+c_{1000}\,|Z\!-\!-+,\!+\!+\!+\rangle  + c_{1001}\,|Z\!-\!-+,\!+\!-\!-\rangle  \nonumber \\ &\hspace{-0.75cm}+c_{1010}\,|Z\!-\!-+,\!-\!-\!+\rangle  + c_{1011}\,|Z\!-\!-+,\!-\!+\!-\rangle  \nonumber \\
&\hspace{-0.75cm}+c_{1100}\,|Z\!-\!+-,\!+\!+\!+\rangle + c_{1101}\,|Z\!-\!+-,\!+\!-\!-\rangle \nonumber \\ &\hspace{-0.75cm}+c_{1110}\,|Z\!-\!+-,\!-\!-\!+\rangle + c_{1111}\,|Z\!-\!+-,\!-\!+\!-\rangle,  
\label{eq:Psi-state}
\end{align}
where the 16 coefficients $c_{q_1q_2q_3q_4}$ ($q_j=\{0,1\}$) together describe the state of {\it four gauge qubits}. The conventional operation of the nine-qubit Bacon-Shor code is characterized by the {\it discrete evolution} of the gauge qubits due to {projective} measurement of the gauge operators. Indeed, after step-1 measurements {($Z$-gauge operators)}, only one of the coefficients $c_{q_1q_2q_3q_4}$ is 1 and {all} others are 0, while after step-2 measurements {($X$-gauge operators)}, all  coefficients are non-zero and equal to $\pm1/4$. {However}, the logical state $(\alpha,\beta)$ is  not affected by the measurements. 

In {continuous operation of the code, this} discrete evolution of the gauge qubits state is replaced by diffusive evolution {as we describe in Section \ref{sec:continuous-operation} below.} 
\newline 
\subsection{Operation with errors}
\label{Section:Operation-with-errors}
While environmental decoherence in physical qubit systems is typically a gradual process, 
we can model it  as the average effect of discrete (instantaneous) $X$, $Y$ and $Z$ errors that occur at random times on the nine physical qubits. This is the jump/no-jump method~\cite{Nielsen-Chuang-Book,Preskill_notes} that we use to describe decoherence---see Section~\ref{Error-models} for a brief description of this method and how it is used to calculate the logical error rates.

In contrast to errors on physical qubits, logical $X$, $Y$ and $Z$ errors are operations on the logical state $(\alpha,\beta)$ that are defined as 
\begin{align}
\label{eq:logical-operators}
X_{\rm L} (\alpha,\beta) =&\, (\beta,\alpha), \;\;\; Z_{\rm L} (\alpha,\beta) = (\alpha,-\beta), \nonumber \\
 &\,Y_{\rm L} (\alpha,\beta) =\inum(\beta,-\alpha). 
\end{align}
Logical errors can only come from two or more physical errors happening in a faulty cycle, since all single-qubit errors are fully correctable after application of the error correcting operation $C_{\rm op}$ (the nine-qubit Bacon-Shor code is a full single-qubit quantum error correcting code~\cite{Bacon2006,Terhal2015,Cross2007}.) For sufficiently small occurrence rate of errors, logical errors are mainly due to two {physical qubit} errors; three errors are much less probable, and {higher order errors {are} increasing{ly} less likely}. We thus focus on two-qubit errors. {In this section we shall also }assume that errors occur at the same time. This assumption is allowed if we are only interested in changes of the logical state due to Pauli-type errors with trivial no-jump evolution. We also assume no errors occur {between the} measurement steps. 

Two-qubit errors can be of two types. {\it Harmless} two-qubit errors are those that leave the logical qubit state ($\alpha,\beta$) unperturbed after a faulty cycle, {although} the state of the {computationally} unimportant gauge qubits is {usually} affected. Examples of harmless two-qubit errors are the gauge operators. In contrast, {\it harmful} two-qubit errors are those that together with $C_{\rm op}$ create a logical error (the state of gauge qubits is {usually also affected in this case.}) That is, the two-qubit error combination $E_1$ and $E_2$ is harmful if 
\begin{align} 
E_1E_2C_{\rm op}\sim X_{\rm L},\, Y_{\rm L}\,{\rm or}\, Z_{\rm L}, 
\end{align}
where ``$\sim$'' indicates equivalence modulo gauge operators. In Appendix~\ref{Appendix-B} we explain how to obtain all the harmful two-qubit error combinations that are listed below in Eqs.~\eqref{eq:two-qubit-X-errors}--\eqref{eq:two-qubit-Y-errors}.

{\subsubsection{Logical X errors}}
There are 90 harmful two-qubit errors that lead to a logical $X$ error after a faulty cycle. {We list these below:}
\begin{widetext}
\begin{align}
\mathcal{Q}_1:&\; X_1X_4,\,X_1X_5,\,X_1X_6,\,X_2X_4,\, X_2X_5,\,X_2X_6,\, X_3X_4,\, 
X_3X_5,\, X_3X_6, \, Y_3Y_6,\, Y_1Y_4, \,Y_2Y_5, \nonumber \\
\mathcal{Q}_2:&\;  X_1Y_6,\, X_2Y_6,\, X_3Y_6, X_4Y_3, \,X_5Y_3,\, X_6Y_3,\,  \nonumber \\
\mathcal{Q}_4:&\; X_7X_4,\,X_7X_5,\,X_7X_6,\,X_8X_4,\, X_8X_5,\,X_8X_6,\, X_9X_4,\, 
X_9X_5,\, X_9X_6,\, Y_6Y_9,\, Y_4Y_7,\,Y_5Y_8, \nonumber \\
\mathcal{Q}_5:&\; X_1X_7,\,X_1X_8,\,X_1X_9,\,X_2X_7,\, X_2X_8,\,X_2X_9,\, X_3X_7,\, X_3X_8,\, X_3X_9, \, Y_3Y_9,\, Y_1Y_7, \,Y_2Y_8, \nonumber \\
\mathcal{Q}_6:&\; X_7Y_3,\, X_8Y_3,\, X_9Y_3, X_1Y_9, \,X_2Y_9,\, X_3Y_9,\,  \nonumber \\
\mathcal{Q}_7:&\; X_7Y_6,\, X_8Y_6,\, X_9Y_6, X_4Y_9, \,X_5Y_9,\, X_6Y_9,\,  \nonumber \\
\mathcal{Q}_8:&\; X_7Y_4,\, X_8Y_4,\, X_9Y_4, X_4Y_7, \,X_5Y_7,\, X_6Y_7,\,  \nonumber \\
\mathcal{Q}_{9}:&\; X_7Y_1,\, X_8Y_1,\, X_9Y_1, X_1Y_7, \,X_2Y_7,\, X_3Y_7,\,  \nonumber \\
\mathcal{Q}_{10}:&\; X_7Y_2,\, X_8Y_2,\, X_9Y_2, X_1Y_8, \,X_2Y_8,\, X_3Y_8,\,  \nonumber \\
\mathcal{Q}_{11}:&\; X_7Y_5,\, X_8Y_5,\, X_9Y_5, X_4Y_8, \,X_5Y_8,\, X_6Y_8,\,  \nonumber \\
\mathcal{Q}_{13}:&\; X_1Y_4,\, X_2Y_4,\, X_3Y_4, X_4Y_1, \,X_5Y_1,\, X_6Y_1,\, \nonumber \\
\mathcal{Q}_{14}:&\; X_1Y_5,\, X_2Y_5,\, X_3Y_5, X_4Y_2, \,X_5Y_2,\, X_6Y_2.
\label{eq:two-qubit-X-errors}
\end{align}
\end{widetext}
The top line in Eq.~\eqref{eq:two-qubit-X-errors} shows the two-qubit errors that map code space states $|\Psi_{\mathcal{Q}_0}\rangle$ [see Eq.~\eqref{eq:Psi-state}] to the error subspace $\mathcal{Q}_1$ (before applying $C_{\rm op}$), and the remaining lines show two-qubit errors that map $|\Psi_{\mathcal{Q}_0}\rangle$ to the error subspaces $\mathcal{Q}_2$, $\mathcal{Q}_4$, $\mathcal{Q}_5$, $\mathcal{Q}_6$, $\mathcal{Q}_7$, $\mathcal{Q}_8$, $\mathcal{Q}_9$, $\mathcal{Q}_{10}$, $\mathcal{Q}_{11}$, $\mathcal{Q}_{13}$ and $\mathcal{Q}_{14}$, respectively. (Note absence of harmful two-error combinations corresponding to subspaces $\mathcal{Q}_0$, $\mathcal{Q}_3$, $\mathcal{Q}_{12}$ and $\mathcal{Q}_{15}$.) {Establishing which subspaces are reached after two-qubit errors is important, since it allows one to determine the appropriate error} correcting operation $C_{\rm op}$ from Table~\ref{table-I}. After application of {the appropriate} $C_{\rm op}$, the system state is returned to the code space. However, {in all these cases} the logical  state suffers from a logical $X$ error (i.e., $\alpha$ and $\beta$ are  exchanged). Note that there are no $X_iZ_{i'}$ combinations in the list~\eqref{eq:two-qubit-X-errors}, since {these} are equivalent to $Y$ errors (modulo gauge operators), which are correctable {and thus harmless, in our categorization}. {The combinations $Z_iZ_{i'}$ and $Z_iY_{i'}$ can only lead to logical $Z$ errors, see below.} 

{\subsubsection{Logical Z errors}}
There are also 90 harmful two-qubit errors that lead to a logical $Z$ error after a faulty cycle.  {These} can be obtained from list~\eqref{eq:two-qubit-X-errors} by applying exchange of $X\leftrightarrow Z$, as well as exchange of  the qubit indices $ 2\leftrightarrow4, 3\leftrightarrow7$ and $ 6\leftrightarrow8$. {We note that these exchanges are possible because of the symmetry properties of the nine-qubit Bacon-Shor code, specifically, the $X-Z$ symmetry and the square symmetry  of the qubit layout (i.e., reflexion in the main diagonal of square of Fig.~\ref{fig:fig1}-(a)).} 
\begin{align}
{\rm Logical}\; Z \; {\text{error:}}&\; {\rm list~\eqref{eq:two-qubit-X-errors}}\; {\rm with}\; {\rm exchanges}\;  X\leftrightarrow Z, \nonumber \\
&\; 2\leftrightarrow4, 3\leftrightarrow7\; {\rm and} \; 6\leftrightarrow8.
\label{eq:two-qubit-Z-errors}
\end{align}
The error combinations of the lines of list~\eqref{eq:two-qubit-X-errors} with the changes indicated in Eq.~\eqref{eq:two-qubit-Z-errors} now provide the two-qubit errors that map $|\Psi_{\mathcal{Q}_0}\rangle$ to the error subspaces $\mathcal{Q}_3$, $\mathcal{Q}_2$, $\mathcal{Q}_{12}$, $\mathcal{Q}_{15}$, $\mathcal{Q}_{14}$, $\mathcal{Q}_{13}$, $\mathcal{Q}_{8}$, $\mathcal{Q}_{11}$, $\mathcal{Q}_{10}$, $\mathcal{Q}_{9}$, $\mathcal{Q}_{7}$, and $\mathcal{Q}_{6}$, respectively.


{\subsubsection{Logical Y errors}}
We list below the 18 harmful two-qubit errors that lead to a logical $Y$ error after a faulty cycle:
\begin{align}
\mathcal{Q}_2: &\; Y_1Y_5,\, Y_2Y_4, \nonumber \\
\mathcal{Q}_6: &\;  Y_1Y_8,\, Y_2Y_7, \nonumber \\
\mathcal{Q}_7: &\; Y_4Y_8,\, Y_5Y_7, \nonumber \\
\mathcal{Q}_8: &\; Y_5Y_9,\, Y_6Y_8, \nonumber \\
\mathcal{Q}_9: &\; Y_2Y_9,\, Y_3Y_8, \nonumber \\
\mathcal{Q}_{10}:  &\; Y_1Y_9,\, Y_3Y_7 ,\nonumber \\
\mathcal{Q}_{11}: &\; Y_4Y_9,\, Y_6Y_7 ,\nonumber \\
\mathcal{Q}_{13}: &\; Y_2Y_6,\, Y_3Y_5 ,\nonumber \\
\mathcal{Q}_{14}: &\; Y_1Y_6,\, Y_3Y_4.
\label{eq:two-qubit-Y-errors}
\end{align}
The lines of list~\eqref{eq:two-qubit-Y-errors}  show the two-qubit errors that map $|\Psi_{\mathcal{Q}_0}\rangle$ to the error subspaces $\mathcal{Q}_2$, $\mathcal{Q}_6$, $\mathcal{Q}_7$, $\mathcal{Q}_8$, $\mathcal{Q}_9$, $\mathcal{Q}_{10}$, $\mathcal{Q}_{11}$, $\mathcal{Q}_{13}$ and $\mathcal{Q}_{14}$, respectively. 

\subsection{Logical error rates}
\label{Error-models}
{We now} calculate the logical error rates for the nine-qubit Bacon-Shor code operating {under} projective measurements. 

We assume that the nine qubits are subject to 27 uncorrelated Markovian errors of $X$, $Y$ and $Z$ type, with occurrence rates $\Gamma_i^{(X)}$, $\Gamma_i^{(Y)}$, $\Gamma_i^{(Z)}$, {respectively}, where the index $i$ denotes the physical qubit. We also assume that $\Gamma_i^{(X,Y,Z)}\Delta t\ll 1$, so {that} single-qubit errors are {the} most probable, followed by two-qubit errors, three-qubit errors, {etc.} {The time duration for a full cycle of measurements is $\Delta t$---see Fig.~\ref{fig:fig1}~(b).} 

In the jump/no-jump method, the actual system evolution, {which is} characterized by  a density matrix $\rho(t)$, is replaced by an ensemble of wavefunction  trajectories $|\psi(t)\rangle$ that are conditional on the error-event realizations. The ensemble average of {these trajectories} $|\psi(t)\rangle\langle \psi(t)|$ produces the mixed state $\rho(t)$ that describes the actual {decohering} evolution, according to the standard Lindblad equation 
\begin{align}\label{eq:rho-ens}
\dot{\rho} =&\; \sum_{i,E}\Gamma_i^{(E)} \mathcal{L}[E_i]\rho,\nonumber \\
\mathcal{L}[A]\rho \equiv&\; A\rho A^\dagger  - \frac{1}{2}(A^\dagger A\rho + \rho A^\dagger A), 
\end{align}
where $E_i$ is the Kraus operator associated with {error of type} $E$ acting on the $i$th qubit. 
 At each infinitesimal  timestep $\delta t$, the wavefunction $|\psi(t)\rangle$ can exhibit a jump that changes it to {the value} $|\psi(t+\delta t)\rangle =E_i|\psi(t)\rangle/\mathcal{N}_{\rm j}$,  where $\mathcal{N}_{\rm j}$ is a normalization factor (for Pauli errors $\mathcal{N}_{\rm j}=1$). The probability of { a jump occurring} in each $\delta t$ is given by $\big(\Gamma_i^{(E)}\delta t\big)\,\langle \psi(t)| E_i^\dagger E_i|\psi(t)\rangle$.  In the case of no jump, the wavefunction changes to $\big(\openone -\sum_{i,E}E^\dagger_i E_i\delta t/2\big)|\psi(t)\rangle/\mathcal{N}_{\rm nj}$, {with normalization factor} $\mathcal{N}_{\rm nj}$. In the particular case where the Kraus operators $E_i$ are Pauli operators ({i.e.}, $E_i^\dagger E_i=\openone$), no-jump evolution is trivial (i.e., no evolution) {while} the jump probability is $\Gamma_i^{(E)}\delta t$, which is state-independent. In this paper we consider decoherence due to {all possible single Pauli errors, {i.e.,}} $X,Y$ or $Z$ errors ($E_i=\{X_i,Y_i,Z_i\}$). 

For {a} sufficiently small occurrence rate of errors, the probability of a logical error after $M$ operation cycles  is equal to  $\sum_{\{E_i,E'_{i'}\}} \left(T_{\rm op}\Gamma_{i}^{(E)}\right)\left(\Gamma_{i'}^{(E')}\,\Delta t\right)$, where $T_{\rm op}=M\Delta t$ is the operation duration and the sum is over all harmful two-qubit errors $E_iE'_{i'}$  [see Eqs.~\eqref{eq:two-qubit-X-errors}--\eqref{eq:two-qubit-Y-errors}] that lead to a logical $X$, $Z$ or $Y$ error. We {then} obtain the discrete-operation logical error rate $\gamma_{L}^{{\rm disc}}$ by dividing this probability by $T_{\rm op}$ {($L=X,Y$ or $Z$)}:
\begin{align}
\label{eq:Gamma-logical-discrete}
\gamma_{L}^{{\rm disc}} = \sum_{\{E_i,E'_{i'}\}} \Gamma_{i}^{(E)}\Gamma_{i'}^{(E')}\,\Delta t.
\end{align}

Since we have in total 198 harmful two-qubit errors {[Eqs.~\eqref{eq:two-qubit-X-errors}--\eqref{eq:two-qubit-Y-errors}], we shall for simplicity evaluate the logical error rate formula~\eqref{eq:Gamma-logical-discrete} for the depolarizing channel, for which all three Pauli error rates are equal:}
\begin{align}
\label{eq:Gi-depolarization}
\Gamma_i^{(X)}=\Gamma_i^{(Y)}=\Gamma_i^{(Z)}=\frac{\Gamma_{\rm d}}{3},
\end{align}
{with $\Gamma_{\rm d}$ the depolarization error rate, which we assume to be the same for all qubits.}  {Taking all of the error channels in Eqs.~\eqref{eq:two-qubit-X-errors}--\eqref{eq:two-qubit-Y-errors} into account, we find that} the logical $X$, $Z$ and $Y$ error rates for the depolarizing channel are {given by}
\begin{align}
\gamma_X^{{\rm disc}} = \gamma_Z^{{\rm disc}} = 10\,\Gamma_{\rm d}^2\Delta t \;\; {\rm and} \;\; \gamma_Y^{{\rm disc}} = \frac{\gamma_X^{{\rm disc}}}{5}=2\,\Gamma_{\rm d}^2\Delta t,
\label{eq:discrete-logical-error-rate}
\end{align}
respectively. The total logical error rate is {then} equal to 
\begin{align}
\label{eq:GL-total-discrete}
\gamma_{\rm disc} = \gamma_X^{{\rm disc}}+ \gamma_Y^{{\rm disc}} + \gamma_Z^{{\rm disc}} =22\,\Gamma_{\rm d}^2\Delta t. 
\end{align}

The full formulae for the logical $X$, $Y$ and $Z$ error rates {in the case of non-equivalent qubits and a general asymmetric error channel} are given in  Appendix~\ref{Appendix-C}. 

\begin{figure}[t!]
\centering
\includegraphics[width=0.9\linewidth, trim =1.5cm 1.7cm 0.4cm 1cm,clip=true]{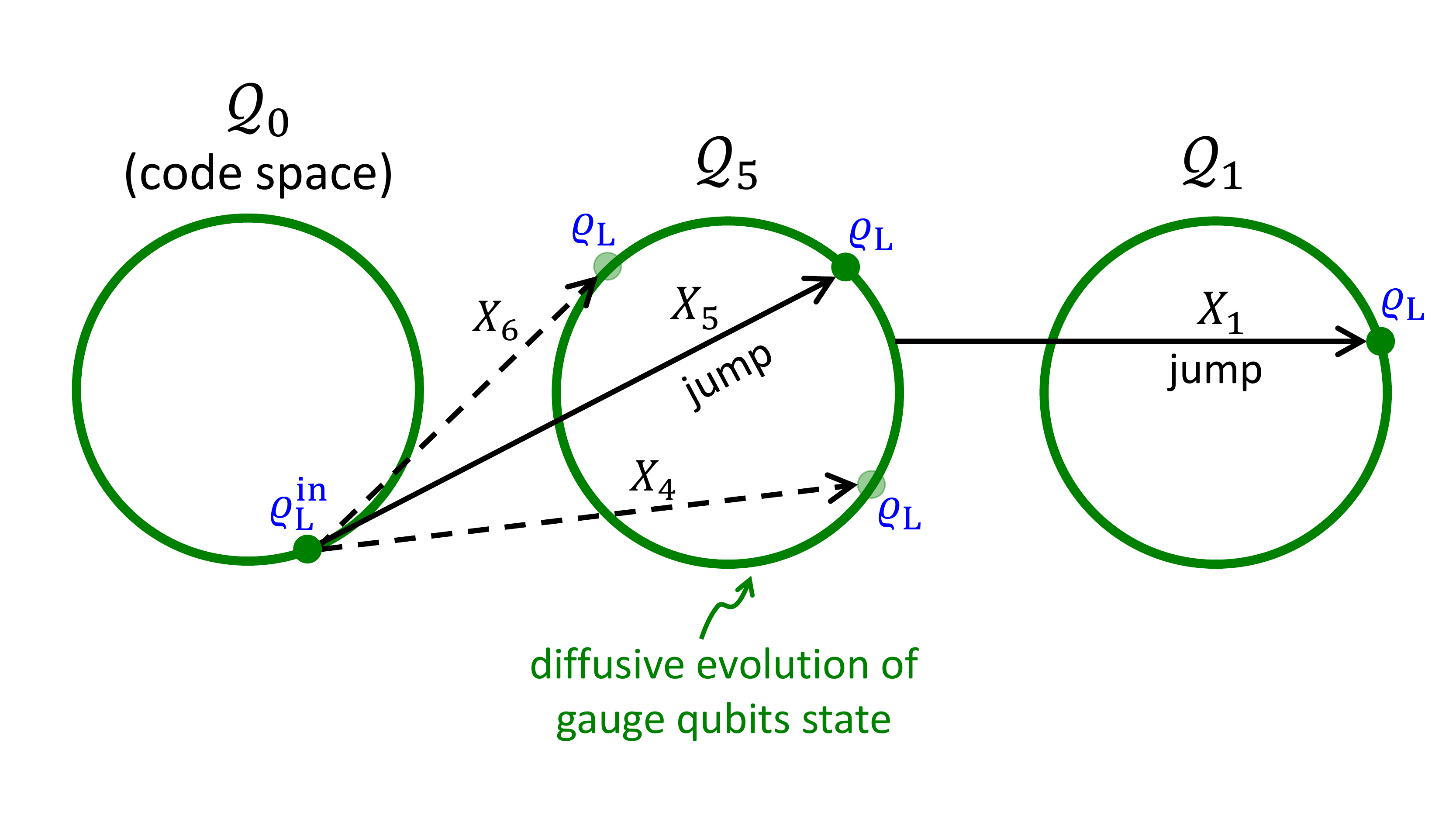}
\caption{Illustration of the state evolution due to continuous measurement and $X_5$ and $X_1$ errors. Green circles represent 16-dimensional spheres where diffusive evolution of the state of the four gauge qubits  takes place. Before the first error ($X_5$) occurs, the gauge qubits diffusively evolve  in the code space $\mathcal{Q}_0$ while the logical state, denoted by $\varrho_{\rm L}^{\rm in}$, is unaffected by the continuous measurement. Immediately after the first error occurs, the system state is in the error subspace $\mathcal{Q}_5$ with a new logical state ($\varrho_{\rm L}=X_{\rm L}\varrho_{\rm L}^{\rm in}X_{\rm L}$) and a new state for the gauge qubits, symbolically indicated by the solid dot on the circle of $\mathcal{Q}_5$. Note that the errors $X_4$ and $X_6$  are equivalent to $X_5$ up to $X$-gauge operators so they map the system state to the same error subspace ($\mathcal{Q}_5$) with the same logical state ($\varrho_{\rm L}$) but a different gauge qubits state; this is indicated by the dashed arrows pointing at different points on the circle of $\mathcal{Q}_5$. During the time between the errors, there is diffusive evolution of the gauge qubits in $\mathcal{Q}_5$ due to measurement while the logical state is constant. The second error ($X_1$) makes the system state jump from $\mathcal{Q}_5$ to $\mathcal{Q}_1$, and, according to Table~\ref{table-II}, this error does not affect the logical state nor the gauge qubits state, so the logical state in $\mathcal{Q}_1$ is also $\varrho_{\rm L}$. The logical error in the code operation occurs if we misidentify two jumps $\mathcal{Q}_0\to \mathcal{Q}_5\to\mathcal{Q}_1$ with a single jump $\mathcal{Q}_0\to \mathcal{Q}_{1}$.}
\label{fig:overview}
\end{figure}

\section{Nine-qubit Bacon-Shor code with continuous measurements}
\label{sec:continuous-operation}
{We now} analyze the continuous operation of the nine-qubit Bacon-Shor code, {in which the} sequential projective measurements of gauge operators are replaced by simultaneous continuous quantum measurements. Before we present our {model of the continuous quantum measurement}, we first qualitatively discuss the evolution of the system state due to {continuous measurement combined with physical qubit errors}.

Continuous measurement of the noncommuting gauge operators changes   the state of the gauge qubits {in a diffusive fashion.  This is characterized by the time evolution of} the $c$-coefficients of Eq.~\eqref{eq:Psi-state} if the system state is in the code space, or, more generally, by the time evolution of the $c$-coefficients of Eq.~\eqref{eq:Psi-state-v2} if the system state is in the subspace $\mathcal{Q}_\ell$. A somewhat similar  diffusive state evolution was discussed in Ref.~\cite{Atalaya2017}, where continuous operation of the error detecting four-qubit Bacon-Shor code (with one logical qubit and one gauge qubit) was analyzed. It turns out that these $c$-coefficients can be regarded as real if they are initially set to real values. We can then think of the measurement-induced diffusive evolution of the gauge qubits state as {taking place 
on} a 16-dimensional sphere (one for each subspace $\mathcal{Q}_\ell$).  {These are represented in Fig.~\ref{fig:overview} by green circles; points on these circles denote different states of the gauge qubits and also carry a label ($\varrho_{\rm L}$) that denotes the logical state{, which is not affected by the continuous measurements.} In our model, the diffusive state evolution of the gauge qubits will be given {by a} stochastic master equation that is described in the following subsection below.}

Let us now {consider} the effect of errors. Errors cause jumps of the system state between two subspaces $\mathcal{Q}_{\ell_{\rm in}}$ and $\mathcal{Q}_{\ell}$. {In general, the logical state is different immediately before the jump and immediately after the jump. The same holds for the state of the gauge qubits (any change of the logical state is {not harmful} as long as we know what this change is).} Table~\ref{table-II} {shows} how errors affect the state of the logical and gauge qubits. In this Table we express all 27 possible errors as a product of three operations.  For instance, the error $X_{5}$ is equivalent (up to a phase factor) to $Q_5 X_{\rm L}X_{14}^{\rm g}$, so this error changes the initial subspace (depending on the operator $Q_5$), also applies a logical $X$ operation ($\alpha$, $\beta$ are exchanged), and also changes the $c$-coefficients as follows:  $c_{q_1q_2q_3q_4}\to X_4^{\rm g}X_1^{\rm g}c_{q_1q_2q_3q_4}$, where $X_j^{\rm g}$ is an effective $X$ operation on the $j$th gauge qubit---see Eq.~\eqref{eq:Xj^g-def}. 
{Table~\ref{table-III} provides the multiplication table for operators $Q_{\ell}$ that is necessary to find out which subspace the system state is mapped to after an error. }

Tables~\ref{table-II} and~\ref{table-III} are then used to figure out {to which} subspace ($\mathcal{Q}_{\ell}$) the system state jumps, after an error. For {the example above}, after the $X_5$ error, the system wavefunction can be spanned in the orthonormal basis $Q_\ell|\phi_j\rangle$ of the subspace $\mathcal{Q}_\ell$, where the operator ${Q}_\ell = Q_5\times Q_{\ell_{\rm in}}$ is obtained from Table~\ref{table-III} and the operator $Q_{\ell_{\rm in}}$ defines the orthonormal basis of the initial subspace $\mathcal{Q}_{\ell_{\rm in}}$ (see Table~\ref{table-I}). Assuming {that the system state is initially in the code space (so that} $Q_{\ell_{\rm in}}=Q_0=\openone$ and ${Q}_\ell={Q}_5$), the system state {jumps from $\mathcal{Q}_0$ to $\mathcal{Q}_5$, as illustrated} in Fig.~\ref{fig:overview}. (We point out that $Q_\ell$ operators play two roles in the continuous operation analysis, {namely i) determination of transitions between subspaces due to errors, and ii) construction} of basis transformations from the code space basis to the error subspace bases.) Note that, for Pauli-type errors, the no-jump evolution is trivial, so after a jump only the gauge qubits {will evolve} due to the measurement until the next jump, and so on. {Note further that overall phase factors in the system wavefunction may be introduced by the errors; however, for Pauli-type errors we can disregard these, since phase factors affect neither the jump/no-jump probabilities nor the temporal correlations of the measurement signals.} The $c$-coefficients can then be regarded as real even in the presence of errors. Figure~\ref{fig:overview} {illustrates} the effect of two errors $X_5$ and $X_1$ on {a system state that is set initially} in the code space ($\mathcal{Q}_0$).   
\newline

\subsection{Evolution due to measurement and errors}
\label{evolution}
We first discuss the measurement-induced state dynamics in the code space {and} in the error subspaces, and then {analyze} the effect of errors using the jump/no-jump method. Analysis of the state dynamics in the error subspaces is important because our continuous QEC protocol does not correct errors at the moment when they are detected; instead, the correction is at the end of the code operation, {as described in Section~\ref{section:QEC-algorithm}}. 

In {both the code and error subspaces} the system wavefunction can be parametrized as in Eq.~\eqref{eq:Psi-state}. {We rewrite this here as}
\begin{widetext}
\begin{align}
&|\Psi_{\mathcal{Q}_\ell}(t)\rangle = \nonumber \\
&\;c_{0000}(t)\,Q_\ell|Z\!+\!++,\!+\!+\!+\rangle + c_{0001}(t)\,Q_\ell|Z\!+\!++,\!+\!-\!-\rangle  + c_{0010}(t)\,Q_\ell|Z\!+\!++,\!-\!-\!+\rangle  + c_{0011}(t)\,Q_\ell|Z\!+\!++,\!-\!+\!-\rangle +  \nonumber \\
&\;c_{0100}(t)\,Q_\ell|Z\!+\!--,\!+\!+\!+\rangle  +  c_{0101}(t)\,Q_\ell|Z\!+\!--,\!+\!-\!-\rangle  +  c_{0110}(t)\,Q_\ell|Z\!+\!--,\!-\!-\!+\rangle  +  c_{0111}(t)\,Q_\ell|Z\!+\!--,\!-\!+\!-\rangle + \nonumber \\
&\;c_{1000}(t)\,Q_\ell|Z\!-\!-+,\!+\!+\!+\rangle  + c_{1001}(t)\,Q_\ell|Z\!-\!-+,\!+\!-\!-\rangle  + c_{1010}(t)\,Q_\ell|Z\!-\!-+,\!-\!-\!+\rangle  + c_{1011}(t)\,Q_\ell|Z\!-\!-+,\!-\!+\!-\rangle + \nonumber \\
&\;c_{1100}(t)Q_\ell|Z\!-\!+-,\!+\!+\!+\rangle + c_{1101}(t)\,Q_\ell|Z\!-\!+-,\!+\!-\!-\rangle + c_{1110}(t)\,Q_\ell|Z\!-\!+-,\!-\!-\!+\rangle + c_{1111}(t)\,Q_\ell|Z\!-\!+-,\!-\!+\!-\rangle,  
\label{eq:Psi-state-v2}
\end{align}
\end{widetext}
where the {additional} operator $Q_\ell$ {now} specifies that we are dealing with a wavefunction that belongs to the subspace $\mathcal{Q}_\ell$---see Table~\ref{table-I}. The  quantum backaction {(see discussion below)} from the simultaneous continuous measurement of the noncommuting gauge operators makes the coefficients $c_{q_1q_2q_3q_4}(t)$ change in a diffusive fashion, while the logical state $(\alpha,\beta)$ that parametrizes $|Z\!\pm\pm\pm,\pm\pm\pm\rangle$ is unchanged by the continuous measurements. 

The $512\times512$ density matrix $\rho(t)$ that corresponds to the wavefunction Eq.~\eqref{eq:Psi-state-v2} has only one nontrivial  $32\times32$ diagonal submatrix $\varrho(t)$: this can be written in a direct-product form 
\begin{align}
\label{eq:direct-prod-form}
\varrho(t) =\varrho_{\rm L} \varotimes \varrho_{\rm g}(t) = \left(\begin{array}{cc} |\alpha^2|\,\varrho_{\rm g}(t)&  \alpha\beta^*\,\varrho_{\rm g}(t)\\ \alpha^*\beta\, \varrho_{\rm g}(t) &|\beta^2|\,\varrho_{\rm g}(t)\end{array} \right), 
\end{align}
where $\varrho_{\rm L}$ represents the logical density matrix 
\begin{align}
\label{eq:rhoL}
\varrho_{\rm L} = \left( \begin{array}{c c} |\alpha^2| & \alpha\beta^* \\ \alpha^*\beta & |\beta^2|\end{array} \right),
\end{align}
and $\varrho_{\rm g}(t)$ represents the $16\times16$ density matrix for the four gauge qubits, with matrix elements given by 
\begin{align}
\label{eq:rho-gauge}
{\varrho_{\rm g}}(t)_{q_1q_2q_3q_4,\,q_1' q_2' q_3' q_4'} =&\; c_{q_1q_2q_3q_4}(t)\,c^*_{q_1'q_2'q_3'q_4'}(t). 
\end{align}

Since the gauge operators $G_k$ do not cause transitions between the subspaces $\mathcal{Q}_\ell$, they must have a block-diagonal matrix representation {over these spaces}. In the orthonormal basis of $\mathcal{Q}_\ell$ {that is} given in Table~\ref{table-I},  the $32\times32$  diagonal submatrices of $G_k$ ($k=1,2,...12$) that correspond to the subspaces $\mathcal{Q}_\ell$ can be written as 
\begin{align}
\label{eq:Gk-Ql}
\left[G_k\right]_{\mathcal{Q}_\ell} =\zeta_k^{(\ell)}\left(\begin{array}{cccc} \mathcal{G}_k & 0\\0 & \mathcal{G}_k \end{array}\right),
\end{align}
where $\zeta_k^{(\ell)}=-1$  ($\zeta_k^{(\ell)}=1$) if $G_k$ anticommutes (commutes) with ${Q}_\ell$ {as} given in Table~\ref{table-I}. In the code space we have $\zeta_k^{(0)}=1$  for all $k$ since $Q_0=\openone$. The $16\times16$ matrices $\mathcal{G}_k$ in Eq.~\eqref{eq:Gk-Ql} read as  
\begin{align}
\label{eq:curli-G_k}
\mathcal{G}_1=&\,{Z}^{\rm g}_{1},\;\; \mathcal{G}_2={Z}^{\rm g}_{12},\;\; \mathcal{G}_3={Z}^{\rm g}_2, \nonumber \\
\mathcal{G}_4=&\,{Z}^{\rm g}_3,\;\; \mathcal{G}_5={Z}^{\rm g}_{34},\;\; \mathcal{G}_6={Z}^{\rm g}_4, \nonumber \\ 
 \mathcal{G}_7 =&\, {X}^{\rm g}_1,\;\; \mathcal{G}_8={X}^{\rm g}_{13},\;\; \mathcal{G}_9 ={X}^{\rm g}_3, \nonumber \\
 \mathcal{G}_{10}=&\,{X}^{\rm g}_2,\;\; \mathcal{G}_{11} = {X}^{\rm g}_{24},\;\; \mathcal{G}_{12}={X}^{\rm g}_4, 
\end{align}
where ${Z}^{\rm g}_j$ and ${X}^{\rm g}_j$ respectively are the $Z$ and $X$ Pauli operators acting on the $j$th gauge qubit. Specifically, the action of ${Z}^{\rm g}_j$ and ${X}^{\rm g}_j$ on the $c$-coefficients is given {below} in Eqs.~\eqref{eq:Xj^g-def}--\eqref{eq:Zj^g-def}. Their matrix representations are $Z_1^{\rm g}=\sigma_z\varotimes\openone_{8}$ ($\openone_8$ is the $8\times8$ identity matrix), $X_1^{\rm g}=\sigma_x\varotimes\openone_{8}$, etc. {For simplicity of notation, we shall} denote $Z_{1}^{\rm g}Z_{2}^{\rm g}=Z_{12}^{\rm g}$, $X_{1}^{\rm g}X_{3}^{\rm g}=X_{13}^{\rm g}$, etc. 

The stochastic master equation that describes the measurement-induced evolution of {the density matrix for the four gauge qubits,} $\varrho_{\rm g}(t)$, is given 
in It\^o form~\cite{RiskenBook} by~\cite{Korotkov2010,CarmichaelBook,WisemanBook}
\begin{align}
\label{eq:rho_g-EOM}
\dot \varrho_{\rm g}(t) =&\; \sum_k\bigg\{\frac{\Gamma_k}{2}({\mathcal{G}}_k\varrho_{\rm g} {\mathcal{G}}_k-\varrho_{\rm g}) + \frac{\xi_k(t)}{\sqrt{\tau_k}}\big(\frac{ {\mathcal{G}}_k\varrho_{\rm g} + \varrho_{\rm g}{\mathcal{G}}_k}{2}  \nonumber \\
&\hspace{0.95cm}-\varrho_{\rm g} {\rm Tr}[ {\mathcal{G}}_k\varrho_{\rm g}]\big)\bigg\}, 
\end{align}
where $\Gamma_k$ is the ensemble average dephasing rate due to measurement of $G_k$, and $\tau_k$ is the ``measurement time'' {that is employed} to distinguish between the two eigenvalues $\pm1$ of $G_k$. These two parameters are related through the quantum efficiency $\eta_k$ of the $k$th detector as follows~\cite{Korotkov2016,Korotkov2001} 
\begin{align}
\tau_k = \frac{1}{2\eta_k\Gamma_k}.  
\end{align}
For ideal detectors $\eta_k=1$, and for nonideal detectors $0<\eta_k<1$. In the Markovian approximation, the independent noises $\mathcal{\xi}_k(t)$ in Eq.~\eqref{eq:rho_g-EOM}  are delta-correlated; their two-time correlation functions are
\begin{align}
\label{eq:xi-corr}
\langle \xi_{k}(t)\xi_{k'}(t')\rangle = \delta_{kk'}\,\delta(t-t'),  
\end{align}
where $\langle\cdot\rangle$ indicates average over an ensemble of noise realizations. Equation~\eqref{eq:rho_g-EOM} is valid for any subspace $\mathcal{Q}_\ell$.  The measurement signal from the $k$th detector is
\begin{align}
\label{eq:Ik-Ql}
I_{G_k}(t)  =\zeta_k^{(\ell)} I_{\mathcal{G}_k}(t),\;\;\;\; I_{\mathcal{G}_k}(t)  = {\rm Tr}[\mathcal{G}_k\varrho_{\rm g}(t)] + \sqrt{\tau_{k}}\, \xi_{k}(t), 
\end{align}
where $\zeta_k^{(\ell)}=\pm1$ is the same sign factor that appears in Eq.~\eqref{eq:Gk-Ql}.  

{Note that the evolution equation~\eqref{eq:rho_g-EOM} keeps the gauge qubits density matrix $\varrho_{\rm g}(t)$ real if it is initially set to a real density matrix at some earlier moment; this is so because  the matrices~\eqref{eq:curli-G_k} are real.} 

Equations~\eqref{eq:rho_g-EOM} and~\eqref{eq:Ik-Ql} are not general. They are applicable only if {i) the full density matrix $\rho(t)$ has support in only one of the subspaces $\mathcal{Q}_\ell$,  and ii) }$\rho(t)$ can be written in a direct-product form that separates the logical and gauge degrees of freedom (this separation is referred to as the subsystem structure of subsystem codes.) If these conditions cannot be fulfilled, we have to use the  evolution equation for full $\rho(t)$: {this} reads as  (in It\^o form)
\begin{align}
\label{eq:rho-EOM}
&\dot \rho(t) = \nonumber \\
&\;\sum_k\frac{\Gamma_k}{2}(G_k\rho G_k-\rho) +  \frac{\xi_k(t)}{\sqrt{\tau_k}}(\frac{G_k\rho + \rho G_k}{2} - \rho {\rm Tr}[G_k\rho]), 
\end{align}
with the detector output signals given by 
\begin{align}
\label{eq:Ik-full}
I_{G_k}(t) = {\rm Tr}[\rho(t){G}_k] + \sqrt{\tau_{k}}\, \xi_{k}(t).  
\end{align}
Although Eqs.~\eqref{eq:rho-EOM}--\eqref{eq:Ik-full} hold for any physical density matrix $\rho(t)$, Eqs.~\eqref{eq:rho_g-EOM} and~\eqref{eq:Ik-Ql} are more convenient for the analysis of the continuous operation of the nine-qubit Bacon-Shor code, because they allow us to effectively reduce the problem complexity from nine to four qubits. Note that the subsystem structure of $\rho(t)$ is preserved by Eq.~\eqref{eq:rho-EOM} because of the block-diagonal form of $G_k$---see Eq.~\eqref{eq:Gk-Ql}. We also point out that in deriving Eqs.~\eqref{eq:rho_g-EOM} and~\eqref{eq:Ik-Ql} from Eqs.~\eqref{eq:rho-EOM}--\eqref{eq:Ik-full}, we have used the trick of replacing $\xi_k(t)$ by $\zeta_k^{(\ell)}\xi_k(t)$~\cite{Atalaya2017}. This is done to make Eq.~\eqref{eq:rho_g-EOM} applicable to any subspace $\mathcal{Q}_\ell$, {and not only to the code space}. A consequence of this trick is that $\zeta_k^{(\ell)}=\pm1$ appears as an overall sign factor in the formulas for the actual measurement signals $I_{G_k}(t)$ in  Eq.~\eqref{eq:Ik-Ql}. In this way, we still preserve the sign of the temporal cross-correlations of $I_{G_k}(t)$, which is important to determine the error syndromes in the continuous operation, {as described in} Section~\ref{Section:monitored-error-syndrome-path}. 

Let us now discuss how errors $X_i$, $Y_i$ and $Z_i$ act on wavefunctions that are parametrized as in Eq.~\eqref{eq:Psi-state-v2}. Such errors preserve this parametrization and map the system state from subspace $\mathcal{Q}_\ell$ to {one of the error} subspaces $\mathcal{Q}_{\ell'}$. In addition, {just as discussed above for the discrete operation,} the errors {can} change the logical state $(\alpha,\beta)$, the state of the gauge qubits (the $c$-coefficients $c_{q_1q_2q_3q_4}$), and introduce an overall phase factor. {As noted earlier,} the latter is actually not important for Pauli-type errors, since no-jump evolution is trivial and the probability of jump is state-independent, so we {may} disregard overall phase factors in the wavefunctions. These phase factors are, however, important for other types of errors such as energy relaxation~\cite{Atalaya2017}. 
{For reference, these factors are explicitly} written in Appendix~\ref{Appendix-B}. 
\begin{table}
\caption{\label{table-II} Equivalence relations for the 27 Pauli errors in terms of subspace-basis-transformation operators $Q_\ell$, given in  Table~\ref{table-I}, logical operations $X_{\rm L}$, $Y_{\rm L}$ and $Z_{\rm L}$, defined in Eq.~\eqref{eq:logical-operators}, and gauge qubit operations $X_j^{\rm g}$, $Y_j^{\rm g}$ and $Z_j^{\rm g}$, defined in Eqs.~\eqref{eq:Xj^g-def}--\eqref{eq:Yj^g-def}. Phase factors are not included. }
\begin{ruledtabular}
\begin{tabular}{l l l }
$X_1\leftrightarrow\; Q_4$  & $Z_1 \leftrightarrow\;  Q_{12}$ & $Y_1 \leftrightarrow\; Q_8$\\
$X_2\leftrightarrow\; Q_4\, X_1^{\rm g}$ & $Z_2\leftrightarrow\;  Q_{15}\, Z_{\rm L}\, Z_{24}^{\rm g}$  & $Y_2 \leftrightarrow\; Q_{11}\, Z_{\rm L}\,Z_{24}^{\rm g}\,X_1^{\rm g}$ \\
$X_3\leftrightarrow\; Q_4 \,  X_{12}^{\rm g}$ & $Z_3 \leftrightarrow\;  Q_{3}\, Z_{24}^{\rm g}$  & $Y_3 \leftrightarrow\; Q_{7}\, X_1^{\rm g}\, Y_2^{\rm g}\, Z_{4}^{\rm g} $ \\
$X_4\leftrightarrow\; Q_5 \,  X_{\rm L}\, X_{34}^{\rm g}$ & $Z_4 \leftrightarrow\;  Q_{12}\, Z_{1}^{\rm g}$  & $Y_4 \leftrightarrow\; Q_{9}\, X_{\rm L}\, X_{34}^{\rm g}\, Z_{1}^{\rm g} $ \\
$X_5\leftrightarrow\; Q_5 \,  X_{\rm L}\, X_{14}^{\rm g}$ & $Z_5 \leftrightarrow\;  Q_{15}\,  Z_{\rm L}\,Z_{14}^{\rm g}$ & $Y_5 \leftrightarrow\; Q_{10}\, Y_{\rm L}\, Y_{14}^{\rm g}$ \\
$X_6\leftrightarrow\; Q_5 \,  X_{\rm L}\, X_{12}^{\rm g}$ & $Z_6 \leftrightarrow\;  Q_{3}\,  Z_{4}^{\rm g}$ & $Y_6 \leftrightarrow\; Q_{6}\, X_{\rm L}\, X_{12}^{\rm g}\, Z_4^{\rm g}$ \\
$X_7\leftrightarrow\; Q_1 \,  X_{34}^{\rm g}$ & $Z_7\leftrightarrow\;  Q_{12} \, Z_{13}^{\rm g}$ & $Y_7 \leftrightarrow\; Q_{13}\, X_{4}^{\rm g}\, Y_{3}^{\rm g}\, Z_1^{\rm g}$ \\
$X_8\leftrightarrow\; Q_1 \,  X_{4}^{\rm g}$ & $Z_8 \leftrightarrow\; Q_{15}\, Z_{\rm L}\, Z_{13}^{\rm g}$ & $Y_8 \leftrightarrow\; Q_{14}\, Z_{\rm L}\, X_{4}^{\rm g} Z_{13}^{\rm g}$ \\
$X_9\leftrightarrow\; Q_1$ & $Z_9\leftrightarrow\;  Q_{3}$ &  $Y_9 \leftrightarrow\;  Q_2$
\end{tabular}
\end{ruledtabular}
\end{table}

Table~\ref{table-II} shows the representation of all 27 Pauli-type errors in terms of operators $Q_\ell$, the logical operations $X_{\rm L}$, $Y_{\rm L}$ and $Z_{\rm L}$ [defined in Eq.~\eqref{eq:logical-operators}]  and gauge-qubit operations $X_{j}^{\rm g}$,  $Y_{j}^{\rm g}$ and $Z_{j}^{\rm g}$. The latter perform the following linear transformations on the $c$-coefficients
\begin{align}
\label{eq:Xj^g-def}
X_1^{\rm g}\,c_{q_1q_2q_3q_4} =&\; c_{\bar q_1q_2q_3q_4},\;\;  X_2^{\rm g}\,c_{q_1q_2q_3q_4} = c_{q_1 \bar q_2q_3q_4},\nonumber \\
X_3^{\rm g}\,c_{q_1q_2q_3q_4} =&\; c_{q_1 q_2 \bar q_3q_4},\;\; X_4^{\rm g}\,c_{q_1q_2q_3q_4} = c_{q_1 q_2 q_3 \bar q_4},
\end{align}
where $\bar0=1$ and $\bar1=0$, and
\begin{align}
\label{eq:Zj^g-def}
Z_j^{\rm g}\,c_{q_1q_2q_3q_4} =&\; (-1)^{q_j}c_{ q_1q_2q_3q_4},  \\
\label{eq:Yj^g-def}
Y_j^{\rm g}\,c_{q_1q_2q_3q_4} =&\; \inum X_j^{\rm g} Z_j^{\rm g} c_{ q_1q_2q_3q_4}.
\end{align}
The imaginary phase factor $\inum$ in Eq.~\eqref{eq:Yj^g-def} is actually not necessary since we are dropping out  phase factors. 
\begin{table}  
\caption{\label{table-III} Multiplication table for error-subspace basis operators $Q_{\ell_1}$ and $Q_{\ell_2}$. The table is symmetric (i.e., $Q_{\ell_1}\times Q_{\ell_2} = Q_{\ell_2}\times Q_{\ell_1}$) if phase factors are disregarded (the table that includes phase factors is given in Appendix~\ref{Appendix-B}.)}  
\begin{center}\scalebox{0.825}{%
\begin{tabular}{c | c c c c  c c c c  c c c c  c c c c } \hline\hline
$\boldsymbol{\times}$& ${Q_0}$ & $\!{Q}_1$ & $\!{Q}_2$ & $\!{Q}_3$ & $\!{Q}_4$ & $\!{Q}_5$ & $\!{Q}_6$ & $\!{Q}_7$ & $\!{Q}_8$ & $\!{Q}_9$ & $\!{Q}_{10}$ & $\!{Q}_{11}$ & $\!{Q}_{12}$ & $\!{Q}_{13}$ & $\!{Q}_{14}$& $\!{Q}_{15}$\\ \hline
$Q_0$& $\!\openone$ & & & & &  &  & &  &  &  &  & &  & & \\ 
${Q}_1$& $Q_1$ & $\!\openone$ & & & & &  &  & &  &  & &  &  &  &  \\ 
${Q}_2$& $Q_2$& $\!Q_3$ & $\!\openone$ & & & & &  &  & &  &  & &  &  &  \\
${Q}_3$ & $Q_3$ & $\!Q_2$ & $\!Q_{1}$& $\!\openone$ & & & &  &  & &  &  & &  &  & \\
${Q}_4$& $Q_4$ & $\!Q_5$ & $\!Q_6$& $\!{Q}_{7}$ & $\!\openone$ & & & &  &  & &  &  & &  & \\
${Q}_5$& $Q_5$ & $\!Q_4$ & $\!Q_7$ & $\!{Q}_{6}$ & $\!{Q}_{1}$& $\!\openone$ & & &  &  & &  &  & &  & \\
${Q}_6$& $Q_6$ & $\!Q_7$ & $\!Q_4$ & $\!{Q}_{5}$ & $\!{Q}_{2}$ & $\!{Q}_{3}$ & $\!\openone$ & &  &  & &  &  & &  & \\
${Q}_7$ & $Q_7$ & $\!Q_6$& $\!Q_5$ & $\!{Q}_{4}$& $\!{Q}_{3}$ & $\!{Q}_{2}$ & $\!{Q}_{1}$ & $\!\openone$ &  &  & &  &  & &  & \\
${Q}_8$ & $Q_8$ & $\!Q_9$ & $\!Q_{10}$ & $\!{Q}_{11}$& $\!{Q}_{12}$ & $\!{Q}_{13}$& $\!{Q}_{14}$ & $\!{Q}_{15}$& $\!\openone$  &  &  & &  &  & & \\
${Q}_9$ & $Q_9$ & $\!Q_8$ & $\!Q_{11}$ & $\!{Q}_{10}$ & $\!{Q}_{13}$ & $\!{Q}_{12}$ & $\!{Q}_{15}$ &$\!{Q}_{14}$ & $\!{Q}_{1}$ & $\!\openone$ &  & &  &  & & \\ 
${Q}_{10}$ & $Q_{10}$& $\!Q_{11}$ & $\!Q_8$& $\!{Q}_{9}$& $\!{Q}_{14}$ &$\!{Q}_{15}$ & $\!{Q}_{12}$& $\!{Q}_{13}$  & $\!{Q}_{2}$ & $\!{Q}_{3}$ & $\!\openone$ & &  &  & & \\
${Q}_{11}$ & $Q_{11}$ & $\!Q_{10}$& $\!Q_9$ & $\!{Q}_{8}$& $\!{Q}_{15}$ & $\!{Q}_{14}$ & $\!{Q}_{13}$ & $\!{Q}_{12}$ & $\!{Q}_{3}$ & $\!{Q}_{2}$& $\!{Q}_{1}$ & $\!\openone$ &  &  & & \\
${Q}_{12}$ & $Q_{12}$ & $\!Q_{13}$& $\!Q_{14}$ & $\!{Q}_{15}$ & $\!{Q}_{8}$& $\!{Q}_{9}$ & $\!{Q}_{10}$& $\!{Q}_{11}$ & $\!{Q}_{4}$ & $\!{Q}_{5}$ &$\!{Q}_{6}$ & $\!{Q}_{7}$ & $\!\openone$ & & & \\
${Q}_{13}$ & $Q_{13}$ & $\!Q_{12}$ & $\!Q_{15}$ & $\!{Q}_{14}$ &$\!{Q}_{9}$ &$\!{Q}_{8}$ & $\!{Q}_{11}$& $\!{Q}_{10}$ & $\!{Q}_{5}$ & $\!{Q}_{4}$ & $\!{Q}_{7}$ & $\!{Q}_{6}$ & $\!{Q}_{1}$& $\!\openone$ & &  \\
${Q}_{14}$ & $Q_{14}$ & $\!Q_{15}$ & $\!Q_{12}$ & $\!{Q}_{13}$ & $\!{Q}_{10}$& $\!{Q}_{11}$& $\!{Q}_{8}$& $\!{Q}_{9}$ & $\!{Q}_{6}$& $\!{Q}_{7}$ & $\!{Q}_{4}$ & $\!{Q}_{5}$&$\!{Q}_{2}$ & $\!{Q}_{3}$ & $\!\openone$ &  \\
${Q}_{15}$ & $Q_{15}$ & $\!Q_{14}$ & $\!Q_{13}$ & $\!{Q}_{12}$ &$\!{Q}_{11}$ &$\!{Q}_{10}$ & $\!{Q}_{9}$ & $\!{Q}_{8}$ & $\!{Q}_{7}$ & $\!{Q}_{6}$ & $\!{Q}_{5}$ &$\!{Q}_{4}$ & $\!{Q}_{3}$& $\!{Q}_{2}$ & $\!{Q}_{1}$& $ \!\openone$ \\ \hline\hline
\end{tabular}
}\end{center}
\end{table}

We now have all {the} necessary elements to describe the dynamics of the code operation. At $t=0$ there is an encoding step, after which the system state is initially set in the code space {$\mathcal{Q}_0$} and parametrized according to Eq.~\eqref{eq:direct-prod-form},
  with some intended logical state $\varrho^{\rm in}_{\rm L}$ [corresponding to Eq.~\eqref{eq:logical-state}] and some arbitrary initial density matrix $\varrho_{\rm g}^{\rm in}$ for the gauge qubits. 
{Subsequently,} simultaneous continuous measurement of the gauge operators induces diffusive evolution of $\varrho_{\rm g}(t)$ according to Eq.~\eqref{eq:rho_g-EOM}, with initial condition $\varrho_{\rm g}(0)=\varrho_{\rm g}^{\rm in}$ (see left  green circle of Fig.~\ref{fig:overview}), while the logical state remains constant. As an example, suppose the first error is $X_5$ (bit-flip in physical qubit 5) and occurs at the moment 
$t_{\rm err}^{(1)}$. From Table~\ref{table-II}, we find that $X_5$ is equivalent to $Q_5\,X_{\rm L}\,X_{14}^{\rm g}$. This means that immediately after this error the system state is in the error subspace $\mathcal{Q}_5$, the logical state is $\varrho_{\rm L} = X_{\rm L}\,\varrho_{\rm L}^{\rm in}\,X_{\rm L}$ (i.e., {the} logical state {undergoes} a logical $X$ operation), and the state of the gauge qubits changes to $X_{14}^{\rm g} \,\varrho_{\rm g}(t_{\rm err}^{(1)} - 0)\, X_{14}^{\rm g}$, {where we have explicitly displayed the elapsed time interval}. {We then again} have diffusive evolution of $\varrho_{\rm g}(t)$ according to Eq.~\eqref{eq:rho_g-EOM} with {new} initial condition $\varrho_{\rm g}(t_{\rm err}^{(1)} + 0) = X_{14}^{\rm g} \,\varrho_{\rm g}(t_{\rm err}^{(1)} - 0)\, X_{14}^{\rm g}$,  until the next error occurs. Suppose the second error is $X_1$ and occurs at moment $t_{\rm err}^{(2)}$. From Table~\ref{table-II}, we see that $X_1$ is equivalent to $Q_4$ and {affects neither} the logical state nor the state of the gauge qubits. We use Table~\ref{table-III} to find out {to} which error subspace the system state jumps. From this table we {see} that $Q_4\times Q_5 = Q_1$. This means that immediately after the second error, the system state is in the error subspace $\mathcal{Q}_1$, while the state of the logical and gauge qubits are the same as before the occurrence of the $X_1$ error.  Then we again have diffusion of {the} $\varrho_{\rm g}(t)$ state {but} now in {subspace} $\mathcal{Q}_1$ until {the} next error occurs, and so on.  

From this example, it is clear that the jump/no-jump method is an efficient method to describe decoherence due to Pauli-type errors in subsystem QEC codes, because it allows us to describe measurement-induced state diffusion using the reduced stochastic master equation Eq.~\eqref{eq:rho_g-EOM} for $\varrho_{\rm g}(t)$. It would be much more expensive computationally to solve the full evolution equation, Eq.~\eqref{eq:rho-EOM}. However, we point out that decoherence due to energy-relaxation (or any other non Pauli-type errors) requires the use of the full stochastic master equation [Eq.~\eqref{eq:rho-EOM}]. The reason is that in this case the nontrivial no-jump evolution does not preserve the subsystem structure {evident} in Eqs.~\eqref{eq:Psi-state-v2} or~\eqref{eq:direct-prod-form}. Nevertheless, we can still use the jump/no-jump method to approximately calculate the logical error rates as {undertaken} in Ref.~\cite{Atalaya2017}. In this case, it is important to keep track of the overall phase factors that errors introduce to the wavefunctions. For this reason, in Appendix~\ref{Appendix-B}, we present the {modified} versions of Tables~\ref{table-II} and~\ref{table-III} that include the phase factors. 

For simplicity, {from now on} we {shall} assume that all detectors have the same measurement strength (symmetric case); i.e., 
\begin{align}
\label{eq:symmetric-measurement}
\tau_k = \tau_{\rm m}, \;\; \Gamma_{k} = \Gamma_{\rm m},\;\; {\rm and} \;\; \eta_{k} = \eta,\;\; {\forall k}.
\end{align}

\subsection{Continuous QEC protocol and the error syndrome path}
\label{section:QEC-algorithm}
In this section we discuss the QEC protocol that we use in the continuous operation of the nine-qubit Bacon-Shor code. 
{The spirit of t}his protocol is somewhat similar to {that of} Mabuchi's QEC protocol for stabilizer codes~\cite{Mabuchi2009}, in the sense that we do not correct errors during the code operation, so the system state can explore all 16 subspaces $\mathcal{Q}_{\ell}$ in the presence of errors. {However, i}n contrast to Ref.~\cite{Mabuchi2009}, we do not estimate the probability that the system state is in the subspaces $\mathcal{Q}_\ell$ during the code operation; instead, we monitor in real time the stabilizer generators $S_x^{(1)}$, $S_z^{(1)}$, $S_x^{(2)}$ and $S_z^{(2)}$, as explained below. {The values of these} in a given realization {of our protocol} determine what we refer to as the {\it error syndrome path}, denoted by $\mathcal{S}(t)$. The latter has values $\mathcal{S}(t)=\ell=0,1,...15$ depending on the error syndrome pattern at a given moment $t$---see Table~\ref{table-I}. For a given realization, $\mathcal{S}(t)$ is a piece-wise function of time. Knowledge of $\mathcal{S}(t)$ is sufficient to determine the logical state at the end of the code operation and to restore it before we {return it to the user}. 

From the error syndrome path $\mathcal{S}(t)$ we can determine the (single-qubit) errors $E_i$ that may have occurred, modulo gauge operators.  Indeed, every time that $\mathcal{S}(t)$ jumps from, say, $\ell_1$ to $\ell_2$, an error has occurred that {causes the system state to jump} from subspace $\mathcal{Q}_{\ell_1}$ to subspace $\mathcal{Q}_{\ell_2}$. To figure out which errors $E_i$ have caused {this} transition, we use Table~\ref{table-III} to find {the ${Q}_\ell$ {operator} satisfying} $Q_{\ell}\times Q_{\ell_1}=Q_{\ell_2}$, and then we use Table~\ref{table-II} to determine all errors $E_i$ that are ``proportional'' to $ Q_{\ell}$. Although this procedure does not tell us the {specific} error that has actually happened ({as noted above,} it gives $E_i$ modulo gauge operators), it
{does} {\it uniquely } {identify} the logical operation (if any) that is induced by all {the} possible errors $E_i$. For instance, for $Q_\ell=Q_5$, Table~\ref{table-II} indicates that $E_i$ could be $X_3$, $X_4$ or $X_5$, which are {all} equivalent modulo $X$-gauge operators and {which} all introduce a logical $X$ operation. The main idea of our continuous QEC algorithm is to continuously monitor $\mathcal{S}(t)$ as {accurately as possible} (see {below}), take note of all logical operations induced by the errors, and undo them at the end of the code operation. 

Let us denote the product of all inferred logical operations from the jumps of $\mathcal{S}(t)$ as $\mathcal{O}$ (the total logical operation in a {single} realization {of the continuous QEC protocol}). For each realization, we can restore the logical state by applying the multi-qubit Pauli operations $X_1X_4X_7$, $Z_1Z_2Z_3$, or $X_1X_4X_7\, Z_1Z_2Z_3$ to the physical system if the total logical operation is $\mathcal{O}=X_{\rm L}$, $Z_{\rm L}$, or $Y_{\rm L}$, respectively. Finally, we apply a decoding step to obtain the restored logical state from the final subspace, where the system state is at the end of each realization; this is the {final} logical state. 

\subsection{The monitored  error syndrome path $\mathcal{S}_{\rm m}(t)$}
\label{Section:monitored-error-syndrome-path}
In this section we discuss how to monitor the error syndrome path in real time. To do this, we introduce the following triple cross-correlators 
\begin{subequations}
\label{eq:cross-correlators}
\begin{align}
\mathcal{C}_{z}^{(1)}(t) = \int_{0}^t dt'\, \frac{e^{-\frac{t-t'}{T_{\rm c}}}}{T_{\rm c}} \,\mathcal{I}_{Z_{14}}(t')\,\mathcal{I}_{Z_{25}}(t')\,\mathcal{I}_{Z_{36}}(t'),
\label{eq:C1z}
\end{align}
\vspace{-0.4cm}
\begin{align}
\mathcal{C}_{z}^{(2)}(t) = \int_{0}^t { dt'} \, \frac{e^{-\frac{t-t'}{T_{\rm c}}}}{T_{\rm c}}\,\mathcal{I}_{Z_{47}}(t')\,\mathcal{I}_{Z_{58}}(t')\,\mathcal{I}_{Z_{69}}(t'),
\label{eq:C2z}
\end{align}
\vspace{-0.4cm}
\begin{align}
\mathcal{C}_{x}^{(1)}(t) = \int_{0}^t{ dt'}\, \frac{e^{-\frac{t-t'}{T_{\rm c}}}}{T_{\rm c}} \, \mathcal{I}_{X_{12}}(t')\,\mathcal{I}_{X_{45}}(t')\,\mathcal{I}_{X_{78}}(t'),
\label{eq:C1x}
\end{align}
\vspace{-0.4cm}
\begin{align}
\mathcal{C}_{x}^{(2)}(t) = \int_{0}^t { dt'}\, \frac{e^{-\frac{t-t'}{T_{\rm c}}}}{T_{\rm c}}\, \mathcal{I}_{X_{23}}(t')\,\mathcal{I}_{X_{56}}(t')\,\mathcal{I}_{X_{89}}(t'),
\label{eq:C2x}
\end{align}
\end{subequations}
where $T_{\rm c}$ is an integration time parameter whose optimal value will be determined later, and $\mathcal{I}_{G_k}(t)$ is the measurement signal from continuous measurement of $G_k$, {smoothed out} by time averaging as follows: 
\begin{align}
\label{eq:I-Gk-avg}
\mathcal{I}_{G_k}(t) =\int_{0}^t dt'\, \frac{e^{-\frac{t-t'}{\tau_{\rm c}}}}{\tau_{\rm c}}\, {I}_{G_k}(t'), 
\end{align}
{Here} $\tau_{\rm c}$ is another integration time parameter that can also be optimized. We point out that, for detectors with different measurement strengths, the bare output signals $I_{G_k}(t)$ should be {smoothed} out using different integration time parameters; similarly, different integration time parameters should also be used in Eqs.~\eqref{eq:C1z}--\eqref{eq:C2x}.  {In this work, we} carry out {the} time averaging with exponential weighting functions. Other functions (e.g., uniform weighting function{s} over a {specified} time window~\cite{Atalaya2017}) may {also} be used. {Choosing the optimal weighting function is a topic for} future work. 

The cross-correlators $\mathcal{C}_z^{(1)}(t)$, $\mathcal{C}_z^{(2)}(t)$, $\mathcal{C}_x^{(1)}(t)$ and $\mathcal{C}_x^{(2)}(t)$ are constructed to continuously monitor the stabilizers $S_z^{(1)}$, $S_z^{(2)}$, $S_x^{(1)}$ and $S_x^{(2)}$ in real time, respectively. This  monitoring is, however, not perfect since the cross-correlators are noisy even when the system state is in a fixed subspace $\mathcal{Q}_\ell$ and they cannot immediately follow abrupt changes of the values of the stabilizers after occurrence of errors---{Eq.~\eqref{eq:cross-correlators} shows that the} response time of cross-correlators is determined by the integration time parameter $T_{\rm c}$. These imperfections {render} the jumps in the monitored error syndrome path, $\mathcal{S}_{\rm m}(t)$, different from those of the actual error syndrome path, $\mathcal{S}(t)$, potentially leading to logical errors since the series of logical operations inferred from {the monitored path} $\mathcal{S}_{\rm m}(t)$ is not the same as the actual one, obtained from {the real error syndrome path} $\mathcal{S}(t)$. Note that, in principle, continuous monitoring of the error syndrome path could be performed via simultaneous measurement of the four commuting stabilizer generators, Eq.~\eqref{eq:stabilizers}, at the same time. However, this operation mode would require measurement of six-qubit operators,  which is much more difficult to realize than measurement of two-qubit operators. Moreover, this is not necessary. We emphasize that the main implementation advantage of Bacon-Shor codes is that they can be operated with {\it only} two-qubit measurements. 

To determine the monitored error syndrome path  we use the following two-error-threshold algorithm. At $t=0$, we set $\mathcal{S}_{\rm m}(0)= \mathcal{S}(0)=0$ because the initial encoding step is assumed perfect and the initial system state is in the code space. We {\it do not update} the monitored error syndrome path at the moment $t$ if at least one of the following four inequalities holds ($q=x,$ $z$ and $n=1$, $2$): 
\begin{align}
\label{eq:error-uncertainty-cond}
1-\Theta_2 <&\;{\tilde{S}_{q}^{(n)}}(t-\delta t)\,\frac{\mathcal{C}_{q}^{(n)} (t)}{\big|\!\big\langle \mathcal{C}_{q}^{(n)}\big\rangle\!\big|}< 1-\Theta_1,
\end{align}
where $\Theta_1$ and $\Theta_2$ are the error threshold parameters that are fixed beforehand such that $0\leq \Theta_{1}\leq1$ and $1\leq \Theta_{2}\leq2$, and $\tilde{S}_{q}^{(n)}(t-\delta t)=\pm1$ is the estimated value of the stabilizer generator $S_{q}^{(n)}$ that corresponds to the monitored error syndrome path at the moment $t-\delta t$ ($\delta t$ is a small timestep). The denominator in Eq.~\eqref{eq:error-uncertainty-cond} is used for normalization of the  cross-correlators~\eqref{eq:cross-correlators}. The {two threshold} parameters ($\Theta_1$ and $\Theta_2$) will be optimized later.  This strategy essentially says that if we are not sure about the values of the stabilizer generators, we hold the previous value of the monitored error syndrome path; i.e., $\mathcal{S}_{\rm m}(t) = \mathcal{S}_{\rm m}(t-\delta t)$. The error threshold parameters determine what we refer to as the ``syndrome uncertainty region''. For detectors with the same measurement strength, the denominators of Eq.~\eqref{eq:error-uncertainty-cond} are equal and depend on the integration time parameter $\tau_{\rm c}$ as follows, 
\begin{align}
\label{eq:avg-correlator-formula}
\big|\!\big\langle \mathcal{C}_{q}^{(n)}\big\rangle\!\big| =&\; \frac{1}{3}\bigg[\frac{1}{(1+\Gamma_{\rm m}\tau_{\rm c})(1 + 2\Gamma_{\rm m}\tau_{\rm c})}  +  \frac{1}{(1+2\Gamma_{\rm m}\tau_{\rm c})^2} + \nonumber \\ 
&\hspace{0.5cm}\,\frac{1}{(1+\Gamma_{\rm m}\tau_{\rm c})(1 + 4\Gamma_{\rm m}\tau_{\rm c})}\bigg].  
\end{align}
Equation~\eqref{eq:avg-correlator-formula} gives the magnitude of the ensemble average value of the cross-correlators in any subspace $\mathcal{Q}_\ell$; this result is derived in the next {sub}section. {The last component is to}  update the value of the monitored error syndrome path {when}   {all} cross-correlators are outside of the ``syndrome uncertainty region''. We {do this as follows.  We} first digitize the cross-correlators, {assigning them values of} $+1$ or $-1$ if $\mathcal{C}_{q}^{(n)}(t)$ is larger than $(1-\Theta_1)|\!\langle \mathcal{C}_{q}^{(n)}\rangle\!|$ or  smaller than $(1-\Theta_2)|\!\langle \mathcal{C}_{q}^{(n)}\rangle\!|$, respectively. The digitized values of $\mathcal{C}_x^{(1)}(t)$, $\mathcal{C}_z^{(1)}(t)$, $\mathcal{C}_x^{(2)}(t)$ and $\mathcal{C}_z^{(2)}(t)$ in this order {constitute} the estimated error syndrome pattern at moment $t$. We then use Table~\ref{table-I} to read out the subspace $\mathcal{Q}_\ell$ that agrees with that error syndrome pattern and update the monitored error syndrome path to $\mathcal{S}_{\rm m}(t) = \ell$, where $\ell=0,1,...15$. 

The QEC {protocol} discussed in Section~\ref{section:QEC-algorithm} works perfectly if we {have}  access to the true error syndrome path, $\mathcal{S}(t)$. However, we actually have at our disposal  {only the monitored path} $\mathcal{S}_{\rm m}(t)$ that  generally differs from $\mathcal{S}(t)$ {because of the time {averaging} and {the noise in the cross-correlators, $\mathcal{C}_q^{(n)}(t)$}}. This discrepancy can lead to different inferred total logical operations for the {individual} realizations  and {hence} to logical errors. Indeed, let us assume that $\mathcal{O}$ and $\mathcal{O}_{\rm m}$ are, respectively, the total logical operations inferred from $\mathcal{S}(t)$ and $\mathcal{S}_{\rm m}(t)$ for a given realization. {The final logical state at the end of this realization is $\mathcal{O}|\Psi_{\rm L}\rangle$; however, the logical state that is returned to the user is $\mathcal{O}_{\rm m}\mathcal{O}|\Psi_{\rm L}\rangle$}. Therefore, if $\mathcal{O}_{\rm m}\mathcal{O}=X_{\rm L}$, $Y_{\rm L}$ or $Z_{\rm L}$, such {a} realization contributes to the probability of {a} logical $X$, $Y$, or $Z$ error, respectively. Assuming that the error rates of the physical qubits are sufficiently small, averaging over realizations {then} leads to {a} probability {for} logical errors of the form ($L=X$, $Y$ or $Z$)
\begin{align}
\label{eq:Prob-logical-err}
P_{L}(T_{\rm op}) = \gamma_{L}^{{\rm cont}}\,T_{\rm op} +  \Delta P_L,
\end{align}
where $\gamma_L^{{\rm cont}}=\gamma_X^{{\rm cont}}$, $\gamma_Y^{{\rm cont}}$ or $\gamma_Z^{{\rm cont}}$ is the logical $X$, $Y$ or $Z$ error rate for the continuous operation of duration $T_{\rm op}$ and $\Delta P_L$ is a small probability offset. The logical error rates are calculated in Section~\ref{section:logical-error-rates-cnt}. The probability offsets are due {primarily} to single-qubit errors that occur {so} close to the end of the continuous operation that there is no enough time for the cross-correlators to switch sign. Such errors therefore remain undetected and their associated logical operations (given by Table~\ref{table-II}) are not accounted for in the total logical operation $\mathcal{O}_{\rm m}$,  obtained from the monitored error syndrome path. These undetected errors also make $\mathcal{S}(T_{\rm op})\neq \mathcal{S}_{\rm m}(T_{\rm op})$; that is, the final system state is in subspace $\mathcal{Q}_{\mathcal{S}{(T_{\rm op})}}$, while, from the monitored error syndrome path, we infer it is in subspace $\mathcal{Q}_{\mathcal{S}_{\rm m}{(T_{\rm op})}}$. We estimate the probability offsets as 
\begin{subequations}
\label{eq:DeltaP}
\begin{align}
\label{eq:DeltaP-x}
\Delta P_X \approx &\left(\Gamma_{X}^{(4)} + \Gamma_{X}^{(5)} + \Gamma_{X}^{(6)} + \Gamma_{Y}^{(4)} + \Gamma_{Y}^{(6)}\right) T_{\rm c}
\end{align} 
\begin{align}
\label{eq:DeltaP-z}
\Delta P_Z \approx &\left(\Gamma_{Z}^{(2)} + \Gamma_{Z}^{(5)} + \Gamma_{Z}^{(8)} + \Gamma_{Y}^{(2)} + \Gamma_{Y}^{(8)}\right) T_{\rm c} 
\end{align}
\begin{align}
\label{eq:DeltaP-y}
\Delta P_Y \approx &\; \Gamma_{Y}^{(5)} T_{\rm c} 
\end{align}
\end{subequations}
 if $T_{\rm op}\gtrsim T_{\rm c}$. The single-qubit errors, whose occurrence rates enter in Eqs.~\eqref{eq:DeltaP-x}--\eqref{eq:DeltaP-y}, are, respectively, those that affect the logical state by a single logical $X_{\rm L}$, $Z_{\rm L}$ and $Y_{\rm L}$ operation, according to Table~\ref{table-II}. For $T_{\rm op}\lesssim T_{\rm c}$,  we can approximate $\Delta P_L$  by Eq.~\eqref{eq:DeltaP} with $T_{\rm c}$ replaced by $T_{\rm op}$. It is, however, possible to significantly reduce the probability offsets by measuring the gauge operators immediately after the end of the continuous operation (e.g., with projective measurements). This additional step would give us the actual value of the error syndrome path at the end of the continuous operation [i.e., $\mathcal{S}(T_{\rm op})$], and then we can infer (from Tables~\ref{table-II} and~\ref{table-III}) the undetected single-qubit error $E\sim Q_{\ell'}{O}'$ (here $O'=X_{\rm L}$, $Y_{\rm L}$ or $Z_{\rm L}$,  and ``$\sim$'' indicates equivalence modulo gauge operators) that induces the jump from the subspace $\mathcal{Q}_{\mathcal{S}_{\rm m}(T_{\rm op})}$ to the subspace $\mathcal{Q}_{\mathcal{S}(T_{\rm op})}$ near the end of the continuous operation. By adjusting {$\mathcal{O}_{\rm m}$ to $\mathcal{O}_{\rm m} O'$}, the probability of logical errors is approximately given by Eq.~\eqref{eq:Prob-logical-err} with $\Delta P_L$ set to zero, i.e., $P_{L}(T_{\rm op}) \approx \gamma_{L}^{\rm cont}\,T_{\rm op}$. 

\begin{figure}[t!]
\centering
\includegraphics[width=\linewidth, trim =5cm 0.5cm 4cm 0.5cm,clip=true]{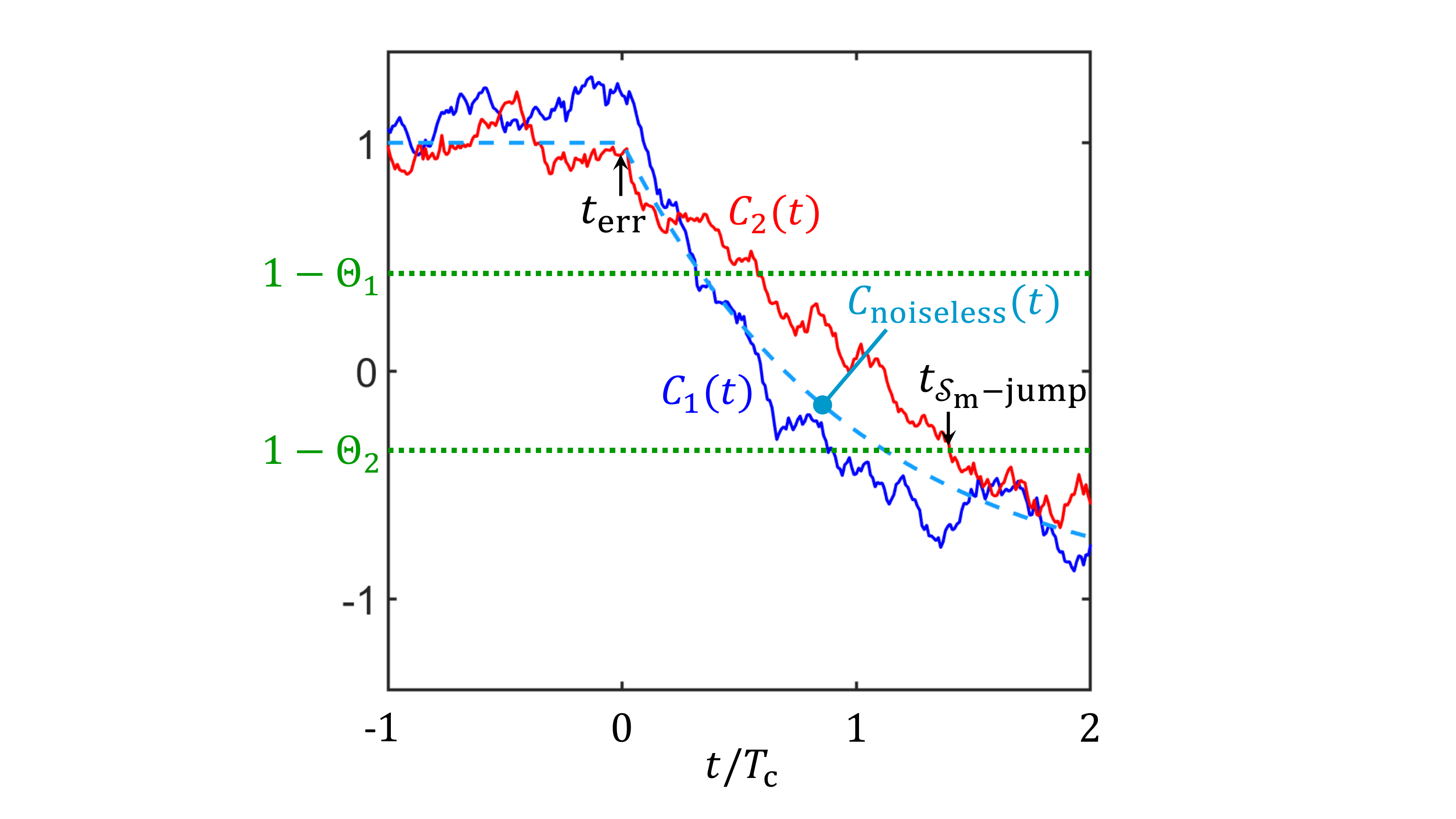}
\caption{Importance of using two error thresholds ($1-\Theta_{1}$ and $1-\Theta_2$) to determine the monitored error syndrome path $\mathcal{S}_{\rm m}(t)$. The blue and red wiggly curves depict two normalized cross-correlators that change sign due to an error happening at moment $t_{\rm err}$ (the other two cross-correlators are assumed not to be affected by the error and not shown). In this scenario, the two-error-threshold algorithm of Section~\ref{Section:monitored-error-syndrome-path} generates a $\mathcal{S}_{\rm m}(t)$ with just one jump at the first moment $t_{\mathcal{S}_{\rm m}\text{-jump}}$ (at this instant, the error is actually detected) when both cross-correlators are below the lower error threshold. This jump in $\mathcal{S}_{\rm m}(t)$ indicates that the error syndrome pattern changes from $++++$ to $--++$, which is the actual error syndrome change due to the error. In contrast, the one-error-threshold algorithm discussed in the main text (with error threshold at, say, $1-\Theta_2$) would lead to $\mathcal{S}_{\rm m}(t)$ with two false jumps since the cross-correlators cross the error threshold at different moments. These false jumps would indicate the error syndrome pattern changing from $++++$ to $-+++$ (at the moment when the blue cross-correlator crosses the error threshold) and then from $-+++$ to $--++$ (at the moment  when the red cross-correlator crosses the error threshold); consequently, two false errors would be detected instead of just one error. }
\label{fig:fig2}
\end{figure}

Before {concluding} this section, {we} discuss the difficulty of using a one-error-threshold {protocol} to extract the monitored error syndrome path. For example, we {might} regard the digitized values of cross-correlators as $-1$ ($+1$) if they are below (above) a certain error threshold, and then update $\mathcal{S}_{\rm m}(t)$ when some of such digitized cross-correlators changes sign, as {described} above. {Unfortunately, such a} one-error-threshold algorithm does not work, {leading} to logical error rates that scale linearly on the error rates of the physical qubits. {Consequently,} for sufficiently small {values of} $\Gamma_i^{(X,Y,Z)}$, continuous operation would perform worse than {the} discrete operation, {for which the} logical error rates scale quadratically {with} $\Gamma_i^{(X,Y,Z)}$, {see} Eq.~\eqref{eq:Gamma-logical-discrete}. The reason for the failure of such one-error-threshold {protocol} is the noise in the cross-correlators: {this} makes a single-qubit error event that affects several cross-correlators induce several false jumps in $\mathcal{S}_{\rm m}(t)$, {thereby} increasing the logical error probabilities. Figure~\ref{fig:fig2} shows an example where the one-error-threshold algorithm leads to $\mathcal{S}_{\rm m}(t)$ with two jumps instead of just one jump. 

\subsubsection{Optimal integration time $\tau_{\rm c}$}
\label{sec:optimal-tauc}
In this section we discuss the statistical properties of the cross-correlators of interest in the absence of errors,  assuming that the system state is in the code space. We specifically consider $\mathcal{C}_x^{(1)}(t)$ since all cross-correlators have the same statistical properties in the case of detectors with the same measurement strength {$\Gamma_{\rm m}$}. We are particularly interested in the limit where the integration time parameter $T_{\rm c}$ is much larger than the collapse time $\tau_{\rm coll}$ due to measurement,  {which is defined as}
\begin{equation}
\tau_{\rm coll}=\frac{1}{\Gamma_{\rm m}}=2\eta\tau_{\rm m}. 
\end{equation}
This limit is relevant for us because {we shall find that} the optimal continuous operation of the nine-qubit Bacon-Shor code requires $T_{\rm c}$ much larger than $\tau_{\rm coll}$---see Section~\ref{cont-operation-optimization}. 

In the large $T_{\rm c}$ limit, the fluctuations of the cross-correlator $\mathcal{C}_x^{(1)}(t)$ are approximately Gaussian because the integrand in Eq.~\eqref{eq:C1x} has  a comparatively short correlation time of order of $\tau_{\rm coll}\ll T_{\rm c}$---see Eq.~\eqref{eq:two-time-corr-Ctilde-result} {below}. The approximate Gaussian statistics of the cross-correlators of interest can be {justified} {by} the central-limit theorem if we consider the integration of Eq.~\eqref{eq:C1x} as a sum of contributions from small nonoverlapping intervals of duration much larger than $\tau_{\rm coll}$ and much shorter than $T_{\rm c}$. These contributions are random and approximately statistically independent, so, in this limit, their sum equal to $\mathcal{C}_x^{(1)}(t)$ can be regarded as a Gaussian random number, characterized by its mean $\big\langle \mathcal{C}_x^{(1)}(t) \big\rangle$ and its variance   
\begin{align}
\label{eq:corr-variance}
{\rm Var}\big[\mathcal{C}_x^{(1)}(t)\big] =\big\langle \big[\mathcal{C}_x^{(1)}(t)\big]^2 \big\rangle - \big\langle \mathcal{C}_x^{(1)}(t)\big\rangle^2. 
\end{align}
We actually characterize the relative size of the fluctuations of $\mathcal{C}_x^{(1)}(t)$ by their Signal-to-Noise Ratio (SNR),
\begin{align}
\label{eq:SNR}
{\rm SNR} = \frac{\big\langle \mathcal{C}_x^{(1)}(t) \big\rangle^2}{{\rm Var}\big[\mathcal{C}_x^{(1)}(t)\big]}.
\end{align}
We find {analytically and confirm} {numerically} that the SNR of the cross-correlators is proportional to $T_{\rm c}$ for sufficiently large $T_{\rm c}$, and it is a nonmonotonic function of the integration time parameter $\tau_{\rm c}$. We  {can} then determine the optimal value of $\tau_{\rm c}$ that maximizes the SNR. 

We are going to calculate the SNR in the stationary regime $t\gg T_{\rm c}$, where the statistical properties of the cross-correlator ${\mathcal{C}}_x^{(1)}(t)$ are time-independent. It is convenient to introduce  {the unfiltered correlator}
\begin{align}
\label{eq:corr-tilde}
\tilde{\mathcal{C}}_x^{(1)}(t) =  \mathcal{I}_{X_{12}}(t)\,\mathcal{I}_{X_{45}}(t)\,\mathcal{I}_{X_{78}}(t). 
\end{align}
In the large $T_{\rm c}$ limit, we can approximate the unfiltered correlator as
\begin{align}
\label{eq:tilde-corr-approx}
\tilde{\mathcal{C}}_x^{(1)}(t) \approx \big\langle \tilde{\mathcal{C}}_x^{(1)}(t) \big\rangle + \sqrt{\mathcal{D}_{\rm c}}\,\tilde\xi_{\rm c}(t), 
\end{align}
where $\tilde\xi_{\rm c}(t)$ is white noise with a two-time correlation function given by [its actual correlation function is given by Eq.~\eqref{eq:two-time-corr-Ctilde-result}]
\begin{align}
\label{eq:xi-tilde-corr}
\langle \tilde\xi_{\rm c}(t)\,\tilde\xi_{\rm c}(t') \rangle  = \delta(t-t'),
\end{align}
and $\mathcal{D}_{\rm c}$ is {an effective diffusion coefficient for the cross-correlator fluctuations,} given by 
\begin{align}
\label{eq:Dc}
\mathcal{D}_{\rm c} = 2\int_0^{\infty} dt\, \big\langle \big[\tilde{\mathcal{C}}_x^{(1)}(t) - \langle \tilde{\mathcal{C}}_x^{(1)}(t) \rangle\big]\! \big[\tilde{\mathcal{C}}_x^{(1)}(0) - \langle\tilde{\mathcal{C}}_x^{(1)}(0)\rangle\big]\big\rangle. 
\end{align}
Using Eq.~\eqref{eq:C1x} and Eqs.~\eqref{eq:corr-tilde}--\eqref{eq:xi-tilde-corr}, we find that the variance of the cross-correlator $\mathcal{C}_x^{(1)}(t)$ is equal to $\mathcal{D}_{\rm c}/2T_{\rm c}$, so { in the large $T_c$ limit,} the SNR of the cross-correlators of interest is equal to 
\begin{align}
\label{eq:SNR-v2}
{\rm SNR} = T_{\rm c}\,\frac{2\big\langle {\mathcal{C}}_x^{(1)}(t) \big\rangle^2}{\mathcal{D}_{\rm c}}.
\end{align}

In the stationary regime, it is easy to {further} see that $\langle {\mathcal{C}}_x^{(1)}(t) \rangle$ is equal to $\langle\tilde{\mathcal{C}}_x^{(1)}(t)\rangle$, since the latter is time-independent {in this regime} and then the integral of Eq.~\eqref{eq:C1x} is trivial.  {Combining} Eqs.~\eqref{eq:I-Gk-avg} and~\eqref{eq:corr-tilde}, we find that the averaged value of the unfiltered correlators is 
\begin{align}
\label{eq:mean-tilde-Corr-derivation-1}
\big\langle \tilde{\mathcal{C}}_x^{(1)}(t) \big\rangle =&\; \int_{-\infty}^tdt_1\int_{-\infty}^tdt_2\int_{-\infty}^tdt_3\, \langle I_{X_{12}}(t_1)I_{X_{45}}(t_2) \times \nonumber \\
&\; I_{X_{78}}(t_3) \rangle\, \frac{e^{-\frac{t-t_1}{\tau_{\rm c}} -\frac{t-t_2}{\tau_{\rm c}} -\frac{t-t_3}{\tau_{\rm c}}}}{\tau_{\rm c}^3},
\end{align}
where we have set the lower integration limits to $-\infty$, which is allowed in the stationary regime. As discussed in Section~\ref{evolution}, the actual measurement signals, $I_{G_k}(t)$, are the same (up to a sign factor) as the measurement signals $I_{\mathcal{G}_k}(t)$ [see Eq.~\eqref{eq:Ik-Ql}] from simultaneous continuous measurement of the noncommuting operators $\mathcal{G}_k$ that only act on the gauge qubits---see Eq.~\eqref{eq:curli-G_k}. The three-time correlator in the integrand of Eq.~\eqref{eq:mean-tilde-Corr-derivation-1} becomes 
\begin{align}
\label{eq:K3}
K_3(t_1,t_2,t_3) =&\; \langle I_{X_3^{\rm g}}(t_3) I_{X_{13}^{\rm g}}(t_2) I_{X_1^{\rm g}}(t_1)\rangle , \nonumber \\
=&\;\langle I_{X_{78}}(t_3)I_{X_{45}}(t_2)I_{X_{12}}(t_1) \rangle, 
\end{align}
since $X_{12}(t) = X_{1}^{\rm g}(t)$, $X_{45}(t) = X_{13}^{\rm g}(t)$ and $X_{78}(t) = X_{3}^{\rm g}(t)$; we assume that the system state is in the code space so the sign factors in Eq.~\eqref{eq:Ik-Ql} are one.  

We {shall} use the following result (shown in the Supplemental Material of Ref.~\cite{Atalaya2018a}, see also Ref.~\cite{Tilloy2018})
\begin{align}
\label{eq:KN-result}
&K_{\upsilon_1\upsilon_2...\upsilon_N}(t_1,t_2,...t_N) = \langle I_{\upsilon_N}(t_N) ...I_{\upsilon_{2}}(t_{2}) I_{\upsilon_{1}}(t_{1}) \rangle= \nonumber \\
& {\rm Tr}\big[\mathcal{M}_{t_N}\mathcal{E}(t_N|t_{N-1})\mathcal{M}_{t_{N-1}}...\mathcal{M}_{t_2}\mathcal{E}(t_2|t_{1}) \mathcal{M}_{t_1}\mathcal{E}(t_1|t_{0})\rho_{\rm 0}\big], 
\end{align}
{with $t_1<t_2<...<t_N$.} 
Equation~\eqref{eq:KN-result}  is a general formula for $N$-time correlators of measurement signals $I_\upsilon(t)$ from simultaneous continuous measurement of an arbitrary number of (commuting or noncommuting) observables $A_{\upsilon}$ of a quantum system. Here, $\mathcal{E}(t|t')$ is the trace-preserving ensemble-averaged evolution from time $t'$ to $t$ due to Lindblad term $\dot \rho_{\rm ens} = \mathcal{L}[\rho_{\rm ens}]$, while $\mathcal{M}_{t_k} \rho = (A_{\upsilon_k}\rho + \rho A_{\upsilon_k})/2$ is a trace-changing operation, related to {the} measurement of observables $A_{\upsilon_k}$. 

In our problem, the ensemble-averaged evolution operation $\mathcal{E}$ is determined by Eq.~\eqref{eq:rho_g-EOM} without the noises:
\begin{align}
\label{eq:ens-avg-EOM}
\dot \varrho_{\rm g, ens}(t) = \sum_k \frac{\Gamma_{\rm m}}{2}\,[\mathcal{G}_k\,\varrho_{\rm g, ens}(t) \,\mathcal{G}_k - \varrho_{\rm g, ens}(t)],
\end{align}
where $ \varrho_{\rm g, ens}(t) = \langle  \varrho_{\rm g}(t)\rangle$. Expanding formula~\eqref{eq:KN-result} with $N=3$, $t_0=t_1$, so {that} $\mathcal{E}(t_1|t_0)$ is the identity, and taking into account that $A_{\upsilon}$ are Hermitian operators, we obtain 
\begin{align}
\label{eq:K3-deriv1}
&K_{\upsilon_1\upsilon_2\upsilon_3}(t_1,t_2,t_3) =\frac{1}{2}\,{\rm Tr}\big[A_{\upsilon_3}\mathcal{E}(t_3|t_2) A_{\upsilon_2} \mathcal{E}(t_2|t_1) A_{\upsilon_1}\times\nonumber \\
&\;\varrho_{\rm g}(t_1)\big] + \frac{1}{2}\,{\rm Tr}\big[A_{\upsilon_3}\mathcal{E}(t_3|t_2) A_{\upsilon_2} \mathcal{E}(t_2|t_1) \varrho_{\rm g}(t_1)A_{\upsilon_1}\big]. 
\end{align}
The trace terms of Eq.~\eqref{eq:K3-deriv1} can be easily calculated from the ensemble-averaged evolution Eq.~\eqref{eq:ens-avg-EOM}. {We show below that} they are actually  independent of  $\varrho_{\rm g}(t_1)$ for {our} cases of interest where $A_{\upsilon_3}A_{\upsilon_2}=A_{\upsilon_1}$ and $A_{\upsilon_{1}}$, $A_{\upsilon_2}$ and $A_{\upsilon_3}$ are Pauli operators. 

Let us calculate Eq.~\eqref{eq:K3-deriv1} for $t_1<t_2<t_3$, so $A_{\upsilon_1} = X_1^{\rm g}$, $A_{\upsilon_2} = X_{13}^{\rm g}$ and $A_{\upsilon_3} =X_{3}^{\rm g}$. We write the first trace term of this equation as 
\begin{equation}
a(t_3) =\frac{1}{2}\, {\rm Tr}[X_3^{\rm g}\,\mathcal{E}(t_3|t_2)\,\tilde\varrho_{\rm g, ens}(t_2)],
\end{equation} 
with 
\begin{equation}
\tilde\varrho_{\rm g, ens}(t_2) = X_{13}^{\rm g}\,\mathcal{E}(t_2|t_1)\,X_1^{\rm g}\,\varrho_{\rm g}(t_1).
 \end{equation} 
By multiplying both sides of Eq.~\eqref{eq:ens-avg-EOM} by $X_3^{\rm g}$ and then taking  trace operations, we obtain 
\begin{equation}
\label{eq:a-eom}
\dot a(t_3) = -2\Gamma_{\rm m}\,a(t_3),
\end{equation} 
with the initial condition: $a(t_2) ={\rm Tr}[X_{3}^{\rm g}\,\tilde\varrho_{\rm g, ens}(t_2)]/2$. The decay rate $2\Gamma_{\rm m}$ in Eq.~\eqref{eq:a-eom} is due to the anticommutation of $X_3^{\rm g}$ with two gauge qubit operations $\mathcal{G}_k$; namely, $Z_{3}^{\rm g}$ and $Z_{34}^{\rm g}$, see Eq.~\eqref{eq:curli-G_k}. We then obtain the solution 
\begin{equation}
\label{eq:a-result}
a(t_3)=\frac{1}{2}\, e^{-2\Gamma_{\rm m}(t_3-t_2)}\,{\rm Tr}[X_{3}^{\rm g}\,\tilde\varrho_{\rm g, ens}(t_2)].
\end{equation}
The trace factor of Eq.~\eqref{eq:a-result} can be written as 
\begin{equation}
b(t_2)= {\rm Tr}[X_{3}^{\rm g}\,\tilde\varrho_{\rm g, ens}(t_2)] ={\rm Tr}[X_{1}^{\rm g}\,\mathcal{E}(t_2|t_1)\,X_{1}^{\rm g}\,\varrho_{\rm g}(t_1)].
\end{equation} 
We then derive the evolution equation for $b(t_2)$ from Eq.~\eqref{eq:ens-avg-EOM} by multiplying both sides of this equation by $X_1$ and then taking trace operations. We obtain 
\begin{equation}
\label{eq:b-eom}
\dot b(t_2) = -2\Gamma_{\rm m}\, b(t_2),
\end{equation} 
where the decay rate $2\Gamma_{\rm m}$ is now due to the anticommutation of $X_1^{\rm g}$ with $\mathcal{G}_k=Z_1^{\rm g}$ and $Z_{12}^{\rm g}$. The initial condition for Eq.~\eqref{eq:b-eom} is $b(t_1)={\rm Tr}[X_1^{\rm g}\,X_1^{\rm g}\,\varrho_{\rm g}(t_1)]=1$. Note that the first trace term of Eq.~\eqref{eq:K3-deriv1} is independent of $\varrho_{\rm g}(t_1)$ because $A_{\upsilon_3}A_{\upsilon_2}=A_{\upsilon_1}$ and $A_{\upsilon_1}=X_1^{\rm g}$ is a Pauli operator; the same holds for the second trace term of Eq.~\eqref{eq:K3-deriv1}. We obtain from Eq.~\eqref{eq:b-eom} 
\begin{equation}
b(t_2) = e^{-2\Gamma_{\rm m}(t_2-t_1)}. 
\end{equation}
The first trace term of Eq.~\eqref{eq:K3-deriv1} is therefore equal to 
\begin{equation}
\label{eq:K3-result-partial}
a(t_3) = \frac{1}{2}\,e^{-2\Gamma_{\rm m}(t_3-t_2)}\,e^{-2\Gamma_{\rm m}(t_2-t_1)}. 
\end{equation}
The calculation of the second trace term of Eq.~\eqref{eq:K3-deriv1} is similar, giving the same contribution~\eqref{eq:K3-result-partial} to $K_3$. The sought three-time correlator~\eqref{eq:K3} is then equal to 
\begin{equation}
\label{eq:K3-result-1}
K_3(t_1,t_2,t_3) = e^{-2\Gamma_{\rm m}(t_3-t_2)}\,e^{-2\Gamma_{\rm m}(t_2-t_1)}, \,{\rm for}\;  t_1<t_2<t_3.
\end{equation}

{We can proceed by similar means to} obtain 
\begin{eqnarray}
\label{eq:K3-result-2}
K_3(t_1,t_2,t_3) &=& e^{-4\Gamma_{\rm m}(t_2-t_3)}\,e^{-2\Gamma_{\rm m}(t_3-t_1)}, \,{\rm for}\;  t_1<t_3<t_2,\nonumber \\
\hspace{-1cm} &=&  e^{-2\Gamma_{\rm m}(t_3-t_1)}\,e^{-4\Gamma_{\rm m}(t_1-t_2)}, \,{\rm for}\;  t_2<t_1<t_3,\nonumber \\
&=& e^{-2\Gamma_{\rm m}(t_1-t_3)}\,e^{-4\Gamma_{\rm m}(t_3-t_2)}, \,{\rm for}\;  t_2<t_3<t_1,\nonumber \\
&=& e^{-4\Gamma_{\rm m}(t_2-t_1)}\,e^{-2\Gamma_{\rm m}(t_1-t_3)}, \,{\rm for}\;  t_3<t_1<t_2,\nonumber \\
&=& e^{-2\Gamma_{\rm m}(t_1-t_2)}\,e^{-2\Gamma_{\rm m}(t_2-t_3)}, \,{\rm for}\;  t_3<t_2<t_1.\nonumber \\
\end{eqnarray}
{Inserting the expressions}~\eqref{eq:K3-result-1}--\eqref{eq:K3-result-2} in Eq.~\eqref{eq:mean-tilde-Corr-derivation-1}, {and} performing the integrals{, we arrive at}
\begin{eqnarray}
\label{eq:mean-Ctilde-result}
&\mbox{}&\big\langle \tilde {\mathcal{C}}_x^{(1)}(t)\big\rangle =  \big\langle {\mathcal{C}}_x^{(1)}(t) \big\rangle =\frac{1}{3}\bigg[\frac{1}{(1+\Gamma_{\rm m}\tau_{\rm c})(1 + 2\Gamma_{\rm m}\tau_{\rm c})} + \nonumber \\
&\mbox{}& \frac{1}{(1+2\Gamma_{\rm m}\tau_{\rm c})^2}\! +\! \frac{1}{(1+\Gamma_{\rm m}\tau_{\rm c})(1 + 4\Gamma_{\rm m}\tau_{\rm c})}\bigg]\! \zeta_{7}^{(\ell)}\!\zeta_{8}^{(\ell)}\!\zeta_{9}^{(\ell)},
\end{eqnarray}
where we have included the sign factor $ \zeta_{7}^{(\ell)}\zeta_{8}^{(\ell)}\zeta_{9}^{(\ell)}$ {from Eq.~\eqref{eq:Ik-Ql}}, {which} can be nontrivial in some error subspaces. {However, this} sign factor does not change the SNR. 

Next, we proceed to calculate $\mathcal{D}_{\rm c}$ given in Eq.~\eqref{eq:Dc}. To do this, we first need to calculate the  two-time correlator of the unfiltered correlator $\tilde{\mathcal{C}}_x^{(1)}(t)$, given by
\begin{eqnarray}
\label{eq:two-time-corr-Ctilde}
&\mbox{}&\big\langle \tilde{\mathcal{C}}_x^{(1)}(t)\,\tilde{\mathcal{C}}_x^{(1)}(0)\big\rangle =\int_{-\infty}^tdt_1 \int_{-\infty}^tdt_2 \int_{-\infty}^tdt_3 \nonumber\\
&\mbox{}&\int_{-\infty}^0dt'_1\int_{-\infty}^0dt'_2 \int_{-\infty}^0dt'_3\; K_6(t_1,t'_1,t_2,t'_2,t_3,t'_3)  \nonumber \\  
&\mbox{}&\times \frac{e^{-\frac{t-t_1}{\tau_{\rm c}}-\frac{t-t_2}{\tau_{\rm c}} -\frac{t-t_3}{\tau_{\rm c}} + \frac{t'_1}{\tau_{\rm c}} + \frac{t'_2}{\tau_{\rm c}} + \frac{t'_3}{\tau_{\rm c}}}}{\tau_{\rm c}^6}, 
\end{eqnarray}
where 
\begin{eqnarray}
\label{eq:C6-def}
&K_6(t_1,t'_1,t_2,t'_2,t_3,t'_3)=\nonumber \\
&\big\langle I_{X_3^{\rm g}}(t_3)\,I_{X_3^{\rm g}}(t'_3)\, I_{X_{13}^{\rm g}}(t_2)\, I_{X_{13}^{\rm g}}(t'_2)\,  I_{X_1^{\rm g}}(t_1)\,I_{X_1^{\rm g}}(t'_1)\big\rangle, 
\end{eqnarray}
is a six-time correlator that we evaluate using formula~\eqref{eq:KN-result}. {The calculation of Eq.~\eqref{eq:two-time-corr-Ctilde} is cumbersome because of the time ordering, needed to evaluate the integrand of this equation using the result~\eqref{eq:KN-result}. We also have to take into account singular contritubitions to $K_6$ that occur when $t_1=t_1'$, $t_2=t_2'$ or $t_3=t_3'$, see Appendix~\ref{Appendix-D}}.  The final result reads as 
\begin{eqnarray}
\label{eq:two-time-corr-Ctilde-result}
&\mbox{}&\big\langle \tilde{\mathcal{C}}_x^{(1)}(t)\,\tilde{\mathcal{C}}_x^{(1)}(0)\big\rangle - \langle \tilde{\mathcal{C}}_x^{(1)}(t)\rangle^2  = \left[\frac{1}{8s^3\eta^3} + R_1\right]e^{-\frac{3|t|}{\tau_{\rm c}}} + \nonumber \\
&\mbox{}&\hspace{0.85cm}R_2\,e^{-\frac{|t|}{\tau_{\rm c}}-\frac{2|t|}{\tau_{\rm coll}}} +  R_3\,e^{-\frac{2|t|}{\tau_{\rm c}}-\frac{2|t|}{\tau_{\rm coll}}} + R_4\,e^{-\frac{|t|}{\tau_{\rm c}}-\frac{4|t|}{\tau_{\rm coll}}} + \nonumber\\
&\mbox{}&\hspace{.85cm}R_5\,e^{-\frac{2|t|}{\tau_{\rm c}}-\frac{4|t|}{\tau_{\rm coll}}},
\end{eqnarray}
where $s=2\tau_{\rm c}\tau_{\rm coll}^{-1}$, $R_l=R_l(s,\eta)$  is a rational function of $s$ and the quantum efficiency parameter $\eta$, see Appendix~\ref{Appendix-D}, and  $\langle \tilde{\mathcal{C}}_x^{(1)}(t)\big\rangle$ is given {explicitly} in Eq.~\eqref{eq:mean-Ctilde-result}. Note that the correlation time of the unfiltered correlator $\tilde{\mathcal{C}}_x^{(1)}(t)$ {in Eq.~\eqref {eq:two-time-corr-Ctilde-result}} is only determined by the integration time parameter $\tau_{\rm c}$ and the collapse time $\tau_{\rm coll}$ due to continuous measurement. We {shall} see below that the optimal integration time parameter $\tau_{\rm c}$ is of the order of $\tau_{\rm coll}$, so our earlier assumption that the fluctuations $\tilde{\xi}_{\rm c}(t)$ of the unfiltered correlators can be regarded as white{, i.e., unstructured,} is justified in the limit of interest {where} $T_{\rm c}/\tau_{\rm coll}\gg 1$. 

{We can now evaluate the effective diffusion coefficient $\mathcal{D}_{\rm c}$ from Eqs.~\eqref{eq:two-time-corr-Ctilde-result} and~\eqref{eq:Dc}, and hence obtain the SNR of the cross-correlators of interest, Eq.~\eqref{eq:SNR-v2}, in the large $T_c$ limit as}
\begin{widetext}
\begin{eqnarray}
\label{eq:SNR-result}
{\rm SNR}  &=&\; \,\frac{16\,(T_{\rm c}/\tau_{\rm coll})\,{\eta}^3\, s^2\left(s+1\right) \left(s+2\right) \left(s+3\right) \left(s+4\right) \left(2 s+1\right) \left(2 s+3\right) \left(8 s^2+15 s+6\right)^2}{\Big[864 + 216\left (55 + 36 {\eta}\right) s + 12\left (6145 + 7902 {\eta} + 2592 {\eta}^2\right) s^2 +  18\left(15211 + 28710 {\eta} + 18512 \eta^2\right) s^3}\nonumber \\
&\mbox{}&\;+3\left (226437 + 555214 \eta + 522576 \eta^2 + 13824 \eta^3\right) s^4 +  3\left (397086 + 1180221 \eta + 1426120 \eta^2 + 116992 \eta^3\right) s^5 \nonumber \\
&\mbox{}&\; +\left (1522503 + 5239407 \eta + 7535364 \eta^2 + 1272544 \eta^3\right) s^6 + 6\left(240069 + 925035 \eta + 1502502 \eta^2 + 430792 \eta^3\right) s^7 \nonumber \\
&\mbox{}&\; +\left(1013421 + 4259496 \eta + 7504140 \eta^2 + 3235624 \eta^3\right) s^8 + 3 \left(175818 + 789057 \eta + 1458060 \eta^2 +  863636 \eta^3\right) s^9   \nonumber \\
&\mbox{}&\;+ \left(199809 + 940479 \eta + 1770048 \eta^2 +   1331548 \eta^3\right) s^{10} + 4 \left(13347 + 64890 \eta + 121128 \eta^2 + 106744\eta^3\right) s^{11} \nonumber \\
&\mbox{}&\; + 24 \left(396 + 1963 \eta + 3548 \eta^2 + 3308 \eta^3\right) s^{12} + 16 \left(63 + 315 \eta + 540\eta^2 + 452 \eta^3\right) s^{13} + 48\left (1 + \eta\right) \left(1 + 2 \eta\right)^2 s^{14}\Big].\nonumber\\
\end{eqnarray}
\end{widetext}
\begin{figure}[b!]
\centering
\includegraphics[width=\linewidth, trim =2.5cm 0.5cm 3.5cm 0.5cm,clip=true]{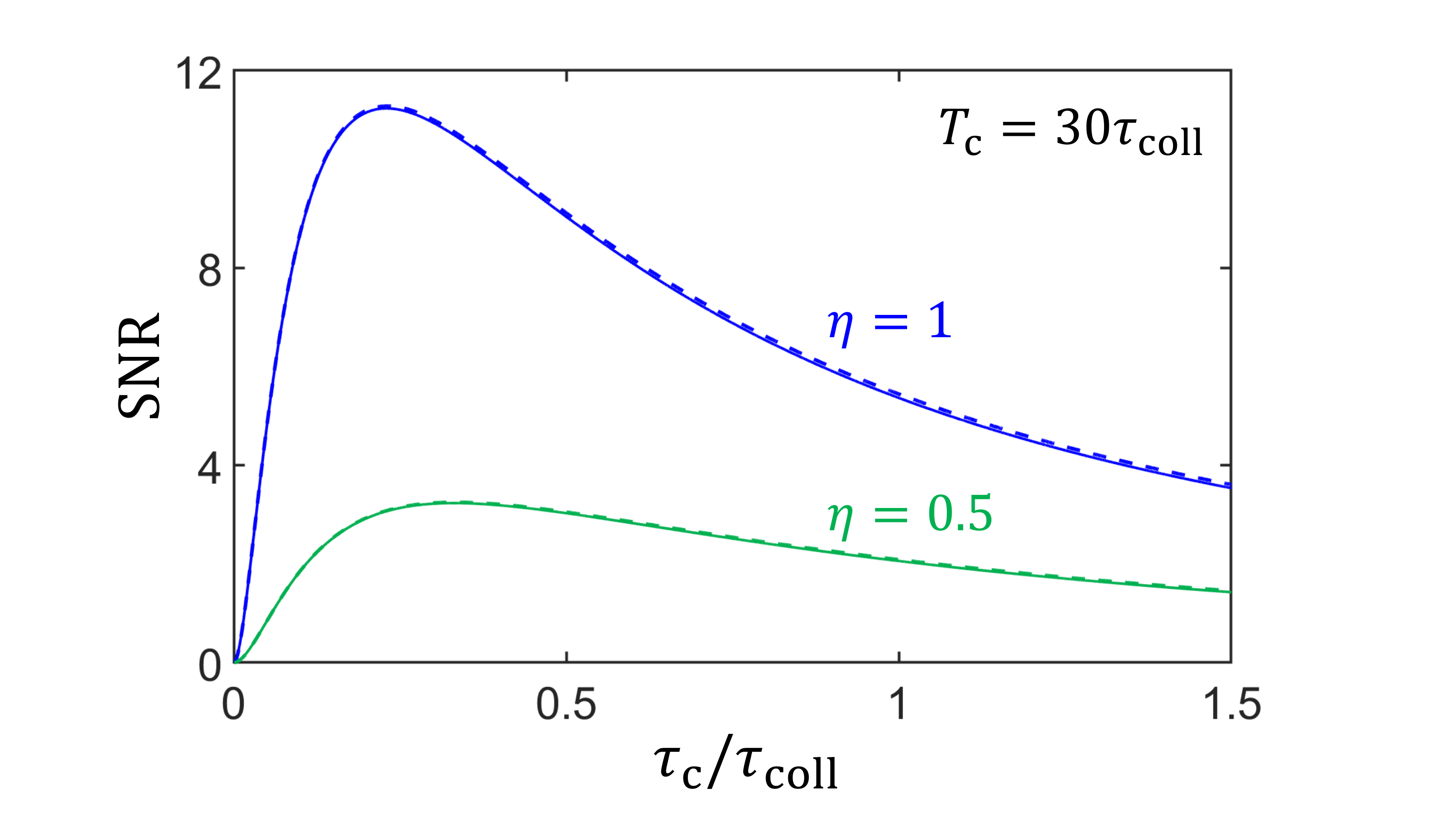}
\caption{SNR of cross-correlators as function of the integration time parameter $\tau_{\rm c}$. Solid lines plot  formula~\eqref{eq:SNR-result} that is valid in the large $T_{\rm c}$ limit for two values of quantum efficiency $\eta$. Dashed lines plot the analytical result that is valid for any value of $T_{\rm c}$ (not given).}
\label{fig:SNR}
\end{figure}

Figure~\ref{fig:SNR} plots {the value of the SNR Eq.}~\eqref{eq:SNR-result}  as function of the integration time parameter $\tau_{\rm c}$ {for two different values of the measurement efficiency $\eta$.} We note that the SNR decreases as $\tau_{\rm c}$ gets smaller. This is expected since the ``signal part'' of $\mathcal{C}_x^{(1)}(t)$ converges to one as $\tau_{\rm c}\to0$ (see Eq.~\eqref{eq:mean-Ctilde-result}), while the variance of the fluctuations of $\mathcal{C}_x^{(1)}(t)$ increases as $\tau_{\rm c}^{-2}$. Indeed, in the small $\tau_{\rm c}$ limit, the leading term of $\langle \tilde{\mathcal{C}}_x^{(1)}(t)\,\tilde{\mathcal{C}}_x^{(1)}(0)\rangle - \langle \tilde{\mathcal{C}}_x^{(1)}(t)\rangle^2$ is $\exp(-3|t|/\tau_{\rm c})\,\tau_{\rm coll}^3/64\eta^3\tau_{\rm c}^3$, so $\mathcal{D}_{\rm c}\approx \tau_{\rm coll}^3/96\eta^3\tau_{\rm c}^2$ and ${\rm SNR}\approx 192\eta^3\tau_{\rm c}^2T_{\rm c}/\tau_{\rm coll}^3$. In the large $\tau_{\rm c}$ limit, the SNR decreases as $\tau_{\rm c}^{-2}$ because the ``signal part'' decreases as $(\tau_{\rm c}^{-2})^2$, see Eq.~\eqref{eq:mean-Ctilde-result}, while the variance of the fluctuations of $\mathcal{C}^{(1)}_x(t)$ decreases as $\tau_{\rm c}^{-2}$. In Fig.~\ref{fig:SNR}, we have also plotted the exact analytical values of SNR (dashed lines),  obtained without taking the large $T_{\rm c}$ limit. We see that, for {an integration time} $T_{\rm c} = 30\tau_{\rm coll}$, the difference between the {estimates of} the SNR with and without taking the large $T_{\rm c}$ limit is small. We do not provide the analytical formula for the SNR {at an arbitrary {integration} time}  $T_{\rm c}$. {However, this} can be  {readily} {obtained from the result~\eqref{eq:two-time-corr-Ctilde-result} when it is used to calculate ${\rm Var}[\mathcal{C}_x^{(1)}(t)]$ in Eq.~\eqref{eq:SNR}.}

By maximizing the SNR with fixed {values of} $T_{\rm c}$ and  $\eta$, we  {arrive at} the optimal value of the integration time parameter $\tau_{\rm c}$
\begin{align}
\tau_{\rm c}^{\rm opt} \approx&\; 0.229\tau_{\rm coll} \;\;(\eta=1), \;\;\;  \tau_{\rm c}^{\rm opt} \approx\; 0.331\tau_{\rm coll} \;\; (\eta=0.5).
\end{align}
The optimal value of the second integration time parameter $T_{\rm c}$ is presented in Section~\ref{cont-operation-optimization}. 

\subsection{Logical error rates for continuous operation of nine-qubit Bacon-Shor code} 
\label{section:logical-error-rates-cnt}
\subsubsection{Large $T_{\rm c}$ limit}
\label{large-Tc-limit}
In this section we calculate the logical error rates for continuous operation of the nine-qubit Bacon-Shor code in the large $T_{\rm c}$ limit. {In this limit all} fluctuations of the  cross-correlators $\mathcal{C}_{q}^{(n)}(t)$ can be neglected ($q= x, z$ and $n = 1,2$).  {The} evolution of the normalized cross-correlators ({the} normalization factor is given by Eq.~\eqref{eq:avg-correlator-formula}; sometimes we {shall omit reference to the normalization}) due to occurrence of a single error at moment $t_{\rm err}$ is  {given} by (for $t>t_{\rm err}$)
\begin{align}
\label{eq:C-noiseless}
C_{\text{noiseless}} (t) = -1 + 2\,e^{-\frac{t-t_{\rm err}}{T_{\rm c}}}. 
\end{align}
Note that previous to the occurrence of the error, $C_{\text{noiseless}}(t)$ has the value of $+1$, and after the error it asymptotically approaches $-1$ at large times---see Fig~\ref{fig:fig2}. The exponential form of $C_{\text{noiseless}}(t)$ comes from the exponential weighting function used in the definition of the the cross-correlators---see Eq.~\eqref{eq:cross-correlators}. 

In the large $T_{\rm c}$ limit and for sufficiently small {physical} error rates $\Gamma_i^{(X,Y,Z)}$, 
 logical errors are due {primarily} to two error combinations ($E_1$ and $E_2$) that are incorrectly diagnosed as a single error ($E_{\rm false}$) by our QEC protocol. This happens when the two errors occur sufficiently close in time, {a situation similar to that in the discrete operation, where logical errors derive} from harmful two-qubit errors that occur within the same cycle. We denote the time window {in which} two errors {are} diagnosed as one error as $\Delta t_{\rm cont}$. {This means that if} the first and second errors occur at moments $t_{\rm err}^{(1)}$ and $t_{\rm err}^{(2)}$, {respectively}, they are not individually diagnosed {when} $t_{\rm err}^{(2)}-t_{\rm err}^{(1)}<\Delta t_{\rm cont}$. We   {shall} determine {the} time window {$\Delta t_{\rm cont}$} that is specific {to} our continuous QEC protocol and use it to calculate the corresponding logical error rates. We will see that the {resulting} formulas for the logical error rates in the continuous operation are similar to those of the discrete operation [Eqs.~\eqref{eq:Appendix-logical-X-error-rate}--\eqref{eq:Appendix-logical-Y-error-rate}], with $\Delta t_{\rm cont}$ playing the role of the cycle time $\Delta t$. 

{Detailed analysis shows that the time window} $\Delta t_{\rm cont}$  can  {take two possible} values, namely,
\begin{subequations}
\label{eq:Deltat-cont}
\begin{align}
\label{eq:Deltat-cont-1}
\Delta t_{\rm cont}^{(1)} =&\; T_{\rm c} \ln\bigg[\frac{2-\Theta_1}{2- \Theta_2}\bigg], \\
\Delta t_{\rm cont}^{(2)} =&\; T_{\rm c} \ln\bigg[\frac{2}{2- \Theta_2} \bigg], \label{eq:Deltat-cont-2} 
\end{align}
\end{subequations}
where $\Delta t_{\rm cont}^{(1)}$ is the time that $C_{\text{noiseless}}(t)$ spends in the ``syndrome uncertainty region'' (i.e., in between the error thresholds $1-\Theta_1$ and $1-\Theta_2$), and $\Delta t_{\rm cont}^{(2)}$ is the time that  $C_{\text{noiseless}}(t)$ takes to reach the lower error threshold ($1-\Theta_2$) after the error occurs. 

\begin{figure}[t!]
\centering
\includegraphics[width=\linewidth, trim =2.5cm 3.25cm 3.5cm 0.cm,clip=true]{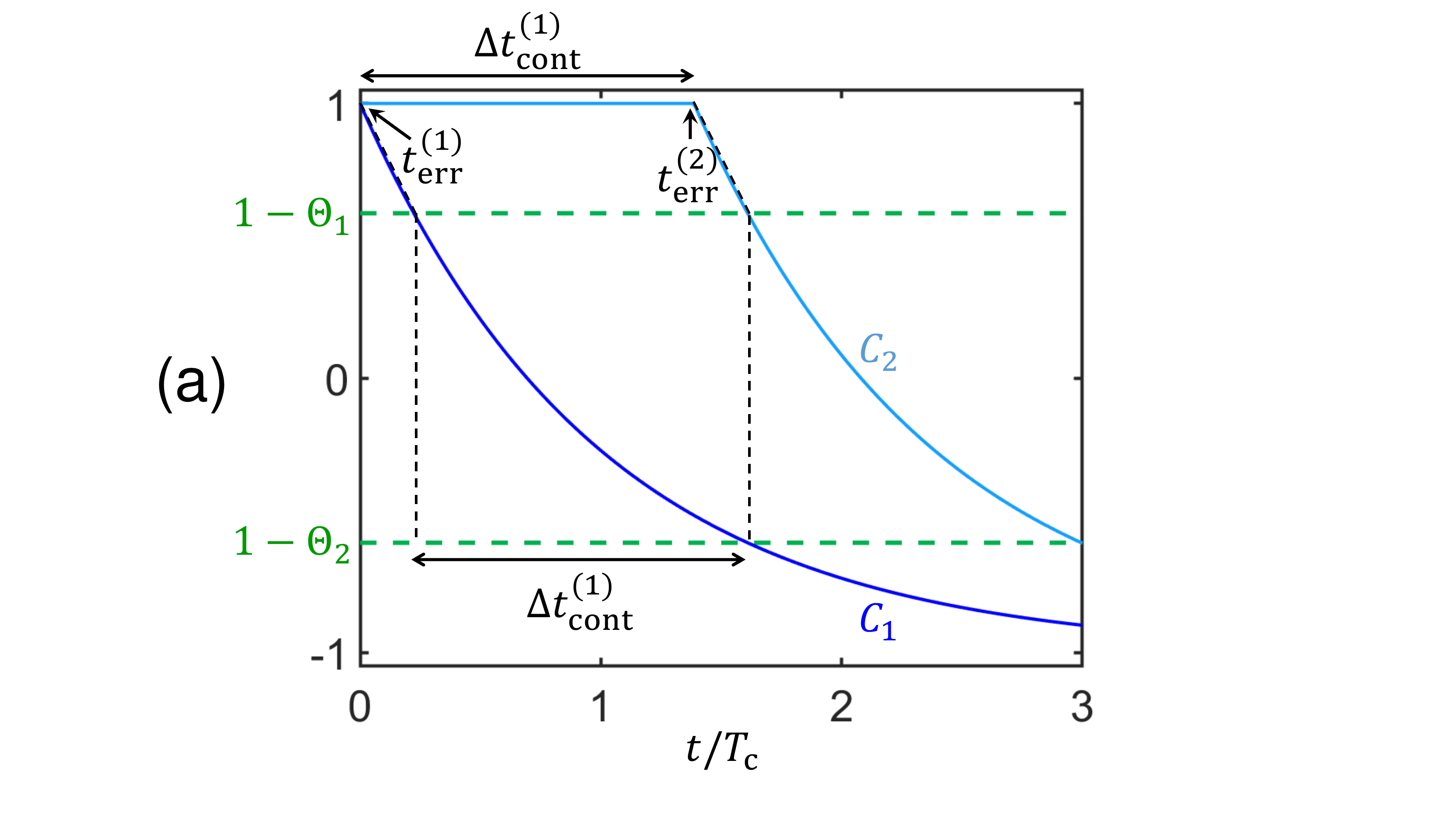}
\includegraphics[width=\linewidth, trim =2.5cm 3.25cm 3.5cm 2.cm,clip=true]{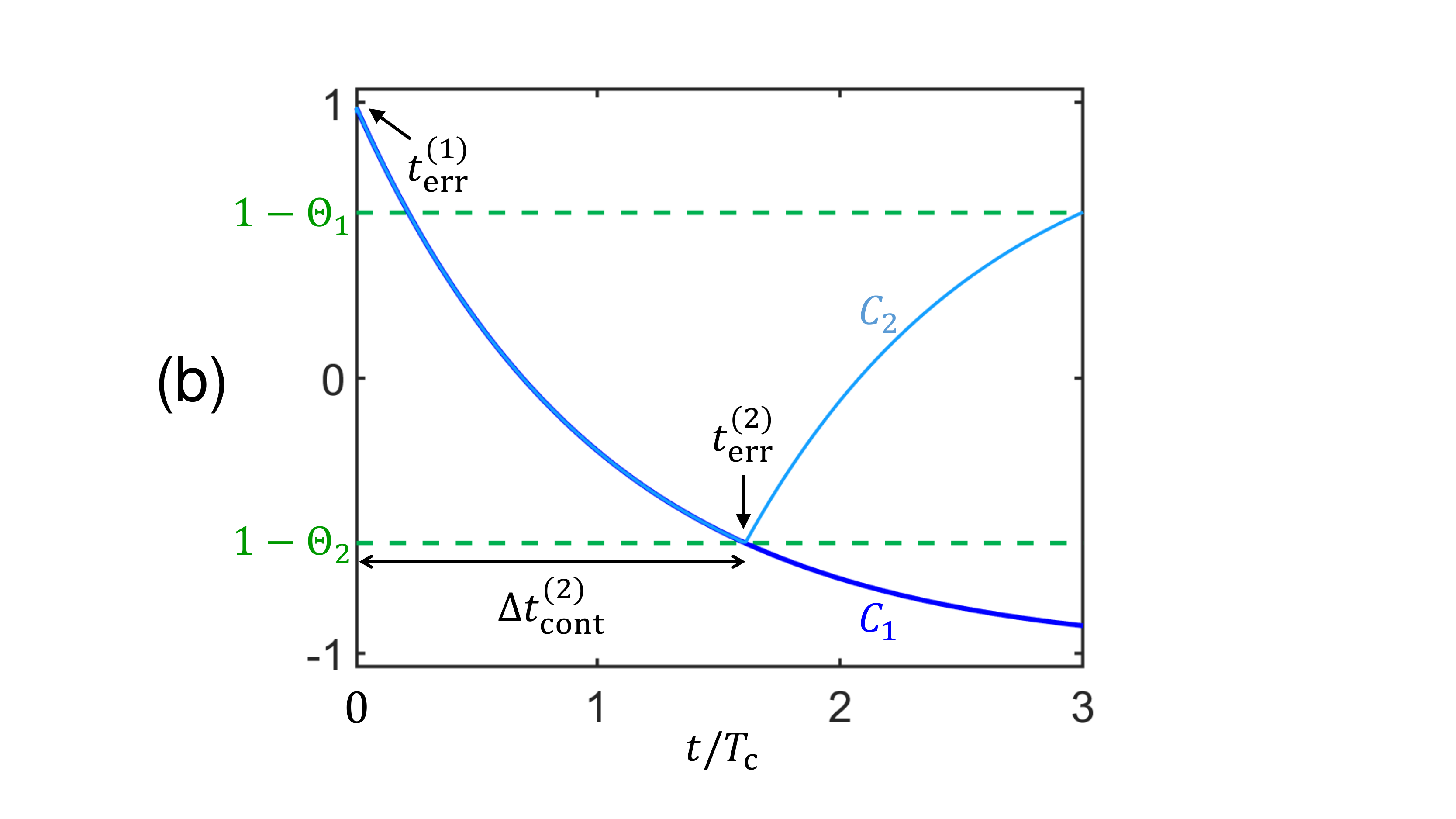}
\includegraphics[width=\linewidth, trim =2.5cm 1cm 3.5cm 2cm,clip=true]{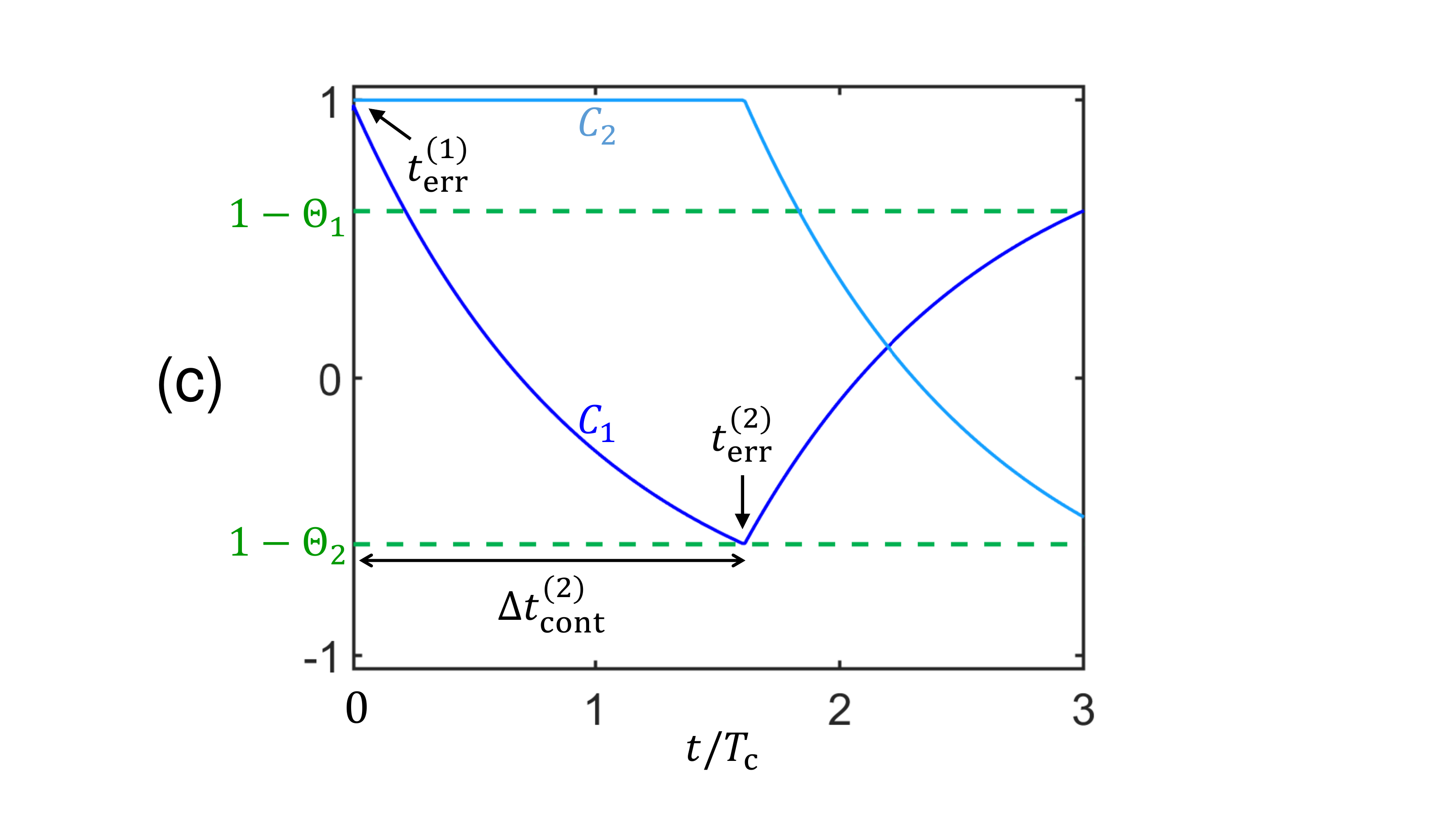}
\caption{Evolution of normalized cross-correlators in the large integration time $T_{\rm c}$ (noiseless) limit after occurrence of two errors. Cross-correlators not affected by errors are constant and equal to $+1$ (not shown). Panel~(a) shows the evolution of two nontrivial cross-correlators $C_1$ and $C_2$ in the case where the first error changes the error syndrome pattern from $++++$ to $-+++$ at moment $t_{\rm err}^{(1)}$ (so only $C_1$ is affected by the first error) and the second error changes the error syndrome pattern from $-+++$ to $--++$ at moment $t_{\rm err}^{(2)}$ (so only $C_2$ is affected by the second error). Both cross-correlators approach $-1$ at large times. Panel~(b) shows the cross-correlators after the errors change the error syndrome pattern from $++++$ to $--++$ and then from $--++$ to $-+++$, so first error affects both correlators and then second error affects only $C_2$. Cross-correlators $C_1$ and $C_2$ approach $-1$ and $+1$ at larger times, respectively. Panel~(c) shows the cross-correlators for the case where the error syndrome pattern changes from $++++$ to $-+++$ and then from $-+++$ to $+-++$. At large times, the correlators $C_1$ and $C_2$ approach $+1$ and $-1$, respectively.}
\label{fig:GL-cont-cases}
\end{figure}

To see why $\Delta t_{\rm cont}$ has the values given in Eq.~\eqref{eq:Deltat-cont}, we analyze the scenarios depicted in Fig.~\ref{fig:GL-cont-cases}. This figure shows the nontrivial noiseless cross-correlators that are affected by the two errors (trivial cross-correlators unchanged by the errors are constant and equal to $+1$). We assume that the first error ($E_1$) occurs at moment $t_{\rm err}^{(1)}=0$ and affects, for simplicity, only one cross-correlator, denoted by $C_1(t)$, and the second error ($E_2$) also affects only one cross-correlator, denoted by $C_2(t)$. In Fig.~\ref{fig:GL-cont-cases}~(a), the first error changes the error syndrome pattern, say, from $++++$ to $-+++$ and then the second error changes the error  syndrome pattern from $-+++$ to $--++$. To detect {both of these} changes in the error syndrome pattern using our continuous QEC protocol, $C_2(t)$ has to cross the upper error threshold ($1-\Theta_1$) later than the moment {at which} $C_1(t)$ exits the ``syndrome uncertainty region''. Otherwise, our algorithm would detect only one error syndrome transition from $++++$ to $--++$. It is easy to see from Fig.~\ref{fig:GL-cont-cases}~(a) and  Eq.~\eqref{eq:C-noiseless} that errors $E_1$ and $E_2$ are not individually detectable if $t_{\rm err}^{(2)} - t_{\rm err}^{(1)} < \Delta t_{\rm cont}^{(1)}$, with $\Delta t_{\rm cont}^{(1)}$ given in Eq.~\eqref{eq:Deltat-cont-1}. 

The scenarios shown in Fig.~\ref{fig:GL-cont-cases}~(b)--(c) are similar, in the sense that {both} lead to the time window $\Delta t_{\rm cont}^{(2)}$.  {However,} the error syndrome transitions are  different in {the two} scenarios. In Fig.~\ref{fig:GL-cont-cases}~(b), the first error $E_1$ changes the error syndrome pattern from $++++$ to $--++$ and then the second error $E_2$ changes it from $--++$ to $-+++$. In Fig.~\ref{fig:GL-cont-cases}~(c), $E_1$ changes the error syndrome pattern from $++++$ to $-+++$ and then $E_2$  changes it from $-+++$ to $+-++$. {Each of these transitions between} error syndromes {values} are individually detectable if the second error occurs after the cross-correlator $C_1(t)$ exits the ``syndrome uncertainty region''.  This condition leads to  {the time window} $\Delta t_{\rm cont}=\Delta t_{\rm cont}^{(2)}$ {that is}  given in Eq.~\eqref{eq:Deltat-cont-2}. 

To get a logical error, the misdiagnosed two error combination $E_1E_2$ has to be {one of the} harmful two-qubit errors given in Eqs.~\eqref{eq:two-qubit-X-errors}--\eqref{eq:two-qubit-Y-errors}---see also Appendix~\ref{Section:Harmful-two-qubit-errors-continuous}. Indeed, for $E_{\rm false}$ (the incorrectly diagnosed single-qubit error) and $E_1E_2$ to produce the same jump of the system state from the code space to the some fixed error subspace, the product $E_1E_2E_{\rm false}$ has to be trivial (i.e., a product of gauge operators) or equivalent ({modulo} gauge operators) to a logical operation ($X_{\rm L}$, $Y_{\rm L}$, or $Z_{\rm L}$). The latter possibility is the condition for a harmful two-qubit error that is given in Eq.~\eqref{eq:Appendix-logical-error-condition} with $E_{\rm false}$ playing the role of the error correcting operation $C_{\rm op}$---see also Appendix~\ref{Section:Harmful-two-qubit-errors-continuous}. The probability that the harmful two-qubit error $E_1E_2$ is misdiagnosed is given by $\big(\Gamma_{E_1}T_{\rm op}\big)\big(\Gamma_{E_2}\Delta t_{\rm cont}\big)$, where the first factor is the probability that first error occurs during the operation duration $T_{\rm op}$ and the second factor is the probability that the second error occurs within the time window $\Delta t_{\rm cont}$. The corresponding logical error rate is obtained by dividing this probability by $T_{\rm op}$.

To find the contributions to the logical $X$, $Z$ and $Y$ error rates from scenario Fig.~\ref{fig:GL-cont-cases}~(a), we look for all harmful two-qubit error combinations from Eqs.~\eqref{eq:two-qubit-X-errors}--\eqref{eq:two-qubit-Y-errors} that lead to cross-correlators evolving as shown in Fig.~\ref{fig:GL-cont-cases}~(a). (Note that we also have to include harmful two-qubit errors where each error affects two or more cross-correlators at the same time; in this case, the blue and cyan curves in Fig.~\ref{fig:GL-cont-cases}~(a) correspond to cross-correlators that evolve in the same fashion without noise.) We obtain 
\begin{subequations}
\label{eq:GL-cont-1}
\begin{align}
&\Delta_1 \gamma_{X} = 2\, \Delta t_{\rm cont}^{(1)}\,\Big[\big(\Gamma^{(X)}_1 + \Gamma^{(X)}_2 +\Gamma^{(X)}_3\big)\big(\Gamma^{(X)}_7 + \Gamma^{(X)}_8 + \nonumber \\
&\;\;\;\Gamma^{(X)}_9\big) + \big(\Gamma^{(Y)}_1 + \Gamma^{(Y)}_2 +\Gamma^{(Y)}_3\big)\big(\Gamma^{(X)}_7 + \Gamma^{(X)}_8 + \Gamma^{(X)}_9\big) +\nonumber \\
&\;\;\; \big(\Gamma^{(X)}_1 + \Gamma^{(X)}_2 +\Gamma^{(X)}_3\big)\big(\Gamma^{(Y)}_7 + \Gamma^{(Y)}_8 +  \Gamma^{(Y)}_9\big) \Big],  \label{eq:GL-cont-1-A}
\end{align}
\begin{align}
&\Delta_1\gamma_{Z} = 2\, \Delta t_{\rm cont}^{(1)}\,\Big[\big(\Gamma^{(Z)}_1 + \Gamma^{(Z)}_4 +\Gamma^{(Z)}_7\big)\big(\Gamma^{(Z)}_3 + \Gamma^{(Z)}_6 + \nonumber \\
&\;\;\;\Gamma^{(Z)}_9\big) + \big(\Gamma^{(Y)}_1 + \Gamma^{(Y)}_4 +\Gamma^{(Y)}_7\big)\big(\Gamma^{(Z)}_3 + \Gamma^{(Z)}_6 + \Gamma^{(Z)}_9\big) +\nonumber \\
&\;\;\;\big(\Gamma^{(Z)}_1 + \Gamma^{(Z)}_4 +\Gamma^{(Z)}_7\big)\big(\Gamma^{(Y)}_3 + \Gamma^{(Y)}_6 +  \Gamma^{(Y)}_9\big) \Big], \label{eq:GL-cont-1-B}
\end{align}
\begin{align}
\Delta_1\gamma_{Y} = 2\, \Delta t_{\rm cont}^{(1)}\,\Big[\Gamma_1^{(Y)}\Gamma_9^{(Y)} + \Gamma_3^{(Y)}\Gamma_7^{(Y)} \Big], \label{eq:GL-cont-1-C}
\end{align}
\end{subequations}
where the factor of two in Eqs.~\eqref{eq:GL-cont-1-A}--\eqref{eq:GL-cont-1-C} is related to the order of the  occurrence of errors $E_1$ and $E_2$ {(two-qubit error combinations ($E_1,E_2$) and $(E_2,E_1)$ give the same contribution for the logical $X$, $Y$ or $Z$ error rates)}.
\newline
We proceed in a similar manner to obtain the contributions to the logical error rates from the scenarios shown in Fig.~\ref{fig:GL-cont-cases}~(b)--(c). We obtain 
\begin{subequations}
\label{eq:GL-cont-2}
\begin{align}
&\Delta_2 \gamma_{X} = 2\,\Delta t_{\rm cont}^{(2)}\,\Big[\big(\Gamma_1^{(X)}  + \Gamma_2^{(X)} + \Gamma_3^{(X)} + \Gamma_7^{(X)} + \Gamma_8^{(X)} \nonumber \\
&\;+ \Gamma_9^{(X)}\big)\big(\Gamma_4^{(X)} + \Gamma_5^{(X)} + \Gamma_6^{(X)} + \Gamma_4^{(Y)} + \Gamma_5^{(Y)} + \Gamma_6^{(Y)} \big) + \nonumber \\
&\,\big(\Gamma_1^{(Y)}  + \Gamma_2^{(Y)} + \Gamma_3^{(Y)} + \Gamma_7^{(Y)} + \Gamma_8^{(Y)}  + \Gamma_9^{(Y)}\big)\big(\Gamma_4^{(X)} + \Gamma_5^{(X)} \nonumber \\
&\;+\Gamma_6^{(X)}\big) + \Gamma_4^{(Y)}\Gamma_7^{(Y)} +  \Gamma_1^{(Y)}\big(\Gamma_4^{(Y)} + \Gamma_7^{(Y)}\big)  +  \Gamma_5^{(Y)}\Gamma_8^{(Y)}  \nonumber \\
&\;+\Gamma_2^{(Y)}  \big(\Gamma_5^{(Y)} + \Gamma_8^{(Y)}\big)+ \Gamma_6^{(Y)}\Gamma_9^{(Y)} + \Gamma_3^{(Y)}\big(\Gamma_6^{(Y)} + \Gamma_9^{(Y)}\big) \Big],
\end{align}
\begin{align}
&\Delta_2 \gamma_{Z} = 2\,\Delta t_{\rm cont}^{(2)}\,\Big[\big(\Gamma_1^{(Z)}  + \Gamma_4^{(Z)} + \Gamma_7^{(Z)} + \Gamma_3^{(Z)} + \Gamma_6^{(Z)} \nonumber \\
&\;+ \Gamma_9^{(Z)}\big)\big(\Gamma_2^{(Z)} + \Gamma_5^{(Z)} + \Gamma_8^{(Z)} + \Gamma_2^{(Y)} + \Gamma_5^{(Y)} + \Gamma_8^{(Y)} \big) + \nonumber \\
&\,\big(\Gamma_1^{(Y)}  + \Gamma_4^{(Y)} + \Gamma_7^{(Y)} + \Gamma_3^{(Y)} + \Gamma_6^{(Y)}  + \Gamma_9^{(Y)}\big)\big(\Gamma_2^{(Z)} + \Gamma_5^{(Z)} \nonumber \\
&\;+\Gamma_8^{(Z)}\big) + \Gamma_2^{(Y)}\Gamma_3^{(Y)} +  \Gamma_1^{(Y)}\big(\Gamma_2^{(Y)} + \Gamma_3^{(Y)}\big)  +  \Gamma_5^{(Y)}\Gamma_6^{(Y)}  \nonumber \\
&\;+\Gamma_4^{(Y)}  \big(\Gamma_5^{(Y)} + \Gamma_6^{(Y)}\big)+ \Gamma_8^{(Y)}\Gamma_9^{(Y)} + \Gamma_7^{(Y)}\big(\Gamma_8^{(Y)} + \Gamma_9^{(Y)}\big) \Big], 
\end{align} 
\begin{align}
&\Delta_2\gamma_{Y} = 2\, \Delta t_{\rm cont}^{(2)}\,\Big[\Gamma_2^{(Y)}\big(\Gamma_4^{(Y)} + \Gamma_7^{(Y)} + \Gamma_6^{(Y)} + \Gamma_9^{(Y)}\big) +\nonumber \\
&\;\;\Gamma_1^{(Y)}\big(\Gamma_5^{(Y)} + \Gamma_8^{(Y)} + \Gamma_6^{(Y)}\big)  + \Gamma_3^{(Y)}\big(\Gamma_4^{(Y)}  + \Gamma_5^{(Y)}  + \Gamma_8^{(Y)}\big) \nonumber \\
&\;\,+\big(\Gamma_5^{(Y)} +\Gamma_6^{(Y)}\big)\Gamma_7^{(Y)} + \big(\Gamma_4^{(Y)} + \Gamma_6^{(Y)}\big)\Gamma_8^{(Y)} \nonumber \\
&\;\,+\big(\Gamma_4^{(Y)} + \Gamma_5^{(Y)}\big)\Gamma_9^{(Y)} \Big]. 
\end{align}
\end{subequations}
The logical $X$, $Z$ and $Y$ error rates are {then} obtained from the sum of {the} corresponding contributions given in Eqs.~\eqref{eq:GL-cont-1}--\eqref{eq:GL-cont-2}, {e.g., $\Delta_1 \gamma_X + \Delta_2 \gamma_X = \gamma_X^{{\rm cont}}$}. We evaluate the logical error rates for the depolarizing channel {of} Eq.~\eqref{eq:Gi-depolarization}.  {For this large $T_{\rm c}$ limit we then obtain}
\begin{subequations}
\label{eq:cont-gammaL-XYZ}
\begin{align}
\gamma^{{\rm cont}}_X = \gamma_Z^{{\rm cont}} = 6T_{\rm c}\Gamma_{\rm d}^2\ln\!\bigg[\frac{2-\Theta_1}{2-\Theta_2} \bigg] + 14T_{\rm c}\Gamma_{\rm d}^2\ln\!\bigg[\frac{2}{2-\Theta_2} \bigg], \label{eq:cont-gammaL-XZ}
\end{align}
\begin{align}
\gamma_Y^{{\rm cont}} = \frac{4}{9}T_{\rm c}\Gamma_{\rm d}^2 \ln\bigg[\frac{2-\Theta_1}{2-\Theta_2} \bigg] + \frac{32}{9}T_{\rm c}\Gamma_{\rm d}^2 \ln\bigg[\frac{2}{2-\Theta_2} \bigg].\label{eq:cont-gammaL-Y}
\end{align}
\end{subequations}
 The total logical error rate, $\gamma_{{\rm cont}} = \gamma_X^{{\rm cont}}+\gamma_Y^{{\rm cont}}+\gamma_Z^{{\rm cont}}$, reads as 
\begin{align}
\label{eq:cont-gammaL-total}
\gamma_{{\rm cont}} =\frac{112}{9}T_{\rm c}\Gamma_{\rm d}^2\ln\!\bigg[\frac{2-\Theta_1}{2-\Theta_2} \bigg] + \frac{284}{9}T_{\rm c}\Gamma_{\rm d}^2\ln\!\bigg[\frac{2}{2-\Theta_2} \bigg],
\end{align}
in the large $T_{\rm c}$ limit. We {thus find} that our continuous QEC protocol leads to logical error rates that scale quadratically on the error rates of the physical qubits {($\Gamma_{\rm d}$ in Eq.~\eqref{eq:cont-gammaL-total})}. This scaling shows that our QEC protocol can successfully detect and correct single-qubit errors if they occur sufficiently far apart in time. A somewhat similar condition also applies in the discrete operation for single-qubit errors to be correctable, {namely, that} they have to occur in different cycles. Note also the similarity between formulas Eq.~\eqref{eq:cont-gammaL-XYZ} and Eq.~\eqref{eq:discrete-logical-error-rate}; this indicates that the integration time parameter $T_{\rm c}$ (up to a proportionality factor) plays the role of the cycle time $\Delta t$ in the discrete operation of the nine-qubit Bacon-Shor code. 

\subsubsection{Small $T_{\rm c}$ limit}
In this limit we cannot neglect fluctuations in the cross-correlators $\mathcal{C}_{q}^{(n)}(t)$  since they can make single-qubit errors appear as two-qubit errors to our continuous QEC protocol, potentially leading to logical errors. That is, the measurement noise present in the continuous operation can  {render} single-qubit errors uncorrectable. {We shall assume that fluctuations of the cross-correlators are Gaussian; this assumption is justified in Section~\ref{sec:optimal-tauc}.}

We now discuss the two most probable scenarios {in which} large fluctuations in the cross-correlators lead to logical errors. 

\noindent {\it Scenario 1:} In scenario one, a single-qubit error $E$ flips the sign of two stabilizer generators at the same time (so error syndrome changes from, e.g., $++++$ to $--++$). If the affected (normalized) cross-correlators, referred to {for simplicity here} as $C_1(t)$ and $C_2(t)$,  do not undergo large fluctuations, they should follow trajectories like those shown in Fig.~\ref{fig:fig2} and then our continuous QEC protocol will detect the actual single-qubit error without problems at the first moment when both cross-correlators are below the lower error threshold. However, our continuous QEC protocol will detect two errors (instead of one error) if one of the cross-correlators crosses the upper error threshold due to a positive large fluctuation while the other  is below the lower error threshold without undergoing large fluctuations. This situation is somewhat similar to the one depicted in Fig.~\ref{fig:GL-cont-cases}~(b), except that the rise of  $C_2(t)$ is {now}  due {not} to a second error but  to a large fluctuation. Naively speaking, the probability that this scenario occurs in one experimental realization is given by 
\begin{align}
\label{eq:Pscn-1}
P_{\text{scn-1}} = 2\,\Gamma_E \,T_{\rm op}\,P_1(\Delta C\geq\Theta_2-\Theta_1),
\end{align}
where the last factor is the probability that the correlator difference $\Delta C(t) = C_2(t)-C_1(t)$ is larger than the difference of the error thresholds that is equal to $\Theta_2-\Theta_1$ and $\Gamma_E$ is the occurrence rate of the actual error. The factor of two in Eq.~\eqref{eq:Pscn-1} is due to the fact that a logical error can come from a large fluctuation in either $C_1(t)$ or $C_2(t)$. The corresponding logical error rate is $P_{\text{scn-1}}/T_{\rm op}$. \newline

To find the probability factor in Eq.~\eqref{eq:Pscn-1}, we consider the following evolution equations for the two normalized cross-correlators $C_1(t)$ and $C_2(t)$ that are affected by the actual error 
\begin{align}
\dot{C}_1(t) =&\; -\frac{C_1(t) - s_1}{T_{\rm c}} +\frac{\sqrt{\mathcal{D}_{\rm c}/ \langle \mathcal{C}\rangle^{2}}}{T_{\rm c}}\, \tilde{\xi}_{\rm c}^{(1)}(t), \label{eq:C1-EOM}\\
\dot{C}_2(t) =&\; -\frac{C_2(t) - s_2}{T_{\rm c}} + \frac{\sqrt{\mathcal{D}_{\rm c}/ \langle \mathcal{C}\rangle^{2}}}{T_{\rm c}}\, \tilde{\xi}_{\rm c}^{(2)}(t), \label{eq:C2-EOM}
\end{align}
where $s_k=\pm1$ are the values of the corresponding stabilizer generators, {the diffusion coefficient} $\mathcal{D}_{\rm c}$ can be obtained from Eqs.~\eqref{eq:SNR-v2} and~\eqref{eq:SNR-result} with $ \langle \mathcal{C}\rangle = \langle \mathcal{C}_x^{(1)}\rangle $ given in Eq.~\eqref{eq:mean-Ctilde-result}, and the uncorrelated white noises $\tilde{\xi}_{\rm c}^{(1,2)}(t)$ have a two-time correlation function given by ($m,n=1,2$) \begin{align}\langle\tilde{\xi}_{\rm c}^{(m)}(t)\tilde{\xi}_{\rm c}^{(n)}(t')\rangle=\delta_{mn}\delta (t-t').\end{align} Note that the factors in front of the noises in Eqs.~\eqref{eq:C1-EOM}--\eqref{eq:C1-EOM} are inversely proportional to $T_{\rm c}$: {thus} the smaller $T_{\rm c}$, the larger the fluctuations of the cross-correlators. Since we consider errors that simultaneously flip the sign of two stabilizer generators, we set $s_1=s_2=-1$. Before the actual error happens, the normalized cross-correlators fluctuate around $+1$ with a typical standard deviation of $\left[\mathcal{D}_{\rm c}/2T_{\rm c}\langle \mathcal{C}\rangle^2\right]^{1/2} = {\rm SNR}^{-1/2}$, where the signal-to-noise ratio SNR is given in ~Eq.~\eqref{eq:SNR-result}. The following evolution equation for the correlator difference $\Delta C(t) $ is obtained from Eqs.~\eqref{eq:C1-EOM}--\eqref{eq:C2-EOM}: 
\begin{align}
\Delta \dot{C}(t) =-\frac{\Delta C(t) }{T_{\rm c}} +\frac{\sqrt{2\mathcal{D}_{\rm c}/ \langle \mathcal{C}\rangle^{2}}}{T_{\rm c}}\, \Delta \tilde{\xi}_{\rm c}(t), \label{eq:DeltaC-EOM}
\end{align}
where the white noise $\Delta\tilde{\xi}_{\rm c}(t)=\big[\tilde{\xi}_{\rm c}^{(2)}(t)-\tilde{\xi}_{\rm c}^{(1)}(t)\big]/\sqrt{2}$ has the same two-time correlation function as $\tilde{\xi}_{\rm c}^{(1)}(t)$ or $\tilde{\xi}_{\rm c}^{(2)}(t)$. Note that $\langle \Delta C(t)\rangle = 0$ before and after the occurrence of the error, so the stochastic process $\Delta C(t)$ is actually stationary and not affected by the error. {The stationary probability distribution function of the correlator difference $\Delta C$ can be  obtained from Eq.~\eqref{eq:DeltaC-EOM}] as:}
\begin{align}
\label{eq:Pst1-DeltaC}
P_{1,{\rm st}}(\Delta C) = \left[\frac{{\rm SNR}}{4\pi}\right]^{1/2}\,\exp\left({-\frac{{\rm SNR}}{4}\,\Delta C^2}\right),
\end{align}
which can be exponentially small as we increase the integration time parameter $T_{\rm c}$ since ${\rm SNR}\propto T_{\rm c}$, for sufficiently large $T_{\rm c}$, see Eq.~\eqref{eq:SNR-result}. Furthermore, the probability that $\Delta C(t)\geq \Theta_2-\Theta_1$ is time-independent and  can be obtained from Eq.~\eqref{eq:Pst1-DeltaC} and  expressed in terms of the error function Erf as follows ($\Delta \Theta = \Theta_2-\Theta_1$): 
\begin{align}
\label{eq:Prob-scn-1-result}
P_1(\Delta C\geq \Theta_2-\Theta_1) = \frac{1}{2}\left [1- {\rm Erf}\left(\frac{\sqrt{\rm SNR}}{2}\, \Delta \Theta \right) \right].
\end{align}
\newline
Next, to determine the logical $X$, $Z$ and $Y$ error rates {for small $T_{\rm c}$,} we need to find those harmful two-qubit errors where the first error only affects two stabilizer generators and the second error only changes one of those affected stabilizer generators. Using Eq.~\eqref{eq:two-qubit-X-errors}, we  {find} that the single-qubit errors that can produce logical errors according to scenario one are $E=X_4$, $X_5$ and $X_6$, so the logical X error rate is given by (from Eqs.~\eqref{eq:Pscn-1} and~\eqref{eq:Prob-scn-1-result})
\begin{align}
\label{eq:scn-1-gammaX}
\gamma_X^{\text{scn-1}} =&\; \left(\Gamma_X^{(4)} + \Gamma_X^{(5)} + \Gamma_X^{(6)}\right)\left[1- 
{\rm Erf}\left(\frac{\sqrt{\rm SNR}}{2}\, \Delta \Theta \right) \right].
\end{align}
The $X_4$, $X_5$ and $X_6$ errors simultaneously affect the cross-correlators $\mathcal{C}_z^{(1)}(t)$ and $\mathcal{C}_z^{(2)}(t)$. A sufficiently large positive fluctuation in either of them will make our continuous QEC algorithm detect two false errors: first $X_1$, $X_2$ or $X_3$ (if the large fluctuation occurs in $\mathcal{C}_z^{(1)}$), followed by $X_7$, $X_8$ or $X_9$ (when the large fluctuation disappears and both $Z$ cross-correlators fluctuate around $-1$), or {\it vice versa}. From Table~\ref{table-II} we see that the product of two false errors, e.g., $X_1X_7\sim Q_5$ (here ``$\sim$'' indicates equivalence modulo gauge operators), maps the system state from the code space to the error subspace (in this example, $\mathcal{Q}_5$)  and does not affect the logical state~$(\alpha,\beta)$. However, the actual error, {in this example} $E\sim Q_5X_{\rm L}$, includes a logical $X$ operation. This discrepancy is the source of logical errors in the small $T_{\rm c}$ limit. {The above analysis shows that this} is due to the  noise from {the} continuous measurements. 

From {the} $X-Z$ symmetry of the nine-qubit Bacon-Shor code, the logical $Z$ error rate due to large fluctuations,  according to scenario one, is {then} given by 
\begin{align}
\label{eq:scn-1-gammaZ}
\gamma_Z^{\text{scn-1}} =&\; \left(\Gamma_Z^{(2)} + \Gamma_Z^{(5)} + \Gamma_Z^{(8)}\right)\left[1- 
{\rm Erf}\left(\frac{\sqrt{\rm SNR}}{2}\, \Delta \Theta \right) \right].
\end{align}
There are no harmful two-qubit errors that can be enacted by {the combination of} a $Y$ error and a large fluctuation in one of the affected cross-correlators by such $Y$ error, so {for scenario one} we have
\begin{align}
\label{eq:scn-1-gammaY}
\gamma_Y^{\text{scn-1}}  =0. 
\end{align}

 \noindent {\it Scenario 2:} {We} now discuss the second likely scenario {in which} large fluctuations in the cross-correlators lead to logical errors. In this scenario, we have a single-qubit error $E$ that affects one, two or three stabilizer generators at the same time and the cross-correlators that are affected by the error do not undergo large fluctuations. A logical error can occur if, at the moment when these cross-correlators cross the lower error threshold ($1-\Theta_2$), some of the other cross-correlators (unaffected by the error $E$) are below this error threshold, due to a negative large fluctuation of magnitude larger than $\Theta_2$. We consider below the situation where a large fluctuation only occurs in one cross-correlator, since this situation is  {the most} likely. This scenario is somewhat similar to the situation shown in Fig.~\ref{fig:GL-cont-cases}~(a), with the important difference that {now} the drop of $C_2(t)$ is due to a large fluctuation and not due to {a physical} error. 
 
\begin{figure}[t!]
\includegraphics[width=\linewidth, trim =9.5cm 6.5cm 7cm 5cm,clip=true]{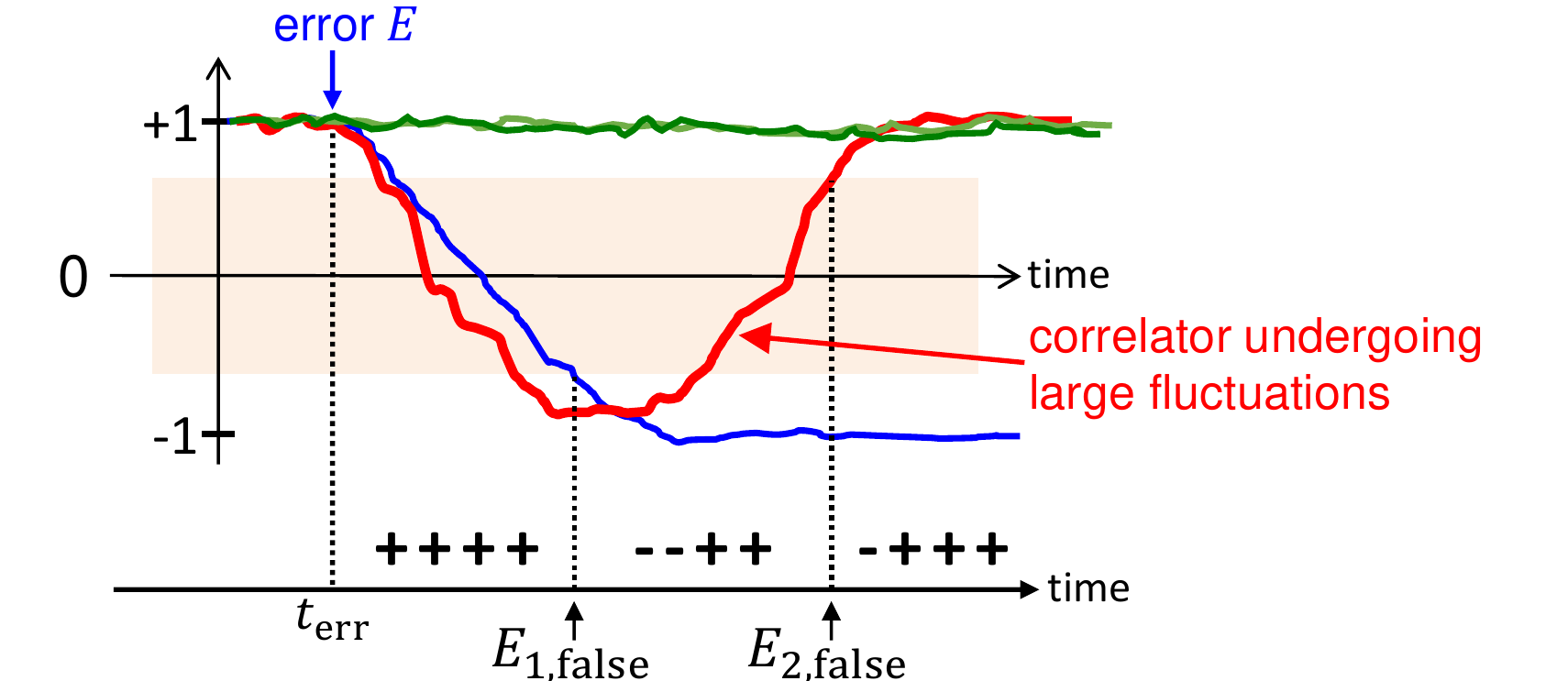}
\caption{Large fluctuations in cross-correlators leading to logical errors. This situation corresponds to scenario 2 of the main text. Actual error $E$ only affects the blue cross-correlator and large fluctuations are only present in the red cross-correlator. The other cross-correlators are depicted by the green lines. Our QEC protocol detects two false errors ($E_{1,{\rm false}}$ and $E_{2,{\rm false}}$) from the indicated error syndrome patterns. Shaded area is the ``syndrome uncertainty region''. }
\label{fig:large_fluctuations_fig}
\end{figure}

 Naively speaking, the probability that this scenario occurs in an experimental realization is given by 
\begin{align}
\label{eq:Pscn-2}
P_{\text{scn-2}} = \Gamma_E \,T_{\rm op}\,P_2(C_2 \leq 1-\Theta_2), 
\end{align}
where the last factor is the probability that $C_2(t)\leq 1-\Theta_2$. The stochastic process $C_2(t)$ is stationary since it is not affected by the actual error; its probability distribution function is obtained from Eq.~\eqref{eq:C2-EOM} with $s_2=1$, {and reads as}
\begin{align}
\label{eq:Pst2-DeltaC}
P_{2,{\rm st}}(C_2) = \left[\frac{{\rm SNR}}{2\pi}\right]^{1/2}\,\exp\left(-\frac{{\rm SNR}}{2}\,(C_2-1)^2\right).
\end{align}
The probability that $C_2$ is below the lower error threshold can {then} be expressed as 
\begin{align}
\label{eq:PC2-leq-lowerEth}
P_2(C_2\leq 1-\Theta_2) =\; \frac{1}{2}\left [1 - {\rm Erf}\left(\sqrt{\frac{\rm SNR}{2}}\, \Theta_2 \right)  \right].
\end{align}
The logical error rates for this scenario are {now} given by (from Eqs.~\eqref{eq:Pscn-2} and~\eqref{eq:PC2-leq-lowerEth})
\begin{align}
\gamma_X^{\text{scn-2}} =&\; \frac{1}{2}\Big(\Gamma_X^{(1)} + \Gamma_X^{(2)} + \Gamma_X^{(3)}+\Gamma_X^{(7)} + \Gamma_X^{(8)} + \Gamma_X^{(9)}+ \nonumber \\
&\;\;\;\;\;\Gamma_Y^{(1)} + \Gamma_Y^{(2)} + \Gamma_Y^{(3)} + \Gamma_Y^{(7)} + \Gamma_Y^{(8)} + \Gamma_Y^{(9)} \Big)\times \nonumber\\
&\;\;\;\; \left [1 - {\rm Erf}\left(\sqrt{\frac{\rm SNR}{2}}\, \Theta_2 \right) \right], \label{eq:scn-2-gammaX}\\
\gamma_Z^{\text{scn-2}}  =&\; \frac{1}{2}\Big(\Gamma_Z^{(1)} + \Gamma_Z^{(4)} + \Gamma_Z^{(7)}+\Gamma_Z^{(3)} + \Gamma_Z^{(6)} + \Gamma_Z^{(9)}+ \nonumber \\
&\;\;\; \Gamma_Y^{(1)} + \Gamma_Y^{(4)} + \Gamma_Y^{(7)} + \Gamma_Y^{(3)} + \Gamma_Y^{(6)} + \Gamma_Y^{(9)}\Big)\times \nonumber \\
&\;\;\;\;\left [1 - {\rm Erf}\left(\sqrt{\frac{\rm SNR}{2}}\, \Theta_2 \right) \right], \label{eq:scn-2-gammaZ}\\
\gamma_Y^{\text{scn-2}} =&\;0. \label{eq:scn2-gammaY}
\end{align}
The errors that contribute to the logical $X$ error rate are $E=\{X_1$, $X_2$, $X_3$, $X_7$, $X_8$, $X_9$, $Y_1$, $Y_2$, $Y_3$, $Y_7$, $Y_8$, $Y_9\}$. From these errors, the $X$ errors only affect one stabilizer generator: {this is} (i) $S_z^{(1)}$ if $E=\{X_1$, $X_2$, $X_3\}$, or (ii) $S_z^{(2)}$ if $E=\{X_7$, $X_8$, $X_9\}$. Then a large fluctuation in the cross-correlator $\mathcal{C}_z^{(2)}(t)$ or $\mathcal{C}_z^{(1)}(t)$, {respectively}, leads to a logical $X$ error, see Fig.~\ref{fig:large_fluctuations_fig}. The two false errors detected by our QEC protocol are (i) $E_{1,{\rm false}}=\{X_4$, $X_5$ or $X_6\}$ (when both $Z$ cross-correlators are below the lower error threshold) and $E_{2,{\rm false}}=\{X_7$, $X_8$ or $X_9\}$ (when large fluctuation disappears) if the error that has actually occurred is $E=\{X_1$, $X_2$, $X_3\}$, or (ii) $E_{1,{\rm false}}=\{X_4$, $X_5$ or $X_6\}$ and $E_{2,{\rm false}}=\{X_1$, $X_2$ or $X_3\}$ if $E=\{X_7$, $X_8$ and $X_9\}$. For $E=\{X_1$, $X_2$, $X_3\}$, a logical $X$ error arises due to the discrepancy between the actual error  $E\sim Q_4$, which does not include a logical $X$ operation, while the product of the two false errors is, e.g., $E_{1,{\rm false}}\,E_{2,{\rm false}}=X_4X_7\sim Q_4X_{\rm L}$ (obtained from Tables~\ref{table-II} and~\ref{table-III}), which does include a logical $X$ operation. A similar discrepancy exists for the other errors $E=\{X_7$, $X_8$, $X_9\}$ that also lead to logical $X$ errors. 

In contrast, the $Y$ errors that contribute to the logical $X$ error rate~\eqref{eq:scn-2-gammaX} affect two [if the error that has occurred is $E=Y_1$, $Y_3$, $Y_7$, $Y_9$] or three [if actual error is $E=Y_2$, $Y_8$] stabilizer generators at the same time. {In this situation,}  a large fluctuation in $\mathcal{C}_z^{(2)}(t)$ {when} $E=\{Y_1$, $Y_2$, $Y_3\}$ or in $\mathcal{C}_z^{(1)}(t)$ {when} $E=\{Y_7$, $Y_8$, $Y_9\}$, leads to a logical $X$ error as well. For example, let us consider the case where the actual error is $E=Y_2$, which affects three cross-correlators; {specifically}, $\mathcal{C}_x^{(1)}(t)$, $\mathcal{C}_x^{(2)}(t)$ and $\mathcal{C}_z^{(1)}(t)$. A sufficiently large negative fluctuation in $\mathcal{C}_z^{(2)}(t)$ will make our continuous QEC algorithm detect two false errors: first $Y_5$ and second $X_7$, $X_8$ or $X_9$. The product of the two false errors is equivalent to $\sim Q_{11}Y_{\rm L}$ (however, the actual error is $Y_2\sim Q_{11}Z_{\rm L}$). Our continuous QEC protocol then says that the system state is in error subspace $\mathcal{Q}_{11}$ and the logical state suffers from a logical $Y$ operation. Before extracting the logical state (with initial probability amplitudes $\alpha,\beta$) from this error subspace we apply the multi-qubit operation $X_{1}X_4X_7Z_1Z_2Z_3$ to the nine-qubit system to undo such apparent logical $Y$ operation; however, this procedure changes the actual logical state from $Z_{\rm L}(\alpha,\beta)$ to $Y_{\rm L}Z_{\rm L}(\alpha,\beta)=\inum X_{\rm L}(\alpha,\beta) = \inum(\beta,\alpha)$. The extracted logical state from the error subspace $\mathcal{Q}_{11}$ is then $(\alpha_{\rm f},\beta_{\rm f})=(\beta,\alpha)$, dropping overall phase factors. A logical $X$ error has therefore arisen. Similarly, the other $Y$ errors lead to a logical $X$ error as well. 

The logical $X$ error rate in the small $T_{\rm c}$ limit is given by $\gamma_X^{{\rm cont}} = \gamma_X^{\text{scn-1}} + \gamma_X^{\text{scn-2}}$, {with} a similar relation for the logical $Z$ error rate. The scenarios discussed above do not contribute to the logical $Y$ error rate in this limit. This does not mean that the latter {error} vanishes. {However,} it is exponentially smaller than the logical $X$ and $Z$ error rates so it can be neglected. The total logical error rate $\gamma_{{\rm cont}}$ for the depolarizing channel {in the small $T_c$ regime is then} equal to   
\begin{align}
\label{eq:cont-gammaL-total-small-Tc}
\gamma_{{\rm cont}} \approx 2\gamma_X^{{\rm cont}}=&\;2\gamma_Z^{{\rm cont}}= 2\,\Gamma_{\rm d}\left[1- {\rm Erf}\left(\frac{\sqrt{\rm SNR}}{2}\, \Delta \Theta \right) \right] \nonumber \\
&\;+ 4\,\Gamma_{\rm d}\left [1 - {\rm Erf}\left(\sqrt{\frac{\rm SNR}{2}}\, \Theta_2 \right) \right].
\end{align}

\section{Optimal Continuous QEC protocol}
\label{cont-operation-optimization}
Our analytical result for the total logical error rate $\gamma_{{\rm cont}}$ is {now} obtained from the sum of Eqs.~\eqref{eq:cont-gammaL-total} and~\eqref{eq:cont-gammaL-total-small-Tc}. {This yields} ($\Delta \Theta = \Theta_2-\Theta_1$)
\begin{widetext}
\begin{align}
\label{eq:GL-final-result}
\gamma_{{\rm cont}} =\frac{112}{9}T_{\rm c}\Gamma_{\rm d}^2\,\ln\bigg[\frac{2-\Theta_1}{2-\Theta_2} \bigg] + \frac{284}{9}T_{\rm c}\Gamma_{\rm d}^2\,\ln\bigg[\frac{2}{2-\Theta_2} \bigg] + 2\Gamma_{\rm d}\left[1- {\rm Erf}\left(\frac{\sqrt{\rm SNR}}{2}\, \Delta \Theta\right) \right] + 4\Gamma_{\rm d}\left [1 - {\rm Erf}\left(\sqrt{\frac{\rm SNR}{2}}\, \Theta_2 \right) \right].
\end{align}
\end{widetext}

\begin{figure}[t!]
\centering
\includegraphics[width=\linewidth, trim =2cm 0cm 2cm 0cm,clip=true]{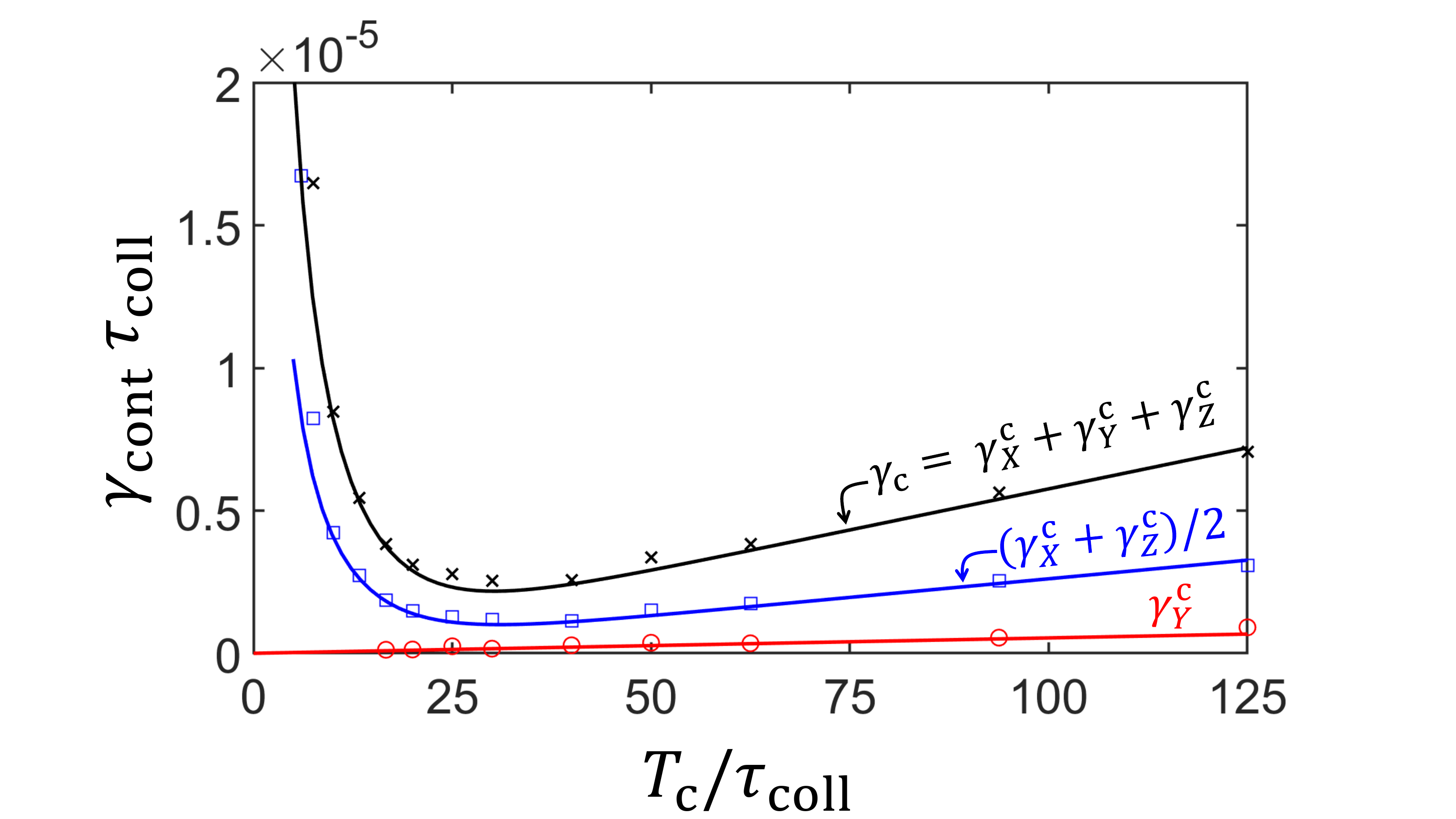}
\caption{Analytical {\it vs.} numerical results for the logical error rates as function of the integration time parameter $T_{\rm c}$. Circles indicate the numerical results for the logical $Y$ error rates; squares show numerical results for the average of the logical $X$ and  $Z$ error rates; and crosses indicate the numerical results for the total logical error rate $\gamma_{{\rm cont}}$. Solid lines show the analytical results. Parameters: $1-\Theta_1=0.56$, $1-\Theta_2=-0.56$, $\Gamma_{\rm d}=3\times 10^{-5}\tau_{\rm coll}^{-1}$,  $\eta=1$ and $\tau_{\rm c}=0.25\tau_{\rm coll}\approx \tau_{\rm c}^{\rm opt}$. There are no fitting parameters. }
\label{fig:theory_vs_numerics}
\end{figure}

Figure~\ref{fig:theory_vs_numerics} compares {our} analytical formula {of} Eq.~\eqref{eq:GL-final-result} against Monte Carlo simulations for the continuous operation of the nine-qubit Bacon-Shor code {under perfect measurement efficiency $\eta = 1$}. In the{se} simulations, the continuous measurements were described using Eq.~\eqref{eq:rho_g-EOM} for the evolution of the gauge qubits and Eq.~\eqref{eq:Ik-Ql} to obtain the measurement signals $I_{G_k}(t)$. Decoherence due to $X$, $Y$ and $Z$ errors was accounted for using the jump/no-jump method {(Section~\ref{Section:Operation-with-errors})}, {with} the action of errors on the system state {specified} by Tables~\ref{table-II} and~\ref{table-III}. We {employ} the {following parameter values}: depolarization error rate $\Gamma_{\rm d}=3\times10^{-5}\tau_{\rm coll}^{-1}$ (the same for all qubits), error thresholds at $1-\Theta_1=0.56$ and $1-\Theta_2=-0.56$, and $\eta=1$ (ideal detectors). 

We find that the {total} logical $X$ and $Z$ error rates are quite similar ({we show only the average and not the individual values}), which agrees with the theoretical prediction that $\gamma^{{\rm cont}}_X=\gamma^{{\rm cont}}_Z$. This is due to the $X-Z$ symmetry of the nine-qubit Bacon-Shor code and the fact that all qubits have the same error rates. The logical $Y$ error rate is roughly five times smaller than the logical $X$ and $Z$ error rates for large values of $T_{\rm c}$ (it was not possible to reliably obtain $\gamma_Y^{{\rm cont}}$ for $T_{\rm c}\leq 15\tau_{\rm coll}$ because {the value} was too small). In general, we find  good agreement, {without any fitting parameters}, between analytics and numerics for the range $T_{\rm c}\geq 10\tau_{\rm coll}$. Most importantly, our analytical result Eq.~\eqref{eq:GL-final-result} is able to estimate the optimal value of the integration time parameter $T_{\rm c}$: {for the assumed parameters in the simulations} we find $T_{\rm c} \simeq 30\tau_{\rm coll}$. 

Next, we use Eq.~\eqref{eq:GL-final-result} to find the optimal operation point ($\Theta_1^{\rm opt},\Theta_2^{\rm opt},T_{\rm c}^{\rm opt}$) for the continuous operation of the nine-qubit Bacon-Shor code by minimizing the total logical error rate $\gamma_{{\rm cont}}$. In this minimization we impose the constraint $\Theta_1\geq \Theta_{1,\min}$ (i.e., the upper error threshold should not be too close to $+1$) and choose $\Theta_{1,\min} = 1.5\,{\rm SNR}^{-1/2}$, where ${\rm SNR}^{-1/2}$ is the standard deviation of the (normalized) cross-correlator fluctuations in the absence of errors. {Equation~\eqref{eq:PC2-leq-lowerEth} then implies that} the probability that a cross-correlator is within the ``syndrome uncertainty region'' is roughly 6.68\%. This constraint guarantees that single-qubit errors are efficiently detected, since their detection requires that the cross-correlators that are unaffected by the errors are above the upper error threshold (i.e., outside the ``syndrome uncertainty region'').  Moreover, it  also guarantees that the window time {intervals $\Delta t_{\rm cont}^{(n)},$ $n=1,2$ that were obtained in} the noiseless limit in Section~\ref{large-Tc-limit}, are approximately correct.

\begin{figure}[t!]
\centering
\includegraphics[width=\linewidth, trim =0cm 3.5cm 2cm 0cm,clip=true]{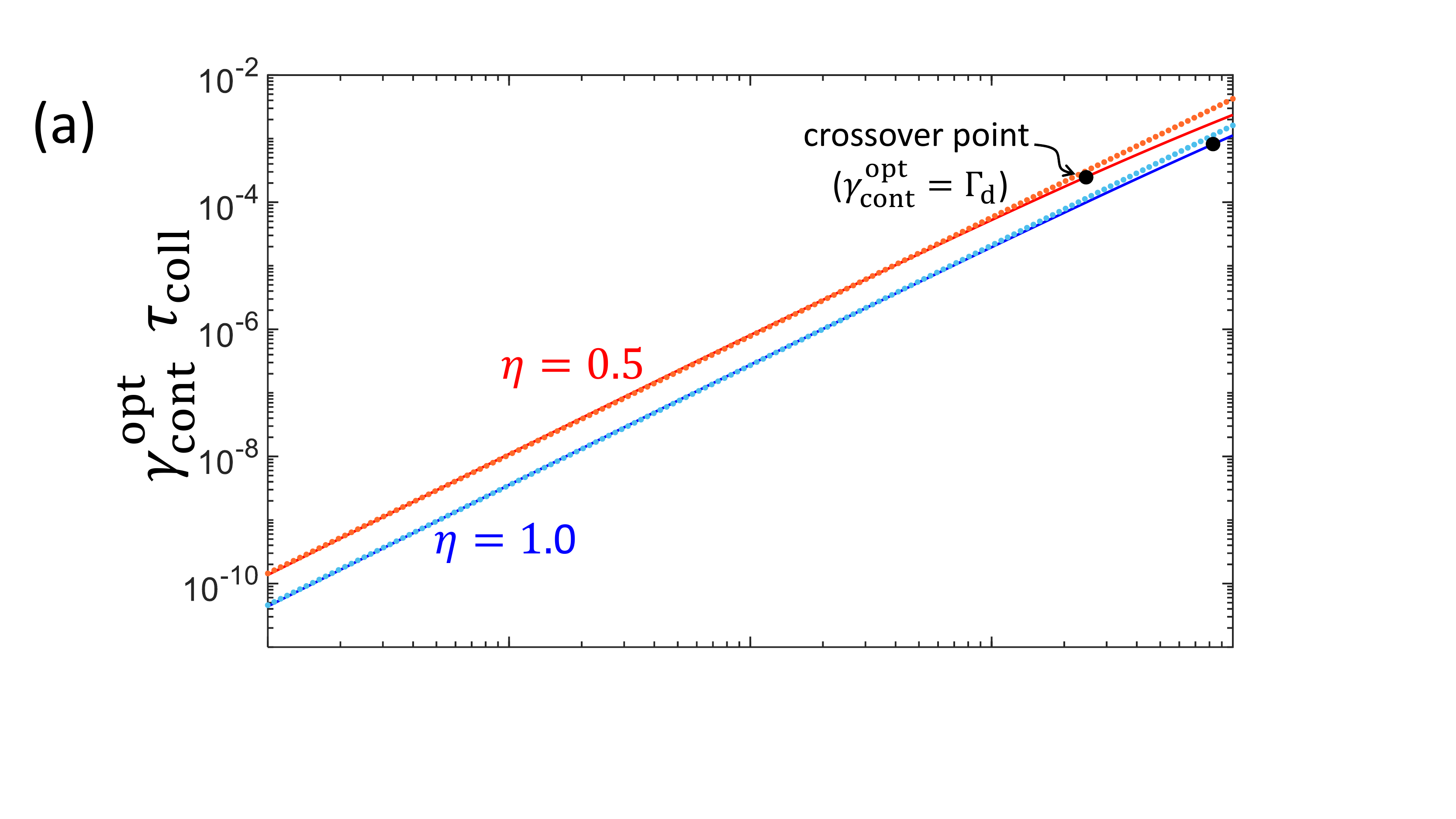}
\includegraphics[width=\linewidth, trim =0cm 0.5cm 2cm 1.5cm,clip=true]{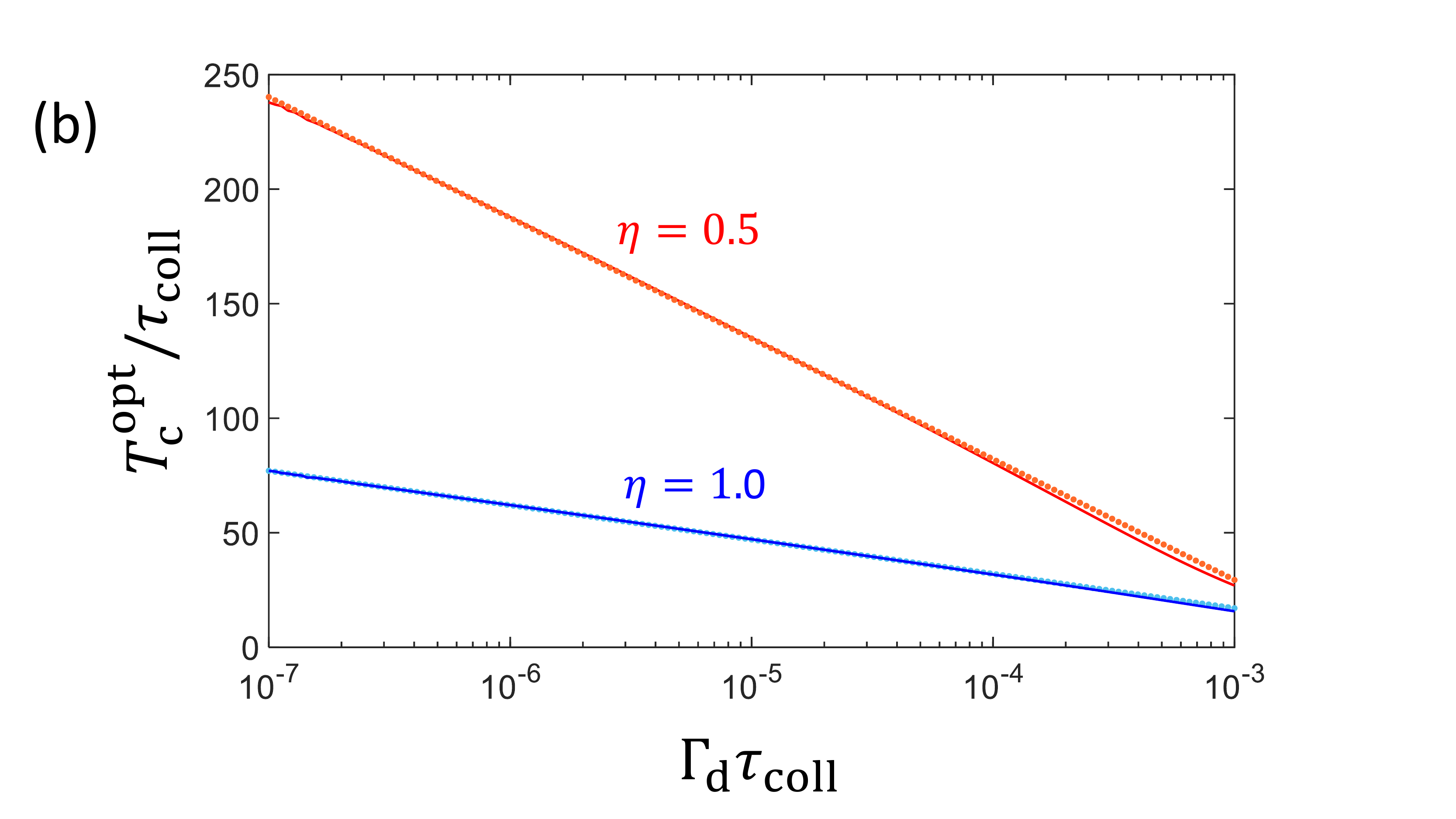}
\caption{Optimized logical error rate $\gamma^{{\rm opt}}_{\rm cont}$ for the continuous operation and optimal integration time parameter $T_{\rm c}^{\rm opt}$ as function of the depolarization error rate $\Gamma_{\rm d}$ and for two values of quantum efficiency $\eta=0.5$ and $\eta=1.0$. Solid lines are obtained from minimization of the total logical error rate formula~\eqref{eq:GL-final-result} and dotted lines show the fitting formulas Eqs.~\eqref{eq:GL-opt-eta-1} and~\eqref{eq:GL-opt-eta-0.5} (top panel) and Eqs.~\eqref{eq:Tc-opt-eta-1} and~\eqref{eq:Tc-opt-eta-0.5} (bottom panel). The  small circles in panel (a) indicate the crossover points for $\Gamma_{\rm d}$, cf. Eqs.~\eqref{eq:GL-crossover-eta-1} and~\eqref{eq:GL-crossover-eta-0.5}. }
\label{fig:Main-result}
\end{figure}

Minimization of the total logical error rate formula~\eqref{eq:GL-final-result} with the above constraint for $\Theta_1$ is carried out numerically. {We}  first discuss our results for the case of ideal detectors ($\eta=1$). We find that $\Theta_1^{\rm opt}=\Theta_{1,\min}$ (i.e., the optimal position of the upper error threshold is as high as allowed by the above constraint), $\Theta_{2}^{\rm opt}\approx 1.40$ (so the optimal position of the lower error threshold is $1-\Theta_2^{\rm opt}\approx -0.40$ and  is weakly dependent on $\Gamma_{\rm d}$ with deviations $\pm0.0075$ from this constant value). 
{Fig.~\ref{fig:Main-result} shows plots of the values of $T_{\rm c}^{\rm opt}$ and $\gamma^{{\rm opt}}_{\rm cont}$ as a function of $\Gamma_{\rm d}\tau_{\rm coll}\in [10^{-7}, 10^{-4}]$ for the optimal values $\Theta_1^{\rm opt}, \Theta_2^{\rm opt}$, where the blue lines refer to the ideal detector $\eta =1$. Fitting these two functions to a simple log {function} ($T_{\rm c}^{\rm opt}$) or power law ($\gamma_{\rm cont}^{{\rm opt}}$), results in the fully optimized formulae}
\begin{align}
 T_{\rm c}^{\rm opt} \approx&\; -6.51\,\ln\left(72.71\,\Gamma_{\rm d}\tau_{\rm coll}\right)\,\tau_{\rm coll}, \label{eq:Tc-opt-eta-1} \\
 \gamma^{{\rm opt}}_{\rm cont}\approx&\; \frac{739.60}{\tau_{\rm coll}} \left(\Gamma_{\rm d}\tau_{\rm coll}\right)^{\nu},\label{eq:GL-opt-eta-1} \;\;\; {\rm with} \;\;\; \nu=1.88.  
\end{align} 
{We find that these fitting formulae also work well for smaller values of} $\Gamma_{\rm d}$. 

The fact that the optimized logical error rate, $\gamma^{{\rm opt}}_{\rm cont}$, for the continuous operation exhibits a power law scaling on $\Gamma_{\rm d}$ (for sufficiently small $\Gamma_{\rm d}$) with exponent $\nu=1.88$ close to 2, which is the expected exponent for a distance-three quantum error correcting code, suggests that our continuous QEC protocol performs well. By equating $\gamma^{{\rm opt}}_{\rm cont}$ and $\Gamma_{\rm d}$, we estimate the crossover value of the depolarization error rate, $\Gamma_{\rm d}^{\rm crossover}$, below which implementation of the nine-qubit Bacon-Shor code is advantageous. {This results in the value} 
\begin{align}
\label{eq:GL-crossover-eta-1}
\Gamma_{\rm d}^{\rm crossover}\approx 10^{-3}\tau_{\rm coll}^{-1} \;\;\;\;\;\; (\eta=1).
\end{align}
\newline 
Moreover, from Eqs.~\eqref{eq:GL-total-discrete} and~\eqref{eq:GL-opt-eta-1}, we find the relationship between the cycle time $\Delta t$ and the collapse time $\tau_{\rm coll}$ such that the continuous and discrete QEC operations have the same performances, i.e., $\gamma_{\rm cont}^{{\rm opt}}=\gamma_{\rm discrete}$.  {This yields} 
\begin{align}
\label{eq:cont-discrete-comparison-performance-eta-1}
\Delta t \approx \frac{33.61 \tau_{\rm coll}}{(\Gamma_{\rm d}\tau_{\rm coll})^{2-\nu}}\;\;\;\;\;\; (\eta=1).
\end{align}
\newline
{We}  now discuss the case of nonideal detectors with $\eta =0.5$. {Numerical optimization of Eq.~\eqref{eq:GL-final-result} yields the optimal parameters} $\Theta_1^{\rm opt} =\Theta_{1,\min}$, $\Theta_2^{\rm opt} \approx 1.40$ (this is weakly dependent on $\Gamma_{\rm d}$ with deviations $\pm0.012$ from this constant value).  {Similarly fitting the corresponding results  over the range $\Gamma_{\rm d}\tau_{\rm coll}\in [10^{-7},10^{-4}]$ (red lines in Fig.~\ref{fig:Main-result}) results in the corresponding formulae for the optimal integration time and logical error rate for $\eta= 0.5$:} 
\begin{align}
T_{\rm c}^{\rm opt} \approx&\;-22.88\,\ln\left(277.27\, \Gamma_{\rm d}\tau_{\rm coll}\right)\, \tau_{\rm coll},\label{eq:Tc-opt-eta-0.5}\\
\gamma^{{\rm opt}}_{\rm cont} \approx&\;  \frac{1690.4}{\tau_{\rm coll}}\left(\Gamma_{\rm d}\tau_{\rm coll} \right)^{\tilde\nu}  \;\; {\rm with} \;\; \tilde\nu= 1.86. \label{eq:GL-opt-eta-0.5}
\end{align}
The {scaling of the logical error result in} Eq.~\eqref{eq:GL-opt-eta-0.5} shows that our continuous QEC protocol also performs well in the case of nonideal detectors, since the exponent $\tilde\nu$ is {still} close to the ideal value of 2. The crossover value of the depolarization error rate is now smaller than that of the ideal case, {which is not surprising given the effect of the measurement inefficiency.}  {Specifically, for $\eta =0.5$, we find}
\begin{align}
\label{eq:GL-crossover-eta-0.5}
\Gamma_{\rm d}^{\rm crossover}\approx 0.2\times 10^{-3}\tau_{\rm coll}^{-1} \;\;\;\;\;\; (\eta=0.5).
\end{align}
Moreover, continuous and discrete operations exhibit the same performances for {inefficiency} $\eta=0.5$ if the cycle time $\Delta t$ from discrete operation and the collapse time $\tau_{\rm coll}$ from continuous measurements are related by Eq.~\eqref{eq:cont-discrete-comparison-performance-eta-1} with the numerical pre-factor equal to $76.83$ and $\nu$ replaced by $\tilde\nu$. We conclude that discrete and continuous operation performances of the nine-qubit Bacon-Shor code can indeed be comparable. 

\section{Summary and Conclusions}
\label{conclusions}
We have analyzed the continuous operation of the error correcting nine-qubit Bacon Shor code, {in which} all noncommuting gauge operators are continuously measured at the same time.  
{Our analysis has} shown that continuous operation of the nine-qubit Bacon-Shor code is not only possible, but {that} it can have a performance that is comparable to that of the conventional operation, {i.e., enforcing a near quadratic scaling of logical errors, while avoiding the use of ancilla qubits and associated circuits to transfer and diagnose errors.  Instead, the errors are passively monitored, e.g., by probe electromagnetic fields as is commonly implemented for superconducting qubits in microwave cavities.} 

Our approach exploits the subsystem structure of the nine-qubit Bacon-Shor code to parametrize the full quantum state in terms of the probability amplitudes of one logical qubit and four gauge qubits. This parametrization is very useful in the analysis of the continuous operation, because it enables us to describe the measurement-induced evolution of the nine-qubit state in terms of an effective evolution of {only} the gauge qubits. {The latter} are subject to simultaneous continuous measurement of the effective noncommuting operators $\mathcal{G}_k$ that  act {only} on these qubits). The effective quantum measurement model for the gauge qubits is also useful to describe the temporal correlations of the actual measurement signals from simultaneous continuous measurement of the gauge operators $G_k$. In this way, our approach reduces the complexity of the problem from nine physical to four {effective} qubits. 

{Due to the continuous measurement of noncommuting operators,} the gauge qubits undergo diffusive state evolution in {both} the code space {and} the error subspaces. Occurrence of an error interrupts the evolution of the gauge qubits in one of these subspaces and moves it to another one, where diffusion continues until the next error occurs. We developed a {general} procedure to figure out which subspace the logical and gauge qubits jump to after an error (Tables~\ref{table-II} and~\ref{table-III}) and also to describe how the logical state and  the state of the gauge qubits are  affected by the errors at the moment of the jump (Table~\ref{table-II}). This procedure can be easily extended to the error analysis of other subsystem codes. 

{Our continuous QEC protocol for quantum memory consists of passively monitoring errors while also diagnosing their associated logical operations (obtained from Table~\ref{table-II}) and recording these, followed by a set of discrete recovery operations that undo the series of logical operations that have occurred during a single  realization at the end of the continuous operation, before the logical state is returned to the user.} We showed that the single qubit errors (modulo gauge operators) can be inferred from the error syndrome path $\mathcal{S}(t)$, which is defined by the {measured} values of the stabilizer generators, and that knowledge of $\mathcal{S}(t)$ is sufficient to determine the entire series of logical operations in a single realization, and hence the required final sequence of recovery operations.  {To monitor the stabilizer generators in real time, we introduced cross-correlators of three measurement signals, defined using time averaging with two integration time parameters $\tau_{\rm c}$ and $T_{\rm c}$.}
{Analytic estimates of the size of the fluctuations of the cross-correlators, characterized by their SNR, were used  to find the optimal value of the filter integration time} $\tau_{\rm c}$ that maximizes the SNR. 
\newline
In order to realize this protocol, it will be necessary to implement continuous measurements of both $XX$ and $ZZ$ operators either simultaneously or with fast alternation. Continuous measurement of $ZZ$ operators has been experimentally demonstrated with superconducting (SC) qubits in Ref. \cite{Riste2013}, where the two qubits were dispersively coupled with a single mode of a cavity to generate an interaction Hamiltonian $H_{\rm int} = \chi (Z_1 + Z_2)n$, where $\chi$ is the dispersive coupling parameter and $n$ the intracavity photon number operator. Under suitable conditions ($\chi \ll \kappa$, with $\kappa$ the cavity decay rate, and the cavity drive frequency set at the average of the resonance frequencies of the four two-qubit states), this results in an effective measurement of $ZZ$.  Such dispersive multi-qubit measurements are now being made in multiple laboratories \cite{Martinis2012, Ustinov2012, Gambeta2016}. In principle, continuous measurement of $XX$ could be made by combining  this approach with that of Vool {\it et} al. in Ref. \cite{Vool2016}, which shows how to engineer continuous measurements of $X$ for a single qubit. Combining these two approaches would allow a continuous measurement of $XX$ to be made.  Implementing both of these simultaneously could be realized by generalization of the single quadrature measurement scheme of Ref. \cite{Hacoen-Gourgy2016}, which implemented simultaneous $X$ and $Z$ measurements on a single qubit.  This will require coupling to two cavity modes and driving of sideband transitions on the two-qubit levels.   Analysis of the details of this scheme present an interesting challenge for future work.

{A key feature of our QEC protocol is} a double error-threshold protocol 
(with error threshold parameters $\Theta_1$ and $\Theta_2$) 
{that is employed} to obtain the monitored error syndrome path $\mathcal{S}_{\rm m}(t)$ from the {measured stabilizer} cross-correlators. {We showed how, in general, $\mathcal{S}_{\rm m}(t)$ can differ from the actual error syndrome path $\mathcal{S}(t)$ by the occurrence of jumps in $\mathcal{S}(t)$ that are missing in $\mathcal{S}_{\rm m}(t)$} {or {\it vice versa}}, {resulting in inference of a different series of logical operations from the measured $\mathcal{S}_{\rm m}(t)$ than from the true $\mathcal{S}(t)$, and identified this discrepancy as the source of logical errors in the continuous QEC protocol.} 

We {derived} analytical expressions for the logical $X$, $Y$ and $Z$ error rates in the continuous QEC protocol {for the limiting situations of small and large correlator integration times $T_{\rm c}$.}
The analytical results are used to find the optimal values of the parameters ($T_{\rm c}$, $\Theta_{1}$ and $\Theta_{2}$) for the continuous QEC protocol by minimization of the total logical error rate.  The analytical formulae for the logical error rates are seen to agree well with Monte Carlo simulations undertaken with the optimized parameters.  
This analysis also allowed us to identify the most likely processes contributing to logical errors in these two limits. 
In the small $T_{\rm c}$ limit, 
we have identified the most likely process for logical errors 
to be the misinterpretation of a single jump in $\mathcal{S}(t)$ produced by an actual error as two false jumps in $\mathcal{S}_{\rm m}(t)$ that are diagnosed as two false errors. 
{This particular mechanism of logical errors is unique to the continuous operation; it does not arise in the conventional operation of the Bacon-Shor code with ideal projective measurements because of the absence of measurement noise.} In the large $T_{\rm c}$ limit, fluctuations in the correlators can be neglected and the main mechanism of logical errors derives then from two errors that are misdiagnosed as a single error by the protocol. This can happen when two errors occur sufficiently close in time, which is now similar to the mechanism for logical errors in the conventional discrete operation of the Bacon-Shor code, in which two errors occurring within the same operation cycle produce a logical error.  

{A primary result of this work is the determination of} the optimized total logical error rate, $\gamma^{{\rm opt}}_{\rm cont}$, for the continuous operation of a nine-qubit Bacon-Shor code. {We found} that $\gamma^{{\rm opt}}_{\rm cont}$ exhibits a power-law dependence on the depolarization error rate $\Gamma_{\rm d}$, with an exponent $\approx 1.9$ that {only} weakly depends on the quantum efficiency. The fact that this exponent is close to the expected value of 2 for a distance-three quantum error correcting code shows that {the} continuous QEC protocol performs well. 

By comparing $\gamma^{{\rm opt}}_{\rm cont}$ with $\gamma_{\rm discrete}$ (total logical error rate for the discrete operation), we find that the discrete and continuous operations exhibit comparable performances {when} the {discrete} cycle time $\Delta t$ is {approximately} $30$ times (if $\eta=1$) or $80$ times (if $\eta=0.5$) the collapse time $\tau_{\rm coll}$ {of the} continuous operation. {This does not  indicate that the discrete operation is faster or better than the continuous operation if we take into account that ``projective measurements'' actually take time to realize, require additional overhead (e.g., application of relatively strong pulses to read out the ancillary qubits) and introduce non-correctable errors at the ancillary qubits.} We point out that the value of the ratio $\Delta t/\tau_{\rm coll}$ that is necessary for comparable performance of the discrete and continuous operations is generally larger when the triple cross-correlator signals, Eq.~\eqref{eq:cross-correlators}, are noisier, for example, due to inefficient detectors. 

Finally, we have estimated the crossover value of {the qubit error rate} $\Gamma_{\rm d}$, below which implementation of the nine-qubit Bacon-Shor code is advantageous in terms of effectiveness, i.e., $\gamma^{{\rm opt}}_{\rm cont}<\Gamma_{\rm d}$. For ideal detectors, we found $\Gamma_{\rm d}^{\rm crossover}\approx 10^{-3}\tau_{\rm coll}^{-1}$, {while for nonideal detectors $\Gamma_{\rm d}^{\rm crossover}$ is approximately five times smaller}. {Assuming $\Gamma_{\rm d}^{\rm crossover}=(500\,\mu{\rm s})^{-1}$~\cite{Oliver2019}, we then obtain that the collapse time of the continuous operation should be $\tau_{\rm coll}\approx 0.5\,\mu$s} in the ideal case or smaller in the nonideal case ($\eta<1$). 

\section{Acknowledgments}
The research is based upon work supported by the Office of
the Director of National Intelligence (ODNI), Intelligence Advanced
Research Projects Activity (IARPA), via the U.S. Army Research Office
contract W911NF-17-C-0050. The views and conclusions contained herein are
those of the authors and should not be interpreted as necessarily
representing the official policies or endorsements, either expressed or
implied, of the ODNI, IARPA, or the U.S. Government. The U.S. Government
is authorized to reproduce and distribute reprints for Governmental
purposes notwithstanding any copyright annotation thereon. 

\appendix
\section{Code space orthonormal basis and nominal collapse states for $X$ gauge operators}
\label{Appendix-A}
\subsection{Code space orthonormal basis}
In this section we describe the procedure to obtain the 32 orthonormal basis vectors $|\phi_j\rangle$, explicitly given in Eq.~\eqref{eq:psi-basis},  that span the code space. The first 16 basis vectors  can be obtained as follows. We start with the initial state $|000\,000\,000\rangle$ and apply the projector 
\begin{align}
\label{eq:Appendix-Projector-step1}
\Pi_{\rm step 1}^{(g_1 g_2g_3, g_4g_5g_6)} =&\; (\frac{\openone + g_1 G_{1}}{2})(\frac{\openone + g_2 G_{2}}{2})(\frac{\openone + g_3 G_{3}}{2})\nonumber \\
&\hspace{-1.0cm}\times(\frac{\openone + g_4 G_{4}}{2})(\frac{\openone + g_5 G_{5}}{2})(\frac{\openone + g_6 G_{6}}{2}),
\end{align}
where $g_1, g_2, ...g_6$ are outcomes of the $Z$ gauge operators $G_1,G_2,...G_6$, respectively, and then apply the projector 
\begin{align}
\label{eq:Appendix-Projector-step2}
\Pi_{\rm step 2}^{(g_7g_8g_9,g_{10}g_{11}g_{12})} =&\;(\frac{\openone + g_7 G_{7}}{2})(\frac{\openone + g_8 G_{8}}{2})(\frac{\openone + g_9 G_{9}}{2})\nonumber\\
& \hspace{-2.25cm}\times (\frac{\openone + g_{10} G_{10}}{2})(\frac{\openone + g_{11} G_{11}}{2})(\frac{\openone + g_{12} G_{12}}{2}),
\end{align}
where $g_{7},g_8,...g_{12}$ are the outcomes  of the $X$ gauge operators $G_7,G_8,...G_{12}$, respectively. The outcomes $g_i$ that should  to be used in Eqs.~\eqref{eq:Appendix-Projector-step1}--\eqref{eq:Appendix-Projector-step2} correspond to any ``good'' outcome configuration that satisfy $g_1g_2g_3=g_4g_5g_6=1$ or $g_7g_8g_9=g_{10}g_{11}g_{12}=1$; we will use $g_1,...g_{12}=1$. Next, the  sought basis vectors are obtained by applying the step-1 projectors that correspond to all 16 ``good'' outcome configurations~($\pm$ indicates $\pm1$) 
\begin{eqnarray}
|\phi_1\rangle &=&\,  {\mathcal{N}_1^{-1}}\, \Pi_{\rm step 1}^{(+++,+++)} |000\,000\,000\rangle_{++}, \nonumber \\
|\phi_2\rangle &=&\, {\mathcal{N}_2^{-1}}\, \Pi_{\rm step 1}^{(+++,+--)} |000\,000\,000\rangle_{++}, \nonumber \\
|\phi_3\rangle &=&\,  {\mathcal{N}_3^{-1}}\, \Pi_{\rm step 1}^{(+++,--+)} |000\,000\,000\rangle_{++}, \nonumber \\
|\phi_4\rangle &=&\,  {\mathcal{N}_4^{-1}}\, \Pi_{\rm step 1}^{(+++,-+-)} |000\,000\,000\rangle_{++}, \nonumber 
\end{eqnarray}
\begin{eqnarray}
|\phi_5\rangle &=&\,  {\mathcal{N}_5^{-1}}\, \Pi_{\rm step 1}^{(+--,+++)} |000\,000\,000\rangle_{++}, \nonumber \\
|\phi_6\rangle &=&\,  {\mathcal{N}_6^{-1}}\, \Pi_{\rm step 1}^{(+--,+--)}  |000\,000\,000\rangle_{++}, \nonumber \\
|\phi_7\rangle &=&\,  {\mathcal{N}_7^{-1}}\, \Pi_{\rm step 1}^{(+--,--+)}  |000\,000\,000\rangle_{++}, \nonumber \\
|\phi_8\rangle &=&\,  {\mathcal{N}_8^{-1}}\, \Pi_{\rm step 1}^{(+--,-+-)}  |000\,000\,000\rangle_{++}, \nonumber \\
|\phi_9\rangle &=&\,  {\mathcal{N}_9^{-1}}\, \Pi_{\rm step 1}^{(--+,+++)} |000\,000\,000\rangle_{++}, \nonumber \\
|\phi_{10}\rangle &=&\, {\mathcal{N}_{10}^{-1}}\, \Pi_{\rm step 1}^{(--+,+--)} |000\,000\,000\rangle_{++}, \nonumber \\
|\phi_{11}\rangle &=&\,  {\mathcal{N}_{11}^{-1}}\, \Pi_{\rm step 1}^{(--+,--+)} |000\,000\,000\rangle_{++}, \nonumber \\
|\phi_{12}\rangle &=&\,  {\mathcal{N}_{12}^{-1}}\, \Pi_{\rm step 1}^{(--+,-+-)} |000\,000\,000\rangle_{++}, \nonumber \\
|\phi_{13}\rangle &=&\,  {\mathcal{N}_{13}^{-1}}\, \Pi_{\rm step 1}^{(-+-,+++)} |000\,000\,000\rangle_{++}, \nonumber \\
|\phi_{14}\rangle &=&\,  {\mathcal{N}_{14}^{-1}}\, \Pi_{\rm step 1}^{(-+-,+--)} |000\,000\,000\rangle_{++}, \nonumber \\
|\phi_{15}\rangle &=&\,  {\mathcal{N}_{15}^{-1}}\, \Pi_{\rm step 1}^{(-+-,--+)} |000\,000\,000\rangle_{++}, \nonumber \\
|\phi_{16}\rangle &=&\,  {\mathcal{N}_{16}^{-1}}\, \Pi_{\rm step 1}^{(-+-,-+-)} |000\,000\,000\rangle_{++}, 
\label{eq:Appendix-phi-states}
\end{eqnarray}
where  $\mathcal{N}_j$ are normalization factors and \begin{align}\label{eq:Appendix-phi-states-2}|000\,000\,000\rangle_{++} =\Pi_{\rm step 2}^{(+++,+++)}\Pi_{\rm step 1}^{(+++,+++)}\,|000\,000\,000\rangle.\end{align} The remaining basis vectors $|\phi_{17}\rangle,...|\phi_{32}\rangle$ are obtained from Eqs.~\eqref{eq:Appendix-phi-states}--\eqref{eq:Appendix-phi-states-2} with initial state $|000\,000\,000\rangle$ replaced by $|111\,111\,111\rangle$. 

\subsection{Nominal collapse states after projective measurement of $X$ gauge operators}
In Section~\ref{sec:operation-w/o-errors} we introduced  $|Xg_7g_8g_9,g_{10}g_{11}g_{12}\rangle$ as the nominal collapse states after step-2 measurements with ``good'' outcomes, satisfying $g_7g_8g_9=g_{10}g_{11}g_{12}=+1$. These collapse states can be written as a linear combination of all nominal step-1 collapse states, given in  Eq.~\eqref{eq:Z-states}, with coefficients $\pm1/4$. The unitary transformation $U$ that relates them  reads as 
\begin{widetext}
\begin{align}
U =\frac{1}{4} \left[
\begin{array}{cccccccccccccccc}
\,1&\,1&\,1&\,1&\,1&\,1&\,1&\,1&\,1&\,1&\,1&\,1&\,1&\,1&\,1&\,1\\
\,1&-1&\,1&-1&\,1&-1&\,1&-1&\,1&-1&\,1&-1&\,1&-1&\,1&-1\\
1&\,1&\,1&\,1&-1&-1&-1&-1&\,1&\,1&\,1&\,1&-1&-1&-1&-1\\
\,1&-1&\,1&-1&-1&\,1&-1&\,1&\,1&-1&\,1&-1&-1&\,1&-1&\,1\\
1&\,1&-1&-1&\,1&\,1&-1&-1&\,1&\,1&-1&-1&\,1&\,1&-1&-1\\
\,1&-1&-1&\,1&\,1&-1&-1&\,1&\,1&-1&-1&\,1&\,1&-1&-1&\,1\\
\,1&\,1&-1&-1&-1&-1&\,1&\,1&\,1&\,1&-1&-1&-1&-1&\,1&\,1\\
\,1&-1&-1&\,1&-1&\,1&\,1&-1&\,1&-1&-1&\,1&-1&\,1&\,1&-1\\
\,1&\,1&\,1&\,1&\,1&\,1&\,1&\,1&-1&-1&-1&-1&-1&-1&-1&-1\\
\,1&-1&\,1&-1&\,1&-1&\,1&-1&-1&\,1&-1&\,1&-1&\,1&-1&\,1\\
\,1&\,1&\,1&\,1&-1&-1&-1&-1&-1&-1&-1&-1&\,1&\,1&\,1&\,1\\
\,1&-1&\,1&-1&-1&\,1&-1&\,1&-1&\,1&-1&\,1&\,1&-1&\,1&-1\\
\,1&\,1&-1&-1&\,1&\,1&-1&-1&-1&-1&\,1&\,1&-1&-1&\,1&\,1\\
\,1&-1&-1&\,1&\,1&-1&-1&\,1&-1&\,1&\,1&-1&-1&\,1&\,1&-1\\
\,1&\,1&-1&-1&-1&-1&\,1&\,1&-1&-1&\,1&\,1&\,1&\,1&-1&-1\\
\,1&-1&-1&\,1&-1&\,1&\,1&-1&-1&\,1&\,1&-1&\,1&-1&-1&\,1
\end{array}\right]. 
\label{eq:Appendix-U-matrix}
\end{align}
\end{widetext}
From Eq.~\eqref{eq:Appendix-U-matrix}, we obtain, e.g., $|X+++,+++\rangle =\big[|Z+++,+++\rangle \,+\, |Z+++,+--\rangle \,+\, |Z+++,--+\rangle + |Z+++,-+-\rangle + |Z+--,+++\rangle +  |Z+--,+--\rangle + |Z+--,--+\rangle + |Z+--,-+-\rangle + |Z--+,+++\rangle + |Z--+,+--\rangle + |Z--+,--+\rangle + |Z--+,-+-\rangle + |Z-+-,+++\rangle + |Z-+-,+--\rangle + |Z-+-,--+\rangle + |Z-+-,-+-\rangle\big]/4$. Note that the unitary matrix $U$ is symmetric.

\section{Harmful two-qubit errors}
{Here we describe the error analysis including the phase factors $\zeta_k^{(\ell)}=\pm1$, $\varsigma_X^{(\ell)}=\pm1$ and $\varsigma_Z^{(\ell)}=\pm1$. }
\label{Appendix-B}
\subsection{Single-qubit errors on wavefunction $|\Psi_{\mathcal{Q}_\ell}\rangle$}
Let us consider a wavefunction $|\Psi_{\mathcal{Q}_\ell}\rangle$ parametrized according to Eq.~\eqref{eq:Psi-state-v2}. We are interested to know how the 27 single-qubit errors change the parameters of  such wavefunction. The results of this section are used later to determine the harmful two-qubit errors for discrete and continuous QEC protocols of the nine-qubit Bacon-Shor code. 

We start by writing the following identities for the $X$, $Z$ and $Y$ errors. For the former we have 
\begin{subequations}
\label{eq:Appendix-X-errors-identities}
\begin{align}
X_1 =&\; Q_4, \label{eq:Appendix-X-errors-identity-1}\\
X_2 =&\; Q_4\, X_{12}, \label{eq:Appendix-X-errors-identity-2}\\
X_3 =&\; Q_4\, X_{12}\, X_{23},\label{eq:Appendix-X-errors-identity-3}\\
X_4 =&\; Q_5\, X_{147} X_{78}\, X_{89},\label{eq:Appendix-X-errors-identity-4}\\
X_5 =&\; Q_5\, X_{147} X_{12}\, X_{89}\,S_x^{\rm (1)},\label{eq:Appendix-X-errors-identity-5}\\
X_6 =&\; Q_5\, X_{147} X_{12}\, X_{23}\,\,S_x^{\rm (1)}\,S_x^{\rm (2)},\label{eq:Appendix-X-errors-identity-6}\\ 
X_7 =&\; Q_1\, X_{78}\, X_{89},\label{eq:Appendix-X-errors-identity-7}\\
X_8 =&\; Q_1\, X_{89},\label{eq:Appendix-X-errors-identity-8}\\ 
X_9 =&\; Q_1,\label{eq:Appendix-X-errors-identity-9}
\end{align}
\end{subequations}
where $Q_1=X_9, Q_4=X_1$ and $Q_5=X_{19}$, cf. Table~\ref{table-I}. For $Z$ errors we have the following identities 
\begin{subequations}
\label{eq:Appendix-Z-errors-identities}
\begin{align}\label{eq:Appendix-Z-errors-identity-1}
Z_1 =& \; Q_{12}, \\
\label{eq:Appendix-Z-errors-identity-2}
Z_2 =& \; Q_{15}\, Z_{123}\, Z_{36}\, Z_{69}, \\
\label{eq:Appendix-Z-errors-identity-3}
Z_3 =& \; Q_3\, Z_{36}\, Z_{69}, \\
\label{eq:Appendix-Z-errors-identity-4}
Z_4 =& \; Q_{12}\, Z_{14}, \\
\label{eq:Appendix-Z-errors-identity-5}
Z_5 =& \; Q_{15}\, Z_{123}\, Z_{14}\,Z_{69}\,S_z^{(1)}, \\
\label{eq:Appendix-Z-errors-identity-6}
Z_6 =& \; Q_{3}\,Z_{69}, \\
\label{eq:Appendix-Z-errors-identity-7}
Z_7 = &\; Q_{12}\,Z_{14}\,Z_{47}, \\
\label{eq:Appendix-Z-errors-identity-8}
Z_8 =& \; Q_{15}\,Z_{123}\, Z_{14}\,Z_{47}\,S_z^{(1)}\,S_z^{(2)},\\
\label{eq:Appendix-Z-errors-identity-9}
Z_9=&\; Q_3, 
\end{align}
\end{subequations}
where $Q_3=Z_9,Q_{12}=Z_1$ and $Q_{15}=Z_{19}$, cf. Table~\ref{table-I}. For $Y$ errors we have 
\begin{subequations}
\label{eq:Appendix-Y-errors-identities}
\begin{align} \label{eq:Appendix-Y-errors-identity-1}
Y_1 = &\; Q_8, \\
\label{eq:Appendix-Y-errors-identity-2}
Y_2 = &\; -Q_{11}\, Z_{123}\, Z_{36}\, Z_{69}\, X_{12}, \\
\label{eq:Appendix-Y-errors-identity-3}
Y_3  = &\; Q_7\, X_{12}\, (iX_{23}\,Z_{36})\, Z_{69}, \\
\label{eq:Appendix-Y-errors-identity-4}
Y_4 = &\; -Q_9\, X_{147}\, X_{78}\, X_{89}\, Z_{14}, \\
\label{eq:Appendix-Y-errors-identity-5}
Y_5 = &\; Q_{10}\, (i X_{147}\,Z_{123})\, (i X_{12}\,Z_{14})\, (i X_{89}\,Z_{69})\,S_x^{(1)}\,S_z^{(1)}, \\
\label{eq:Appendix-Y-errors-identity-6}
Y_6 = &\; -Q_6 \, X_{147}\, X_{12}\, X_{23}\, Z_{69}\, S_x^{(1)}\,  S_x^{(2)}, \\
\label{eq:Appendix-Y-errors-identity-7}
Y_7 = &\; Q_{13}\, X_{89}\, (iX_{78}\,Z_{47})\, Z_{14}, \\
\label{eq:Appendix-Y-errors-identity-8}
Y_8 = &\; -Q_{14}\, Z_{123}\, Z_{14}\,Z_{47}\, X_{89}\, S_{z}^{(1)}\, S_{z}^{(2)}, \\
Y_9 = &\; Q_2, \label{eq:Appendix-Y-errors-identity-9}
\end{align}
\end{subequations}
where $Q_7=X_1Z_9$, $Q_8 = Y_1$, $Q_9 = Y_1X_9$, $Q_{10}=Y_{19}$, $Q_{11}=Y_1Z_9$, $Q_{13}=Z_1X_9$, and $Q_{14}=Z_1Y_9$, cf. Table~\ref{table-I}. 

The above identities express  single-qubit errors in terms of the operators $Q_\ell$, which define the orthonormal bases of the error subspaces (given in Table~\ref{table-I}), the operators $X_{147}$ and $Z_{123}$, the gauge operators $G_k$ and the stabilizer generators $S_{q}^{(n)}$, $q=x,z$ and $n=1,2$. The operators $X_{147}$ and $Z_{123}$ respectively perform logical $X$ ($\alpha \leftrightarrow \beta$) and $Z$ ($\alpha\to \alpha$, $\beta\to -\beta$) operations on the logical state ($\alpha,\beta$) if $\ell=0$ (i.e., when the wavefunction  Eq.~\eqref{eq:Psi-state-v2} belongs to the code space). For $\ell=1,2,...15,$ the  operators $X_{147}$ and $Z_{123}$ additionally introduce an overall sign factor, denoted by $\varsigma_X^{(\ell)}$ and $\varsigma_Z^{(\ell)}$ (respectively), if they anticommute with $Q_\ell$. 
\newline
For $\ell=0$, we replace $G_k\to \mathcal{G}_k$ [since $\zeta_k^{(0)}=1$ in Eq.~\eqref{eq:Gk-Ql}],  $X_{147}\to X_{\rm L}$, $Z_{123}\to Z_{\rm L}$, and $S_{q}^{(n)}\to 1$ (by definition of code space, see Table~\ref{table-I}) in Eqs.~\eqref{eq:Appendix-X-errors-identities}--\eqref{eq:Appendix-Y-errors-identities}. Then, $X$ errors can be rewritten as 
\begin{subequations}
\label{eq:Appendix-X-errors-equivalence}
\begin{align}
\label{eq:Appendix-X-errors-equivalence-1}
X_1\xleftrightarrow{\mathcal{Q}_0}&\; Q_4, \\
\label{eq:Appendix-X-errors-equivalence-2}
X_2\xleftrightarrow{\mathcal{Q}_0}&\; Q_4\, X_1^{\rm g}, \\
\label{eq:Appendix-X-errors-equivalence-3}
X_3\xleftrightarrow{\mathcal{Q}_0}&\; Q_4\,X_{12}^{\rm g}, \\
\label{eq:Appendix-X-errors-equivalence-4}
X_4\xleftrightarrow{\mathcal{Q}_0}&\; Q_5\, X_{\rm L}\, X_{34}^{\rm g}, \\
\label{eq:Appendix-X-errors-equivalence-5}
X_5\xleftrightarrow{\mathcal{Q}_0}&\; Q_5\,X_{\rm L}\, X_{14}^{\rm g}, \\
\label{eq:Appendix-X-errors-equivalence-6}
X_6\xleftrightarrow{\mathcal{Q}_0}&\; Q_5\,  X_{\rm L}\, X_{12}^{\rm g}, \\
\label{eq:Appendix-X-errors-equivalence-7}
X_7\xleftrightarrow{\mathcal{Q}_0}&\; Q_1\,  X_{34}^{\rm g}, \\
\label{eq:Appendix-X-errors-equivalence-8}
X_8\xleftrightarrow{\mathcal{Q}_0}&\; Q_1\,  X_{4}^{\rm g}, \\
\label{eq:Appendix-X-errors-equivalence-9}
X_9\xleftrightarrow{\mathcal{Q}_0}&\; Q_1, 
\end{align}
\end{subequations}
where $\xleftrightarrow{\mathcal{Q}_0}$ indicates that the equivalence relations~\eqref{eq:Appendix-X-errors-equivalence} applies to wavefunctions of the form of Eq.~\eqref{eq:Psi-state-v2} with $\ell=0$. Similarly, for $Z$ and $Y$ errors we have 
\begin{subequations}
\label{eq:Appendix-Z-errors-equivalence}
\begin{align}
\label{eq:Appendix-Z-errors-equivalence-1}
Z_1 \xleftrightarrow{\mathcal{Q}_0}&\;  Q_{12}, \\
\label{eq:Appendix-Z-errors-equivalence-2}
Z_2 \xleftrightarrow{\mathcal{Q}_0}&\;  Q_{15}\,Z_{\rm L}\, Z_{24}^{\rm g}, \\
\label{eq:Appendix-Z-errors-equivalence-3}
Z_3 \xleftrightarrow{\mathcal{Q}_0}&\;  Q_{3}\,Z_{24}^{\rm g}, \\
\label{eq:Appendix-Z-errors-equivalence-4}
Z_4 \xleftrightarrow{\mathcal{Q}_0}&\;  Q_{12}\, Z_{1}^{\rm g},  \\
\label{eq:Appendix-Z-errors-equivalence-5}
Z_5 \xleftrightarrow{\mathcal{Q}_0}&\;  Q_{15}\,Z_{\rm L}\, Z_{14}^{\rm g}, \\
\label{eq:Appendix-Z-errors-equivalence-6}
Z_6 \xleftrightarrow{\mathcal{Q}_0}&\;  Q_{3}\,  Z_{4}^{\rm g}, \\
\label{eq:Appendix-Z-errors-equivalence-7}
Z_7 \xleftrightarrow{\mathcal{Q}_0}&\;  Q_{12}\,Z_{13}^{\rm g},  \\
\label{eq:Appendix-Z-errors-equivalence-8}
Z_8 \xleftrightarrow{\mathcal{Q}_0}&\;  Q_{15}\,  Z_{\rm L}\, Z_{13}^{\rm g},  \\
\label{eq:Appendix-Z-errors-equivalence-9}
Z_9 \xleftrightarrow{\mathcal{Q}_0}&\;  Q_{3}, 
\end{align}
\end{subequations}
and 
\begin{subequations}
\label{eq:Appendix-Y-errors-equivalence}
\begin{align}
\label{eq:Appendix-Y-errors-equivalence-1}
Y_1 \xleftrightarrow{\mathcal{Q}_0}&\; Q_8,  \\
\label{eq:Appendix-Y-errors-equivalence-2}
Y_2 \xleftrightarrow{\mathcal{Q}_0}&\; -Q_{11}\, Z_{\rm L}\,Z_{24}^{\rm g}\,X_1^{\rm g}, \\
\label{eq:Appendix-Y-errors-equivalence-3}
Y_3 \xleftrightarrow{\mathcal{Q}_0}&\; Q_{7}\,X_1^{\rm g}\, Y_2^{\rm g}\, Z_{4}^{\rm g}, \\
\label{eq:Appendix-Y-errors-equivalence-4}
Y_4 \xleftrightarrow{\mathcal{Q}_0}&\; -Q_{9}\, X_{\rm L}\, X_{34}^{\rm g}\, Z_{1}^{\rm g} \\
\label{eq:Appendix-Y-errors-equivalence-5}
Y_5 \xleftrightarrow{\mathcal{Q}_0}&\; Q_{10}\,Y_{\rm L}\, Y_{14}^{\rm g},  \\
\label{eq:Appendix-Y-errors-equivalence-6}
Y_6 \xleftrightarrow{\mathcal{Q}_0}&\; -Q_{6}\, X_{\rm L}\, X_{12}^{\rm g}\, Z_4^{\rm g},  \\
\label{eq:Appendix-Y-errors-equivalence-7}
Y_7 \xleftrightarrow{\mathcal{Q}_0}&\; Q_{13}\,X_{4}^{\rm g}\, Y_{3}^{\rm g}\, Z_1^{\rm g},  \\
\label{eq:Appendix-Y-errors-equivalence-8}
Y_8 \xleftrightarrow{\mathcal{Q}_0}&\; -Q_{14}\, Z_{\rm L}\, X_{4}^{\rm g}\, Z_{13}^{\rm g}, \\
\label{eq:Appendix-Y-errors-equivalence-9}
Y_9 \xleftrightarrow{\mathcal{Q}_0}&\;  Q_2, 
\end{align}
\end{subequations}
where $Y_j^{\rm g} = i X_j^{\rm g}Z_j^{\rm g}$. 
\newline
To get the above equivalence relations for the error subspaces ($\ell=1,2,...15$), we replace   $G_k\to \zeta_k^{(\ell)}\, \mathcal{G}_k$, $X_{147}\to \varsigma_X^{(\ell)}\,X_{\rm L}$, $Z_{123}\to \varsigma_Z^{(\ell)}\,Z_{\rm L}$, and $S_{q}^{(n)}\to\pm1$ (according to Table~\ref{table-I}) in Eqs.~\eqref{eq:Appendix-X-errors-identities}--\eqref{eq:Appendix-Y-errors-identities}. As mentioned above, $\zeta_k^{(\ell)}=-1\,(+1)$ if $G_k$ anticommutes (commutes) with operator $Q_\ell$, and $\varsigma_X^{(\ell)}=-1\,(+1)$ and $\varsigma_Z^{(\ell)}=-1\,(+1)$ if $X_{147}$ and $Z_{123}$ anticommute (commute) with operator $Q_\ell$, respectively. 

\begin{table*}
\caption{\label{Appendix-table-IV} Multiplication table between error-subspace basis operators ${Q}_\ell$. Note that this table is nonsymmetric. } 
\begin{ruledtabular}
\begin{tabular}{c| c c c c c c c c c c c c c c c c c c c}
$\times$& $Q_0$ & ${Q}_1$ & ${Q}_2$ & ${Q}_3$ & ${Q}_4$ & ${Q}_5$ & ${Q}_6$ & ${Q}_7$ & ${Q}_8$ & ${Q}_9$ & ${Q}_{10}$ & ${Q}_{11}$ & ${Q}_{12}$ & ${Q}_{13}$ & ${Q}_{14}$& ${Q}_{15}$\\ \hline
$Q_0=\openone$& $\openone$ & ${Q}_1$ & ${Q}_2$ & ${Q}_3$ & ${Q}_4$ & ${Q}_5$ & ${Q}_6$ & ${Q}_7$ & ${Q}_8$ & ${Q}_9$ & ${Q}_{10}$ & ${Q}_{11}$ & ${Q}_{12}$ & ${Q}_{13}$ & ${Q}_{14}$& ${Q}_{15}$\\ 
${Q}_1$& $Q_1$ & $\openone$ & $i{Q}_3$ & $-i{Q}_2$ & ${Q}_5$ & ${Q}_4$ & $i{Q}_7$ & $-i{Q}_6$ & ${Q}_9$ & ${Q}_8$ & $i{Q}_{11}$ & $-i{Q}_{10}$ & ${Q}_{13}$ & ${Q}_{12}$ & $i{Q}_{15}$ & $-i{Q}_{14}$ \\ 
${Q}_2$& $Q_2$& $-iQ_3$ & $\openone$ & $i{Q}_1$ & ${Q}_6$ & $-i{Q}_7$ & ${Q}_4$ &  $i{Q}_5$ & ${Q}_{10}$ & $-i{Q}_{11}$ & ${Q}_8$  & $i{Q}_9$ & ${Q}_{14}$ & $-i{Q}_{15}$ & ${Q}_{12}$ & $i{Q}_{13}$\\
${Q}_3$ & $Q_3$ & $iQ_2$ & $-iQ_{1}$& $\openone$ & ${Q}_7$ & $i{Q}_6$ & $-i{Q}_5$ & ${Q}_4$ & ${Q}_{11}$ & $i{Q}_{10}$ & $-i{Q}_9$ & ${Q}_8$ & ${Q}_{15}$ & $i{Q}_{14}$ & $-i{Q}_{13}$ & ${Q}_{12}$ \\
${Q}_4$& $Q_4$ & $Q_5$ & $Q_6$& ${Q}_{7}$ & $\openone$& ${Q}_1$ & ${Q}_2$ & ${Q}_3$ & $i{Q}_{12}$ & $i{Q}_{13}$ & $i{Q}_{14}$ & $i{Q}_{15}$ & $-i{Q}_8$ & $-i{Q}_9$ & $-i{Q}_{10}$ & $-i{Q}_{11}$  \\
${Q}_5$& $Q_5$ & $Q_4$ & $iQ_7$ & $-i{Q}_{6}$ & ${Q}_{1}$& $\openone$ & $i{Q}_3$ & $-i{Q}_2$ & $i{Q}_{13}$ & $i{Q}_{12}$ &  $-{Q}_{15}$ & ${Q}_{14}$ & $-i{Q}_9$ & $-i{Q}_8$ & ${Q}_{11}$ & $-{Q}_{10}$ \\
${Q}_6$& $Q_6$ & $-iQ_7$ & $Q_4$ & $i{Q}_{5}$ & ${Q}_{2}$ & $-i{Q}_{3}$ & $\openone$ & $i{Q}_{1}$ & $i{Q}_{14}$ & ${Q}_{15}$ & $i{Q}_{12}$ & $-{Q}_{13}$ & $-i{Q}_{10}$ & $-{Q}_{11}$ & $-i{Q}_{8}$ & ${Q}_{9}$ \\
${Q}_7$ & $Q_7$ & $iQ_6$& $-iQ_5$ & ${Q}_{4}$& ${Q}_{3}$ & $i{Q}_{2}$ & $-i{Q}_{1}$ & $\openone$ & $i{Q}_{15}$ & $-{Q}_{14}$ & ${Q}_{13}$ & $i{Q}_{12}$ & $-i{Q}_{11}$ & ${Q}_{10}$ & $-{Q}_{9}$ & $-i{Q}_{8}$ \\
${Q}_8$ & $Q_8$ & $Q_9$ & $Q_{10}$ & ${Q}_{11}$& $-i{Q}_{12}$ & $-i{Q}_{13}$& $-i{Q}_{14}$ & $-i{Q}_{15}$& $\openone$ & ${Q}_{1}$ & ${Q}_{2}$ & ${Q}_{3}$ & $i{Q}_{4}$ & $i{Q}_{5}$ & $i{Q}_{6}$ & $i{Q}_{7}$ \\
${Q}_9$ & $Q_9$ & $Q_8$ & $iQ_{11}$ & $-i{Q}_{10}$ & $-i{Q}_{13}$ & $-i{Q}_{12}$ & ${Q}_{15}$ &$-{Q}_{14}$ & ${Q}_{1}$ & $\openone$ & $i{Q}_{3}$ & $-i{Q}_{2}$ & $i{Q}_{5}$ & $i{Q}_{4}$ & ${-Q}_{7}$ & ${Q}_{6}$ \\ 
${Q}_{10}$ & $Q_{10}$& $-iQ_{11}$ & $Q_8$& $i{Q}_{9}$& $-i{Q}_{14}$ &$-{Q}_{15}$ & $-i{Q}_{12}$& ${Q}_{13}$  & ${Q}_{2}$ & $-i{Q}_{3}$ & $\openone$ & $i{Q}_{1}$ & $i{Q}_{6}$ & ${Q}_{7}$ & $i{Q}_{4}$ & $-{Q}_{5}$ \\
${Q}_{11}$ & $Q_{11}$ & $iQ_{10}$& $-iQ_9$ & ${Q}_{8}$& $-i{Q}_{15}$ & ${Q}_{14}$ & $-{Q}_{13}$ & $-i{Q}_{12}$ & ${Q}_{3}$ & $i{Q}_{2}$& $-i{Q}_{1}$ & $\openone$ & $i{Q}_{7}$ &  $-{Q}_{6}$ & ${Q}_{5}$ & $i{Q}_{4}$ \\
${Q}_{12}$ & $Q_{12}$ & $Q_{13}$& $Q_{14}$ & ${Q}_{15}$ & $i{Q}_{8}$& $i{Q}_{9}$ & $i{Q}_{10}$& $i{Q}_{11}$ & $-i{Q}_{4}$ & $-i{Q}_{5}$ &$-i{Q}_{6}$ & $-i{Q}_{7}$ & $\openone$ & ${Q}_{1}$ & ${Q}_{2}$ & ${Q}_{3}$ & \\
${Q}_{13}$ & $Q_{13}$ & $Q_{12}$ & $iQ_{15}$ & $-i{Q}_{14}$ &$i{Q}_{9}$ &$i{Q}_{8}$ & $-{Q}_{11}$& ${Q}_{10}$ & $-i{Q}_{5}$ & $-i{Q}_{4}$ & ${Q}_{7}$ & $-{Q}_{6}$ & ${Q}_{1}$& $\openone$ & $i{Q}_{3}$ & $-i{Q}_{2}$ \\
${Q}_{14}$ & $Q_{14}$ & $-iQ_{15}$ & $Q_{12}$ & $i{Q}_{13}$ & $i{Q}_{10}$& ${Q}_{11}$& $i{Q}_{8}$& $-{Q}_{9}$ & $-i{Q}_{6}$& $-{Q}_{7}$ & $-i{Q}_{4}$ & ${Q}_{5}$&${Q}_{2}$ & $-i{Q}_{3}$ & $\openone$ & $i{Q}_{1}$ \\
${Q}_{15}$ & $Q_{15}$ & $iQ_{14}$ & $-iQ_{13}$ & ${Q}_{12}$ &$i{Q}_{11}$ &$-{Q}_{10}$ & ${Q}_{9}$ & $i{Q}_{8}$ & $-i{Q}_{7}$ & ${Q}_{6}$ & $-{Q}_{5}$ &$-i{Q}_{4}$ & ${Q}_{3}$& $i{Q}_{2}$ & $-i{Q}_{1}$& $\openone$
\end{tabular}
\end{ruledtabular}
\end{table*}

There is one more thing that we need to know to fully determine the effect of single-qubit errors on the  wavefunction $|\Psi_{\mathcal{Q}_\ell}\rangle$. This is the multiplication Table~\ref{Appendix-table-IV} between two operators $Q_{\ell_1}$ and $Q_{\ell_2}$ (Table~\ref{table-II} is a simplified version of this table that does not include phase factors.) Note that such a table is nonsymmetric since some $Q_{\ell_1}$ and $ Q_{\ell_2}$ anticommute. To explain how to use this table, let us consider the following example. We wish to find the new logical state ($\tilde\alpha,\tilde\beta$) and new state of the gauge qubits (coefficients $\tilde c_{q_1q_2q_3q_4}$), as well as the overall phase factor that the  wavefunction $|\Psi_{\mathcal{Q}_4}\rangle=Q_4|\Psi_0\rangle$ acquires after applying error $Z_2$ (here, $|\Psi_0\rangle$ is an auxiliary code space wavefunction that only depends on $\alpha,\beta$ and coefficients $c_{q_1q_2q_3q_4}$.) The wavefunction after this error is $Z_2|\Psi_{\mathcal{Q}_4}\rangle=-Q_{15}Z_{\rm L}Z_{24}^{\rm g}\times Q_4|\Psi_0\rangle$, where the sign factor comes from replacing $Z_{123}$ by $-Z_{\rm L}$ in Eq.~\eqref{eq:Appendix-Z-errors-identity-2}, since $Z_{123}$ anticommutes with $Q_4=X_1$. Next, we use Table~\ref{Appendix-table-IV} to obtain $Q_{15}\times Q_4=iQ_{11}$, so $Z_2|\Psi_{\mathcal{Q}_4}\rangle = -i Z_{\rm L}Z_{24}^{\rm g} Q_{11}|\Psi\rangle$. Thus, the wavefunction after the error is in the error subspace  $\mathcal{Q}_{11}$ with $\tilde\alpha=\alpha, \tilde\beta =-\beta$, $\tilde c_{0000}=c_{0000}$, $\tilde c_{0001}=-c_{0001}$, $\tilde c_{0010}=c_{0010}$, $\tilde c_{0011}=-c_{0011}$, $\tilde c_{0100}=-c_{0100}$, $\tilde c_{0101}=c_{0101}$, $\tilde c_{0110}=-c_{0110}$, $\tilde c_{0111}=c_{0111}$, $\tilde c_{1000}=c_{1000}$, $\tilde c_{1001}=-c_{1001}$, $\tilde c_{1010}=c_{1010}$, $\tilde c_{1011}=-c_{1011}$, $\tilde c_{1100}=-c_{1100}$, $\tilde c_{1101}=c_{1101}$, $\tilde c_{1110}=-c_{1110}$, $\tilde c_{1111}=c_{1111}$, and the overall phase factor is $-i$. Note that $Z_{\rm L}$ and $Z_{14}^{\rm g}$ act directly on the probability amplitudes, so they commute with any operator $Q_\ell$, which acts  on the basis vectors.

Note that if we ignore overall phase factors in the equivalence relations~\eqref{eq:Appendix-X-errors-equivalence}--\eqref{eq:Appendix-Y-errors-equivalence}, we obtain the equivalence relations of Table~\ref{table-II}. 

\subsection{Harmful two-qubit errors in conventional operation of nine-qubit Bacon-Shor code} \label{subsec:harmful-two-qubit-errors}
In this section we want to determine the harmful two error combinations ($E_1$ and $E_2$) that induce a logical error after an operation cycle. By definition, these two errors  occur in the same operation cycle and, after application of the error correcting operation $C_{\rm op}$ at the end of the cycle, the logical state suffers from a logical operation ($X_{\rm L}$, $Y_{\rm L}$ or $Z_{\rm L}$). That is, the two errors satisfy %
\begin{align}
\label{eq:Appendix-logical-error-condition}
E_1\,E_2\,C_{\rm op} \sim X_{\rm L}, Y_{\rm L} \; {\rm or}\; Z_{\rm L},
\end{align}
 where ``$\sim$'' means equivalence modulo gauge operations, and we  disregard overall phase factors in this section. We assume that the  system state is initially in the code space; however, this assumption is not crucial.  It is convenient in what follows to simplify the equivalence relations~\eqref{eq:Appendix-X-errors-equivalence}--\eqref{eq:Appendix-Y-errors-equivalence} by disregarding the gauge operations:
\begin{eqnarray}
\label{eq:simplified-factorization-relations}
X_7, X_8, X_9 \sim&\; {Q}_1,  \nonumber \\
Y_9               \sim&\; {Q}_{2}, \nonumber \\ 
Z_3, Z_6, Z_9 \sim&\; {Q}_{3}, \nonumber \\
X_1, X_2, X_3 \sim&\; {Q}_4, \nonumber \\
X_4, X_5, X_6 \sim&\; {Q}_5\, X_{\rm L}, \nonumber \\
Y_6                \sim&\; {Q}_{6}\,X_{\rm L}, \nonumber \\
Y_3 			\sim&\; {Q}_{7}, \nonumber \\
Y_1 			\sim&\; {Q}_{8}, \nonumber \\
Y_4 			\sim&\; {Q}_{9}\, X_{\rm L}, \nonumber \\
Y_5 			\sim&\; {Q}_{10}\, Y_{\rm L},\nonumber \\
Y_2 			\sim&\; {Q}_{11}\,Z_{\rm L},\nonumber \\
Z_1, Z_4, Z_7  \sim&\; {Q}_{12}, \nonumber \\
Y_7 			\sim&\; {Q}_{13},\nonumber \\ 
Y_8 			\sim&\; {Q}_{14}\, Z_{\rm L},\nonumber \\ 
Z_2, Z_5, Z_8 	\sim&\; {Q}_{15}\, Z_{\rm L}. 
\end{eqnarray}
Note that the left-hand sides of Eq.~\eqref{eq:simplified-factorization-relations} are the error correcting operations $C_{\rm op}$ given in Table~\ref{table-I}; this is due to the fact that $Q_{\ell}^2$ is trivial (identity).

Let us assume that $E_1\sim Q_{\ell_1}\, O_1$ and $E_2\sim Q_{\ell_2}\, O_2$ so $E_1E_2\sim Q_\ell \, (O_1O_2)$, where $Q_\ell=Q_{\ell_1}Q_{\ell_2}$, and $O_1$ and $O_2$ can be trivial (identity) or a logical operation ($X_{\rm L}$, $Y_{\rm L}$ or $Z_{\rm L}$). The fact that $E_1E_2$ is "proportional" to $Q_\ell$ implies that such two-qubit error induces a jump from the code space to the error subspace $\mathcal{Q}_\ell$. Since the error correcting operation must bring the system state back to the code space, it must also be "proportional" to $Q_\ell$. To proceed, we fix $Q_{\ell}$ (in this way, $C_{\rm op}$ is also fixed up to gauge operations) and then use Table~\ref{table-II} to find all possible pairs $Q_{\ell_1}$ and $Q_{\ell_2}$ whose product is $Q_\ell$; we obtain: 
\begin{widetext}
\begin{align}
\label{eq:Qell1-2-list}
{Q}_1 =&\, \big\{\, {Q}_{2\,3},\, {Q}_{4\,5}\,[X_{\rm L}],\, {Q}_{6\,7}\,[X_{\rm L}],\, {Q}_{8\,9}\,[X_{\rm L}],%
{Q}_{10\,11}\,[X_{\rm L}],\, {Q}_{12\,13},\,{Q}_{14\,15}\big\}, \nonumber \\
{Q}_2 =&\,\big \{{Q}_{1\,3},\, {Q}_{4\,6}\,[X_{\rm L}],\, {Q}_{5\,7}\,[X_{\rm L}],\, {Q}_{8\,10}\,[Y_{\rm L}],%
{Q}_{9\,11}\,[Y_{\rm L}],\, {Q}_{12\,14}\,[Z_{\rm L}], {Q}_{13\,15}\,[Z_{\rm L}]\big\}, \nonumber \\
{Q}_3 =&\,\big \{{Q}_{1\,2},\, {Q}_{4\,7},\, {Q}_{5\,6},\, {Q}_{8\,11}\,[Z_{\rm L}],%
{Q}_{9\,10}\,[Z_{\rm L}],\, {Q}_{12\,15}\,[Z_{\rm L}], {Q}_{13\,14}\,[Z_{\rm L}]\big\},\nonumber \\
{Q}_4 =&\,\big \{{Q}_{1\,5}\,[X_{\rm L}],\, {Q}_{2\,6}\,[X_{\rm L}],\, {Q}_{3\,7},\, {Q}_{8\,12},%
{Q}_{9\,13}\,[X_{\rm L}],\, {Q}_{10\,14}\,[X_{\rm L}], {Q}_{11\,15}\big\},\nonumber \\
{Q}_5 =&\,\big \{{Q}_{1\,4}\,[X_{\rm L}],\, {Q}_{2\,7}\,[X_{\rm L}],\, {Q}_{3\,6},\, {Q}_{8\,13}\,[X_{\rm L}],%
 {Q}_{9\,12},\, {Q}_{10\,15}, {Q}_{11\,14}\,[X_{\rm L}]\big\},\nonumber \\
{Q}_6 =&\,\big \{{Q}_{1\,7}\,[X_{\rm L}],\, {Q}_{2\,4}\,[X_{\rm L}],\, {Q}_{3\,5},\, {Q}_{8\,14}\,[Y_{\rm L}],%
 {Q}_{9\,15}\,[Z_{\rm L}],\, {Q}_{10\,12}\,[Z_{\rm L}], {Q}_{11\,13}\,[Y_{\rm L}]\big\},\nonumber \\
{Q}_7 =&\,\big \{{Q}_{1\,6}\,[X_{\rm L}],\, {Q}_{2\,5}\,[X_{\rm L}],\, {Q}_{3\,4},\, {Q}_{8\,15}\,[Z_{\rm L}],%
 {Q}_{9\,14}\,[Y_{\rm L}],\, {Q}_{10\,13}\,[Y_{\rm L}], {Q}_{11\,12}\,[Z_{\rm L}]\big\},\nonumber \\
{Q}_8 =&\,\big \{{Q}_{1\,9}\,[X_{\rm L}],\, {Q}_{2\,10}\,[Y_{\rm L}],\, {Q}_{3\,11}\,[Z_{\rm L}],\, {Q}_{4\,12},%
 {Q}_{5\,13}\,[X_{\rm L}],\, {Q}_{6\,14}\,[Y_{\rm L}], {Q}_{7\,15}\,[Z_{\rm L}]\big\},\nonumber \\
{Q}_9 =&\,\big \{{Q}_{1\,8},\, {Q}_{2\,11}\,[Y_{\rm L}],\, {Q}_{3\,10}\,[Z_{\rm L}],\, {Q}_{4\,13}\,[X_{\rm L}],%
 {Q}_{5\,12},\, {Q}_{6\,15}\,[Z_{\rm L}], {Q}_{7\,14}\,[Y_{\rm L}]\big\},\nonumber \\
{Q}_{10} =&\,\big \{{Q}_{1\,11}\,[X_{\rm L}],\, {Q}_{2\,8}\,[Y_{\rm L}],\, {Q}_{3\,9}\,[Z_{\rm L}],\, {Q}_{4\,14}\,[X_{\rm L}],%
 {Q}_{5\,15},\, {Q}_{6\,12}\,[Z_{\rm L}], {Q}_{7\,13}\,[Y_{\rm L}]\big\},\nonumber \\
 {Q}_{11} =&\,\big \{{Q}_{1\,10}\,[X_{\rm L}],\, {Q}_{2\,9}\,[Y_{\rm L}],\, {Q}_{3\,8}\,[Z_{\rm L}],\, {Q}_{4\,15},%
  {Q}_{5\,14}\,[X_{\rm L}],\, {Q}_{6\,13}\,[Y_{\rm L}], {Q}_{7\,12}\,[Z_{\rm L}]\big\},\nonumber \\
{Q}_{12} =&\,\big \{{Q}_{1\,13},\, {Q}_{2\,14}\,[Z_{\rm L}],\, {Q}_{3\,15}\,[Z_{\rm L}],\, {Q}_{4\,8},%
 {Q}_{5\,9},\, {Q}_{6\,10}\,[Z_{\rm L}], {Q}_{7\,11}\,[Z_{\rm L}]\big\},\nonumber \\
{Q}_{13} =&\,\big \{{Q}_{1\,12},\, {Q}_{2\,15}\,[Z_{\rm L}],\, {Q}_{3\,14}\,[Z_{\rm L}],\, {Q}_{4\,9}\,[X_{\rm L}],%
{Q}_{5\,8}[X_{\rm L}],\, {Q}_{6\,11}\,[Y_{\rm L}], {Q}_{7\,10}\,[Y_{\rm L}]\big\},\nonumber \\
{Q}_{14} =&\,\big \{{Q}_{1\,15},\, {Q}_{2\,12}\,[Z_{\rm L}],\, {Q}_{3\,13}\,[Z_{\rm L}],\, {Q}_{4\,10}\,[X_{\rm L}],%
{Q}_{5\,11}[X_{\rm L}],\, {Q}_{6\,8}\,[Y_{\rm L}], {Q}_{7\,9}\,[Y_{\rm L}]\big\},\nonumber \\
{Q}_{15} =&\,\big \{{Q}_{1\,14},\, {Q}_{2\,13}\,[Z_{\rm L}],\, {Q}_{3\,12}\,[Z_{\rm L}],\, {Q}_{4\,11},%
{Q}_{5\,10},\, {Q}_{6\,9}\,[Z_{\rm L}],\, {Q}_{7\,8}\,[Z_{\rm L}]\big\},
\end{align}
\end{widetext}
where $Q_{\ell_1\,\ell_2} = Q_{\ell_1}Q_{\ell_2}$. Next, we use Eq.~\eqref{eq:simplified-factorization-relations} to find the (single-qubit) errors $E_1$ and $E_2$ that are ``proportional'' to $Q_{\ell_1}$ and $Q_{\ell_2}$, respectively. Finally, for each $Q_\ell$-line of Eq.~\eqref{eq:Qell1-2-list}, we obtain the corresponding $C_{\rm op}$ from Table~\ref{table-I} and check the condition~\eqref{eq:Appendix-logical-error-condition} to determine whether $E_1E_2$ induces a logical $X$, $Y$ or $Z$ error; the type of logical error is indicated inside the square brakets of Eq.~\eqref{eq:Qell1-2-list}. The lists of harmful two-qubit errors that lead to logical errors are given in Eqs.~\eqref{eq:two-qubit-X-errors}--\eqref{eq:two-qubit-Y-errors}. 

\subsection{Harmful two-qubit errors in continuous operation of nine-qubit Bacon-Shor code}
\label{Section:Harmful-two-qubit-errors-continuous}
It turns out that harmful two-qubit errors in  the discrete operation are also harmful two-qubit errors in the continuous operation, assuming large time  averaging parameter $T_{\rm c}$ (noiseless cross-correlators limit). To realize this let us consider realizations with only two errors; namely, first $E_1\sim Q_{\ell_1}O_1$ and second $E_2\sim Q_{\ell_2}O_2$ that occur at moments $t_{\rm err}^{(1)}$ and $t_{\rm err}^{(2)}$, respectively. Here, $O_1$ and $O_2$ can be trivial (identity) or logical operations ($X_{\rm L}$, $Y_{\rm L},$ or $Z_{\rm L}$); let us also assume that $Q_{\ell_1}\, Q_{\ell_2} = Q_{\ell}$. 

The actual error syndrome path $\mathcal{S}(t)$ exhibits two jumps: the first one  from $\mathcal{S}=0$ to $\mathcal{S}=\ell_1$ at the moment $t_{\rm err}^{(1)}$, and the second one from $\mathcal{S}=\ell_1$ to $\mathcal{S}=\ell$ at the moment $t_{\rm err}^{(2)}$. In contrast, the monitored error syndrome path $\mathcal{S}_{\rm m}(t)$ exhibits only one jump from $\mathcal{S}_{\rm m}=0$ to $\mathcal{S}_{\rm m}=\ell$ if the errors occur close in time such that $t_{\rm err}^{(2)} - t_{\rm err}^{(1)} < \Delta t_{\rm cont}$, where  $\Delta t_{\rm cont}$  is given in Eq.~\eqref{eq:Deltat-cont}. In this situation, the continuous QEC protocol assigns the single-qubit error $E_{\rm false}=Q_{\ell}\,O_{\rm false}$, where $O_{\rm false}$ can be trivial or a logical operation. The total logical operations from $\mathcal{S}(t)$ and $\mathcal{S}_{\rm m}(t)$ are $\mathcal{O}=O_1O_2$ and $\mathcal{O}_{\rm m} = O_{\rm false}$, respectively, and a logical error occurs if $\mathcal{O}_{\rm m}\mathcal{O}$ is a logical operation--- see discussion above Eq.~\eqref{eq:Prob-logical-err}. This condition for logical error in our continuous QEC protocol is equivalent to say that $E_{1}E_2E_{\rm false}\sim X_{\rm L},$ $Y_{\rm L}$ or $Z_{\rm L}$, which is the condition for harmful two-qubit errors in the conventional operation, cf.  Eq.~\eqref{eq:Appendix-logical-error-condition}, with $E_{\rm false}$ playing the role of $C_{\rm op}$. 

\section{Logical error rates for discrete operation}
\label{Appendix-C}
In this section we present the formulas for the logical error rates of the nine-qubit Bacon-Shor code where the nine physical qubits are subject to $X$, $Y$ and $Z$ errors with occurrence rates $\Gamma_i^{(X)}, \Gamma_i^{(Y)}$ and $\Gamma_i^{(Z)}$. The logical $X$ error rate is equal to 
\begin{widetext}
\begin{eqnarray}
\gamma_X^{\rm cont} &=&\; \Delta t\Big[(\Gamma_1^{(X)}+ \Gamma_2^{(X)} + \Gamma_3^{(X)})(\Gamma_4^{(X)}+ \Gamma_5^{(X)} + \Gamma_6^{(X)}) + \Gamma_3^{(Y)}\Gamma_6^{(Y)} + \Gamma_1^{(Y)}\Gamma_4^{(Y)} + \Gamma_2^{(Y)}\Gamma_5^{(Y)} + \nonumber \\
&\mbox{}&\hspace{0.7cm} (\Gamma_1^{(X)} + \Gamma_2^{(X)} + \Gamma_3^{(X)})\Gamma_6^{(Y)} +  (\Gamma_4^{(X)} + \Gamma_5^{(X)} + \Gamma_6^{(X)})\Gamma_3^{(Y)} + \nonumber \\
&\mbox{}&\hspace{0.7cm}  (\Gamma_7^{(X)}+ \Gamma_8^{(X)} + \Gamma_9^{(X)})(\Gamma_4^{(X)}+ \Gamma_5^{(X)} + \Gamma_6^{(X)}) +  \Gamma_6^{(Y)}\Gamma_9^{(Y)}  + \Gamma_4^{(Y)}\Gamma_7^{(Y)} + \Gamma_5^{(Y)}\Gamma_8^{(Y)}  + \nonumber \\
&\mbox{}&\hspace{0.7cm}   (\Gamma_1^{(X)}+ \Gamma_2^{(X)} + \Gamma_3^{(X)})(\Gamma_7^{(X)}+ \Gamma_8^{(X)} + \Gamma_9^{(X)}) +  \Gamma_3^{(Y)}\Gamma_9^{(Y)}  + \Gamma_1^{(Y)}\Gamma_7^{(Y)} + \Gamma_2^{(Y)}\Gamma_8^{(Y)}  + \nonumber \\
&\mbox{}&\hspace{0.7cm}  \Gamma_3^{(Y)} (\Gamma_7^{(X)}+ \Gamma_8^{(X)} + \Gamma_9^{(X)}) + \Gamma_9^{(Y)} (\Gamma_1^{(X)}+ \Gamma_2^{(X)} + \Gamma_3^{(X)}) + \nonumber \\
&\mbox{}&\hspace{0.7cm}   \Gamma_6^{(Y)} (\Gamma_7^{(X)}+ \Gamma_8^{(X)} + \Gamma_9^{(X)}) + \Gamma_9^{(Y)} (\Gamma_4^{(X)}+ \Gamma_5^{(X)} + \Gamma_6^{(X)}) + \nonumber \\
&\mbox{}&\hspace{0.7cm}   \Gamma_4^{(Y)} (\Gamma_7^{(X)}+ \Gamma_8^{(X)} + \Gamma_9^{(X)}) + \Gamma_7^{(Y)} (\Gamma_4^{(X)}+ \Gamma_5^{(X)} + \Gamma_6^{(X)}) + \nonumber \\
&\mbox{}&\hspace{0.7cm}   \Gamma_1^{(Y)} (\Gamma_7^{(X)}+ \Gamma_8^{(X)} + \Gamma_9^{(X)}) + \Gamma_7^{(Y)} (\Gamma_1^{(X)}+ \Gamma_2^{(X)} + \Gamma_3^{(X)}) + \nonumber \\
&\mbox{}&\hspace{0.7cm}   \Gamma_2^{(Y)} (\Gamma_7^{(X)}+ \Gamma_8^{(X)} + \Gamma_9^{(X)}) + \Gamma_8^{(Y)} (\Gamma_1^{(X)}+ \Gamma_2^{(X)} + \Gamma_3^{(X)}) + \nonumber \\
&\mbox{}&\hspace{0.7cm}   \Gamma_5^{(Y)} (\Gamma_7^{(X)}+ \Gamma_8^{(X)} + \Gamma_9^{(X)}) + \Gamma_8^{(Y)} (\Gamma_4^{(X)}+ \Gamma_5^{(X)} + \Gamma_6^{(X)}) + \nonumber \\
&\mbox{}&\hspace{0.7cm}   \Gamma_4^{(Y)} (\Gamma_1^{(X)}+ \Gamma_2^{(X)} + \Gamma_3^{(X)}) + \Gamma_1^{(Y)} (\Gamma_4^{(X)}+ \Gamma_5^{(X)} + \Gamma_6^{(X)}) + \nonumber \\
&\mbox{}&\hspace{0.7cm}   \Gamma_5^{(Y)} (\Gamma_1^{(X)}+ \Gamma_2^{(X)} + \Gamma_3^{(X)}) + \Gamma_2^{(Y)} (\Gamma_4^{(X)}+ \Gamma_5^{(X)} + \Gamma_6^{(X)}) \Big],
\label{eq:Appendix-logical-X-error-rate}
\end{eqnarray}
\end{widetext}
where the terms in the lines of formula~\eqref{eq:Appendix-logical-X-error-rate} correspond to the two-qubit errors in   the lines of list~\eqref{eq:two-qubit-X-errors}. The formula for the logical $Z$ error rate is obtained from Eq.~\eqref{eq:Appendix-logical-X-error-rate} with upper labels $X$ and $Z$ in $\Gamma_i^{(E)}$ exchanged and the qubit numbering subscripts $2\leftrightarrow4$, $3\leftrightarrow7$ and $6\leftrightarrow8$ also exchanged. The formula for the logical $Y$ error rate reads as 
\begin{eqnarray}
\gamma_Y^{\rm cont} &=&\; \Delta t \Big[\Gamma_1^{(Y)}\Gamma_5^{(Y)} + \Gamma_2^{(Y)}\Gamma_4^{(Y)} + \nonumber \\
&\mbox{}&\hspace{0.7cm} \Gamma_1^{(Y)}\Gamma_8^{(Y)} + \Gamma_2^{(Y)}\Gamma_7^{(Y)} + \nonumber \\
&\mbox{}&\hspace{0.7cm} \Gamma_4^{(Y)}\Gamma_8^{(Y)} + \Gamma_5^{(Y)}\Gamma_7^{(Y)} + \nonumber \\
&\mbox{}&\hspace{0.7cm} \Gamma_9^{(Y)}\Gamma_5^{(Y)} + \Gamma_6^{(Y)}\Gamma_8^{(Y)} + \nonumber \\
&\mbox{}&\hspace{0.7cm} \Gamma_2^{(Y)}\Gamma_9^{(Y)} + \Gamma_3^{(Y)}\Gamma_8^{(Y)} + \nonumber \\
&\mbox{}&\hspace{0.7cm} \Gamma_1^{(Y)}\Gamma_9^{(Y)} + \Gamma_3^{(Y)}\Gamma_7^{(Y)} + \nonumber \\
&\mbox{}&\hspace{0.7cm} \Gamma_4^{(Y)}\Gamma_9^{(Y)} + \Gamma_6^{(Y)}\Gamma_7^{(Y)} + \nonumber \\
&\mbox{}&\hspace{0.7cm} \Gamma_2^{(Y)}\Gamma_6^{(Y)} + \Gamma_3^{(Y)}\Gamma_5^{(Y)} + \nonumber \\
&\mbox{}&\hspace{0.7cm} \Gamma_1^{(Y)}\Gamma_6^{(Y)} + \Gamma_3^{(Y)}\Gamma_4^{(Y)}  \Big]. 
\label{eq:Appendix-logical-Y-error-rate}
\end{eqnarray}

\section{Calculation of two-time correlator Eq.~\eqref{eq:two-time-corr-Ctilde}}
\label{Appendix-D}
We discuss the calculation of the following correlator 
\begin{eqnarray}
\label{eq:Appendix-two-time-corr-Ctilde}
&\mbox{}&\big\langle \tilde{\mathcal{C}}_x^{(1)}(t)\,\tilde{\mathcal{C}}_x^{(1)}(0)\big\rangle =\int_{-\infty}^tdt_1 \int_{-\infty}^tdt_2 \int_{-\infty}^tdt_3 \int_{-\infty}^0dt'_1\nonumber\\
&\mbox{}&\hspace{2cm}\int_{-\infty}^0dt'_2 \int_{-\infty}^0dt'_3\; K_6(t_1,t'_1,t_2,t'_2,t_3,t'_3) \nonumber\\
&\mbox{}&\hspace{2cm}\times \frac{e^{-\frac{t-t_1}{\tau_{\rm c}}-\frac{t-t_2}{\tau_{\rm c}} -\frac{t-t_3}{\tau_{\rm c}} + \frac{t'_1}{\tau_{\rm c}} + \frac{t'_2}{\tau_{\rm c}} + \frac{t'_3}{\tau_{\rm c}}}}{\tau_{\rm c}^6}, 
\end{eqnarray}
where 
\begin{eqnarray}
\label{eq:Appendix-K6-def}
&\mbox{}&\hspace{-0.75cm}K_6(t_1,t'_1,t_2,t'_2,t_3,t'_3)=\nonumber\\
&\mbox{}&\hspace{-0.75cm} \big\langle I_{X_3^{\rm g}}(t_3)\,I_{X_3^{\rm g}}(t'_3)\, I_{X_{13}^{\rm g}}(t_2)\, I_{X_{13}^{\rm g}}(t'_2)\,  I_{X_1^{\rm g}}(t_1)\,I_{X_1^{\rm g}}(t'_1)\big\rangle 
\end{eqnarray}
\newline
is a six-time correlator for the measurement signals $I_{X_1^{\rm g}}(t)$, $I_{X_{13}^{\rm g}}(t)$ and $I_{X_{3}^{\rm g}}(t)$, which are defined in Eq.~\eqref{eq:Ik-Ql}. We remind that the noises for such measurement signals are, respectively, denoted by $\xi_7(t)$, $\xi_8(t)$ and $\xi_9(t)$---see Section~\ref{evolution}.

To evaluate $K_{6}$, we use formula~\eqref{eq:KN-result}. This formula, however, does not include singular contributions that arise when the times of two measurement signals (from the same detector) in $K_6$ coincide~\cite{Atalaya2018a}. Specifically, such singular contributions occur when the pair of times in each group $(t_1,t_1')$, $(t_2,t_2')$ or $(t_3, t_3')$  coincide. For example, if only $t_3$ and $t_3'$ coincide, $K_6= (\tau_{\rm m}/\delta t) \langle I_{X_{13}^{\rm g}}(t_2)\,I_{X_{13}^{\rm g}}(t'_2) I_{X_1^{\rm g}}(t_1)\,I_{X_1^{\rm g}}(t'_1)\rangle_{\rm nc.t.}$  since we may approximately replace the product $I_{X_3^{\rm g}}(t_3)\,I_{X_3^{\rm g}}(t'_3)$ in Eq.~\eqref{eq:Appendix-K6-def} by $\tau_{\rm m}[\xi_9(t_3)]^2$, which is equal to $\tau_{\rm m}/\delta t$ after averaging over the noise $\xi_9(t)$. Here, $\delta t$ is an infinitesimal discretization timestep and the subscript ``nc.t.'' means ``not coinciding times''; that is, the times of the output signals (from the same detector) inside the angular bracket ($\langle \cdot \rangle_{\rm nc.t.}$) do not coincide. The remaining (non-singular) four-time correlator $\langle I_{X_{13}^{\rm g}}(t_2)\,I_{X_{13}^{\rm g}}(t'_2) I_{X_1^{\rm g}}(t_1)\,I_{X_1^{\rm g}}(t'_1)\rangle_{\rm nc.t.}$ is evaluated using formula~\eqref{eq:KN-result}. Similarly, $K_6= (\tau_{\rm m}/\delta t)^2\langle I_{X_1^{\rm g}}(t_1)\,I_{X_1^{\rm g}}(t'_1)\rangle_{\rm nc.t.}$ if $(t_2,t_2')$ and $(t_3,t_3')$ are the only pairs of coinciding times, and $K_6= (\tau_{\rm m}/\delta t)^3$ if $(t_1,t_1')$,  $(t_2,t_2')$ and $(t_3,t_3')$ are all pairs of coinciding times. Note that each singular factor $(\tau_{\rm m}/\delta t)$ replaces one of the integrals in Eq.~\eqref{eq:Appendix-two-time-corr-Ctilde} by a factor $\tau_{\rm m}$; for example, if only $t_3$ and $t_3'$ coincide, the integral over $t_3'$  is replaced by the factor $\tau_{\rm m}$, while the integral over $t_3$ becomes trivial and gives an additional factor equal to $\int_{-\infty}^0dt_3\, \exp(2t_3/\tau_{\rm c})=\tau_{\rm c}/2$. In other words, one may say that each singular factor $(\tau_{\rm m}/\delta t)$ due to pairs of coinciding times $(t_n,t_n')$ in Eq.~\eqref{eq:Appendix-K6-def} effectively replaces the integrals over $t_n$ and $t_n'$ by the factor $\tau_{\rm m}\tau_{\rm c}/2$. Taking into account the considerations just described, the sought correlator~\eqref{eq:Appendix-two-time-corr-Ctilde} can be written as 
\begin{widetext}
\begin{align}
\label{eq:Appendix-derivation-1}
\big\langle \tilde{\mathcal{C}}_x^{(1)}(t)\,\tilde{\mathcal{C}}_x^{(1)}(0)\big\rangle =&\; \int_{-\infty}^tdt_1 \int_{-\infty}^tdt_2 \int_{-\infty}^tdt_3 \int_{-\infty}^0dt'_1\;\int_{-\infty}^0dt'_2 \int_{-\infty}^0dt'_3\; 
\big\langle I_{X_3^{\rm g}}(t_3)\,I_{X_3^{\rm g}}(t'_3)\, I_{X_{13}^{\rm g}}(t_2)\, I_{X_{13}^{\rm g}}(t'_2)\,  I_{X_1^{\rm g}}(t_1)\nonumber\\
&\;\times I_{X_1^{\rm g}}(t'_1)\big\rangle_{\rm nc.t.}\,\frac{e^{-\frac{t-t_1}{\tau_{\rm c}}-\frac{t-t_2}{\tau_{\rm c}} -\frac{t-t_3}{\tau_{\rm c}} + \frac{t'_1}{\tau_{\rm c}} + \frac{t'_2}{\tau_{\rm c}} + \frac{t'_3}{\tau_{\rm c}}}}{\tau_{\rm c}^6}\nonumber\\
& \hspace{-1cm}+ \frac{\tau_{\rm m}\tau_{\rm c}}{2}\int_{-\infty}^tdt_1 \int_{-\infty}^tdt_2 \int_{-\infty}^0dt'_1\;\int_{-\infty}^0dt'_2\; \big\langle I_{X_{13}^{\rm g}}(t_2)\, I_{X_{13}^{\rm g}}(t'_2)\,  I_{X_1^{\rm g}}(t_1)\, I_{X_1^{\rm g}}(t'_1)\big\rangle_{\rm nc.t.}\,\frac{e^{ -\frac{t}{\tau_{\rm c}}-\frac{t-t_1}{\tau_{\rm c}}-\frac{t-t_2}{\tau_{\rm c}} + \frac{t'_1}{\tau_{\rm c}} + \frac{t'_2}{\tau_{\rm c}}}}{\tau_{\rm c}^6}\nonumber\\
& \hspace{-1cm}+ \frac{\tau_{\rm m}\tau_{\rm c}}{2}\int_{-\infty}^tdt_3 \int_{-\infty}^tdt_2 \int_{-\infty}^0dt'_3\;\int_{-\infty}^0dt'_2\; \big\langle I_{X_{13}^{\rm g}}(t_2)\, I_{X_{13}^{\rm g}}(t'_2)\,  I_{X_3^{\rm g}}(t_3)\, I_{X_3^{\rm g}}(t'_3)\big\rangle_{\rm nc.t.}\,\frac{e^{ -\frac{t}{\tau_{\rm c}}-\frac{t-t_3}{\tau_{\rm c}}-\frac{t-t_2}{\tau_{\rm c}} + \frac{t'_3}{\tau_{\rm c}} + \frac{t'_2}{\tau_{\rm c}}}}{\tau_{\rm c}^6}\nonumber\\
& \hspace{-1cm}+ \frac{\tau_{\rm m}\tau_{\rm c}}{2}\int_{-\infty}^tdt_1 \int_{-\infty}^tdt_3 \int_{-\infty}^0dt'_1\;\int_{-\infty}^0dt'_3\; \big\langle I_{X_{3}^{\rm g}}(t_3)\, I_{X_{3}^{\rm g}}(t'_3)\,  I_{X_1^{\rm g}}(t_1)\, I_{X_1^{\rm g}}(t'_1)\big\rangle_{\rm nc.t.}\,\frac{e^{-\frac{t}{\tau_{\rm c}}-\frac{t-t_1}{\tau_{\rm c}} -\frac{t-t_3}{\tau_{\rm c}} + \frac{t'_1}{\tau_{\rm c}} + \frac{t'_3}{\tau_{\rm c}}}}{\tau_{\rm c}^6}\nonumber\\
& \hspace{-1cm}+ \left(\frac{\tau_{\rm m}\tau_{\rm c}}{2}\right)^2\int_{-\infty}^tdt_1 \int_{-\infty}^0dt'_1\; \big\langle I_{X_1^{\rm g}}(t_1)\, I_{X_1^{\rm g}}(t'_1)\big\rangle_{\rm nc.t.}\,\frac{e^{-\frac{2t}{\tau_{\rm c}}-\frac{t-t_1}{\tau_{\rm c}}+ \frac{t'_1}{\tau_{\rm c}}}}{\tau_{\rm c}^6}\nonumber\\
& \hspace{-1cm}+ \left(\frac{\tau_{\rm m}\tau_{\rm c}}{2}\right)^2\int_{-\infty}^tdt_3 \int_{-\infty}^0dt'_3\; \big\langle I_{X_3^{\rm g}}(t_3)\, I_{X_3^{\rm g}}(t'_3)\big\rangle_{\rm nc.t.}\,\frac{e^{-\frac{2t}{\tau_{\rm c}}-\frac{t-t_3}{\tau_{\rm c}}+ \frac{t'_3}{\tau_{\rm c}}}}{\tau_{\rm c}^6}\nonumber\\
& \hspace{-1cm}+ \left(\frac{\tau_{\rm m}\tau_{\rm c}}{2}\right)^2\int_{-\infty}^tdt_2 \int_{-\infty}^0dt'_2\; \big\langle I_{X_{13}^{\rm g}}(t_2)\, I_{X_{13}^{\rm g}}(t'_2)\big\rangle_{\rm nc.t.}\,\frac{e^{-\frac{2t}{\tau_{\rm c}}-\frac{t-t_2}{\tau_{\rm c}}+ \frac{t'_2}{\tau_{\rm c}}}}{\tau_{\rm c}^6}\nonumber\\
& \hspace{-1cm}+  \left(\frac{\tau_{\rm m}\tau_{\rm c}}{2}\right)^3\frac{e^{-\frac{3t}{\tau_{\rm c}}}}{\tau_{\rm c}^6}.
\end{align}
\end{widetext}
The averages inside the integrands of Eq.~\eqref{eq:Appendix-derivation-1} are evaluated again using the result~\eqref{eq:KN-result}. The calculation of the first four lines in Eq.~\eqref{eq:Appendix-derivation-1} is particularly cumbersome since we have to divide the integration domains according to all possible time orderings of the integration variables since application of Eq.~\eqref{eq:KN-result} requires time ordering, as discussed in Section~\ref{sec:optimal-tauc} for the calculation of the three-time correlator $K_3$.  Note that lines 2 and 3 in Eq.~\eqref{eq:Appendix-derivation-1} yield the same contribution in the case of symmetric measurement strengths for all detectors. We eventually find the result~\eqref{eq:two-time-corr-Ctilde-result} with $R_i$ given by (see Mathematica file at Supplemental Information~\cite{SM})

\begin{subequations}
\label{eq:Appendix-derivation-2}
\begin{widetext} 
\begin{align}
&R_1 =\; \frac{6 s (s+1)^2 (2 s+1)}{D(s,\eta)} \big[(12 (4 {\eta}^2+8 {\eta}+5) s^{12}+8 (208 {\eta}^2+234 {\eta}+135) s^{11}+3 (4748 {\eta}^2+4756 {\eta}+2507) s^{10} \nonumber\\
&\hspace{1cm} +(47680 {\eta}^2+51468 {\eta}+23598) s^9+(48220 {\eta}^2+72963 {\eta}+18132) s^8-6 (12824 {\eta}^2+9657 {\eta}+14832) s^7 \nonumber\\
&\hspace{1cm} -(198464 {\eta}^2+302379 {\eta}+250755) s^6-18 (5520 {\eta}^2+14779 {\eta}+9321) s^5+(36352 {\eta}^2+61260 {\eta}+202458) s^4\nonumber\\
&\hspace{1cm} +12(2080 {\eta}^2+10858 {\eta}+27801) s^3-24 (720 {\eta}-2741) s^2-720 (40 {\eta}+141) s-43200\big], 
\end{align}
\begin{align}
&R_2=\; -\frac{12s{\eta}(s-1)^2(s+2)^2}{D(s,\eta)} \big[8 s^5+84 s^4+278 s^3+279 s^2-70 s-75\big]\big[2 (34 {\eta}+9) s^5+(254 {\eta}+147) s^4\nonumber \\
&\hspace{1cm} +18 (16 {\eta}+23) s^3 + (96 {\eta}+531) s^2+318 s+72\big],  \end{align}
\begin{align}
&R_3 =\; \frac{6 (s+1)^2}{D(s,\eta)} \big[8 s^4+4 s^3-42 s^2-s+10\big] \big[4 (28 {\eta}^2-3) s^9+4 (172 {\eta}^2-27 {\eta}-54) s^8-3 (96 {\eta}^2+526 {\eta}+523) s^7 \nonumber \\
&\hspace{1cm}-16 (605 {\eta}^2+552 {\eta}+366) s^6-6 (3800 {\eta}^2+4017 {\eta}+1925) s^5-12 (1656 {\eta}^2+2819 {\eta}+812) s^4 \nonumber\\
&\hspace{1cm} -3(1920 {\eta}^2+7664 {\eta}-1173) s^3-24 (240 {\eta}-569) s^2+9612 s+2160\big],
\end{align}
\begin{align}
&R_4=\; \frac{6 {\eta} s (s+1)^2}{D(s,\eta)} \big[8 s^6+84 s^5+226 s^4-181 s^3-1197 s^2-590 s+600\big] \big[4 (7 {\eta}+3) s^5+4 (47 {\eta}+24) s^4+3 (88 {\eta}+93) s^3\nonumber \\
&\hspace{1cm} +(96 {\eta}+357) s^2+192 s+36\big], 
\end{align}
\begin{align}
&R_5 =\; -\frac{3 (s-1)^2}{D(s,\eta)} \big[2 s^5+17 s^4+30 s^3-53 s^2-152 s-60\big] \big[12 (2 {\eta}+1)^2 s^8 + 8 (76 {\eta}^2+72 {\eta}+21) s^7+(1876 {\eta}^2+2676 {\eta}+981) s^6 \nonumber \\
&\hspace{1cm} +(2024 {\eta}^2+6120 {\eta}+3138) s^5+(672 {\eta}^2+7236 {\eta}+6042) s^4+96 (44 {\eta}+75) s^3+15 (64 {\eta}+347) s^2  +2094 s+360\big],
\end{align}
\end{widetext}
\end{subequations}

where $s=2\tau_{\rm c}\tau_{\rm coll}^{-1}$, $\tau_{\rm coll}=\Gamma_{\rm m}^{-1}=2\tau_{\rm m}\eta$, and the denominator in Eqs.~\eqref{eq:Appendix-derivation-2} reads as 
\begin{widetext}
\begin{align}
D(s,\eta) = 36 s^2{\eta}^2(s-1)^2  (s+1)^4 (s-2) (s+2)^2 (s+3) (s+4) (s+5) (2 s-1) (2 s+1)^2 (2 s+3) (2 s+5).
\end{align}
\end{widetext}


\begin{thebibliography}{30}
\bibitem{SM} Mathematica file that calculates the integrals of Eq.~\eqref{eq:Appendix-derivation-1}.
\end{thebibliography}
\end{document}